\documentclass[rmp,aps,showpacs,twocolumn,nofootinbib,amsmath,amssymb]{revtex4}

\usepackage{graphicx}

\usepackage{amsmath}

\usepackage{color}

\def\inbar{\,\vrule height1.5ex width.4pt depth0pt}
\def\IR{\relax{\rm I\kern-.18em R}}
\def\IC{\relax\hbox{$\inbar\kern-.3em{\rm C}$}}

\newcommand{\ket}[1]{|{#1}\rangle}
\newcommand{\bra}[1]{\langle{#1}|}                     
 
 \newcommand{\kbold}{\mathbf{k}}
 \def\mum{$\mu$m}

\def\kbold{\mathbf{k}}
\def\rjbold{\mathbf{r}_j}

\def\Rb87{$^{87}\text{Rb}$}
\def\0{\ket{0}}
\def\1{\ket{1}}

\def \expect#1{{\left \langle #1 \right\rangle}}

\def\forster{F\"{o}rster}

\def\bshift{{\sf B}}

\newcommand\rmat[2]{{\expect{r}_{#1}^{#2}}}

\def\be{\begin{equation}}
\def\ee{\end{equation}}
\newcommand \bse{\begin{subequations}}
\newcommand \ese{\end{subequations}}
\newcommand \bml{\begin{subequations}\begin{eqnarray}}
\newcommand \eml{\end{eqnarray}\end{subequations}}
\newcommand \bfepsilon{{\boldsymbol \epsilon}}

\begin{document}
\title{Quantum information with Rydberg atoms}

\author{M. Saffman and T. G. Walker}
\affiliation{Department of Physics,
University of Wisconsin, 1150 University Avenue,  Madison, Wisconsin 53706}
\author{K. M{\o}lmer}
\affiliation{Lundbeck Foundation Theoretical Center for
Quantum System Research, Department of Physics and Astronomy,
University of Aarhus, DK-8000 \AA rhus C, Denmark }

\begin{abstract}
Rydberg atoms with principal quantum number $n\gg 1$ have exaggerated atomic properties including
dipole-dipole interactions that scale as $n^4$ and radiative lifetimes that scale as $n^3.$
It was proposed a decade ago  to take advantage of these properties to implement quantum gates between neutral atom qubits.  The availability of a strong, long-range interaction that can be coherently turned on and off
is  an enabling resource for a wide range of quantum information tasks stretching far beyond the original gate proposal. Rydberg enabled capabilities include long-range two-qubit gates,
collective encoding of multi-qubit registers, implementation of robust light-atom quantum interfaces,
and the potential for  simulating quantum many body physics. We review the advances of the last decade, covering both theoretical and experimental aspects of Rydberg mediated quantum information processing.
\end{abstract}


\date{\today}
\maketitle
\tableofcontents


\section{Introduction}
\label{sec.introduction}

The field of quantum information processing is currently attracting intense interest. This is fueled by the promise of applications and by rapid experimental progress. The most advanced experimental demonstrations at this time include trapped ions\cite{Blatt2008}, linear optics\cite{Kok2007}, superconductors\cite{Clarke2008,DiCarlo2009}, and  quantum dots in semiconductors\cite{Li2003,Petta2005,Barthel2009}.
Trapped ion qubits have reached the most advanced state of sophistication and have been used to demonstrate high fidelity gates and small algorithms\cite{Blatt2008}.
Neutral atom qubits represent another promising
approach\cite{Bloch2008}. They share many features in common with trapped ion systems
including long lived encoding of quantum information in atomic hyperfine states, and the possibility of manipulating and measuring the qubit state using resonant laser pulses.

Neutral atoms distinguish themselves from ions when we consider  their state dependent interaction properties, which are essential for implementing two-qubit quantum gates.  Figure  1
shows the dependence of the two-particle interaction strength on separation $R$ for singly charged ions, ground state neutral atoms, and
Rydberg atoms. The interaction of ground state atoms is dominated by $1/R^6$ van der Waals forces at short range, and $1/R^3$ magnetic dipole-dipole forces beyond about 30 nm. At spacings greater than
 $1~\mu\rm m$ the interaction is weak, less than $1~\rm Hz$ in frequency units, which implies that an array of neutral atom qubits
can be structurally stable. On the other hand, excitation of Rb atoms to the $100s$ Rydberg level results in a very strong interaction that
has resonant dipole-dipole character, scaling as $1/R^3$, at short distances and van der Waals character, scaling as $1/R^6$, at long distances. As will be discussed in more detail in Sec. \ref{sec.interactions} the characteristic length scale $R_c$ where the Rydberg interaction changes character depends
on the principal quantum number $n$. For the $100s$ state the crossover length is close to $R_c=9.5~\mu\rm m$, and at this length scale the ratio of the Rydberg interaction to the ground state interaction is approximately $10^{12}$.

\begin{figure}[!t]
\centering {\includegraphics[width=8.5cm]{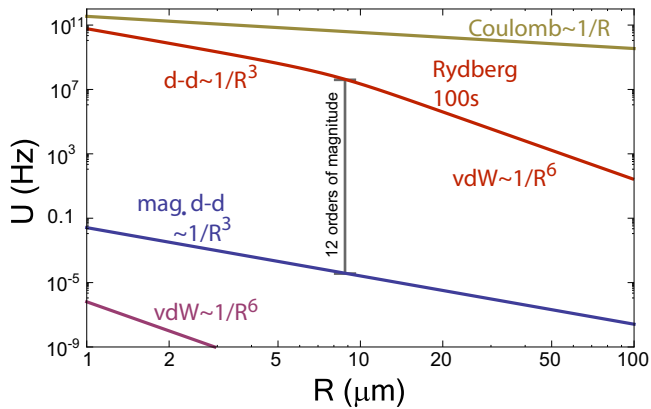}
\caption{(Color online) Two-body interaction strength for  ground state Rb atoms, Rb atoms excited to the $100s$ level, and ions. }\label{fig.Uatom}}
\end{figure}

The applicability of Rydberg atoms for quantum information processing, which is the central topic of this review, can be traced to the fact that
the two-atom interaction  can be turned on and off with a contrast of 12 orders of magnitude. The ability to control the interaction strength over such a wide range appears unique to the Rydberg system. We may compare this with trapped ions whose Coulomb interaction  is much stronger, but is always present.
The strong Coulomb interaction is beneficial for implementing high fidelity
gates\cite{Benhelm2008b} but the always on character of the interaction makes the task of establishing a many qubit register appear more difficult than it may be for an array of weakly interacting neutral atoms.
Several approaches to scalability in trapped ion systems are being explored including the development of
complex, multi-zone trap technologies \cite{Seidelin2006}, and anharmonic traps\cite{Lin2009}.
We note that some of the attractive features of Rydberg mediated interactions, may also be applicable to trapped ion
systems\cite{Muller2008}.

\subsection{Rydberg mediated quantum gates}

\begin{figure}[!t]
\centering {\includegraphics[width=8.5cm]{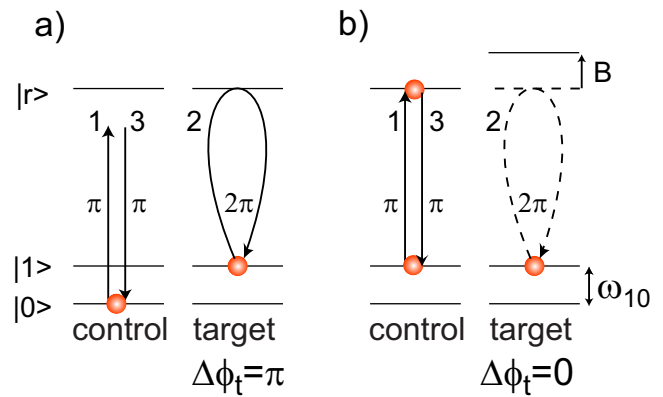}
\caption{(Color online) Rydberg blockade controlled phase gate operating on input states a) $\ket{01}$ and b) $\ket{11}.$  Quantum information is stored in the basis states $\ket{0}, \ket{1}$ and state $\ket{1}$ is coupled to a Rydberg level $\ket{r}$ with excitation Rabi frequency $\Omega$. The controlled phase gate is implemented with a three pulse sequence: 1) $\pi$ pulse on
control atom $\ket{1}\rightarrow \ket{r},$ 2) $2\pi$ pulse on target atom $\ket{1}\rightarrow\ket{r}\rightarrow\ket{1}$ and
3) $\pi$ pulse on control atom $\ket{r}\rightarrow\ket{1}.$ Panel a) shows the case where the control atom starts in $\ket{0}$ and is not Rydberg excited so there is no blockade, while panel b) shows the case where the control atom is in $\ket{1}$ which is Rydberg excited leading to blockade of the target atom excitation.}  \label{fig.cphase}}
\end{figure}

The idea of using dipolar Rydberg interactions for neutral atom quantum gates  was introduced in 2000 \cite{Jaksch2000} and quickly extended to a mesoscopic regime of many atom ensemble qubits \cite{Lukin2001}.
The basic idea of the Rydberg blockade two-qubit gate is shown in Fig. \ref{fig.cphase}. When the initial two-atom state is $\ket{01}$ (Fig. \ref{fig.cphase}a) the control atom is not coupled to the Rydberg level and the target atom picks up a $\pi$ phase shift. When
the initial state is $\ket{11}$ both atoms are coupled to the Rydberg level. In the ideal case when the two-atom ``blockade" shift $\bshift$ due to the Rydberg interaction is large compared to the excitation Rabi frequency $\Omega$, excitation of the target atom is blocked and it  picks up no phase shift.
The evolution matrix expressed in the computational basis $\{\ket{00},\ket{01},\ket{10},\ket{11}\}$ is
\be
U=\begin{pmatrix}
1&0&0&0\\
0&-1&0&0\\
0&0&-1&0\\
0&0&0&-1
\end{pmatrix}
\label{eq.CZ}
\ee
which is a controlled Z ($C_Z$) gate. As is well known\cite{Nielsen2000} the  $C_Z$ gate can be readily
converted into a controlled-NOT (CNOT) gate  by including $\pi/2$ rotations between $\ket{0}\leftrightarrow\ket{1}$ on the  target atom before and after the interaction. The
CNOT gate together with  single qubit operations form a set of universal gates for quantum computing\cite{Nielsen2000}.

If the Rydberg levels $\ket{r}$ were stable the approach of Fig. \ref{fig.cphase} would enable
gates with arbitrarily high fidelity. Rydberg states of real atoms have a finite lifetime $\tau$ due to radiative decay. This leads to a tradeoff between fast excitation which minimizes spontaneous emission and slow excitation which maximizes blockade effectiveness. We will show in Sec. \ref{sec.gatesexp} that the minimum  error  $E$
of the blockade gate scales
as\cite{Saffman2005a} $E\sim 1/(\bshift\tau)^{2/3}$, and also analyze the performance of several alternative Rydberg-based gate protocols. We show that a demonstration of gates with
 errors below $E=0.001$ appears to be a realistic goal.

The blockade concept, while not the only means of performing Rydberg-mediated quantum information processing, is attractive for several reasons.  First, the gate fidelity is to first order independent of the blockade shift in the limit of large shifts.  It is therefore not necessary to control the value of the blockade shift, beyond ensuring that it is sufficiently large.  Second, the fidelity of the entanglement protocol depends only weakly on the
center of mass atomic motion so that sub-Doppler temperatures at the level of $\sim 50~\mu\rm K$ are sufficient for a high fidelity quantum gate\cite{Saffman2005a}.
Undesired entanglement with external, motional degrees of freedom is thereby suppressed.
 Third, the interactions are of sufficiently long range to allow gates between optically resolvable atoms without having to physically move them from place to place.

Numerous alternative proposals exist for two-atom quantum gates including
short range dipolar interactions\cite{Brennen1999}, ground state collisions\cite{Jaksch1999}, coupling of atoms to photons\cite{Pellizzari1995},
magnetic dipole-dipole interactions\cite{You2000}, optically controlled dipolar interactions\cite{Lukin2000} and
gates with delocalized qubits\cite{Mompart2003}. Many particle
entanglement mediated by collisional interactions
has been observed in optical lattice based experiments\cite{Mandel2003,Anderlini2007}, but to date
only the Rydberg interaction has been successfully applied to demonstration of a quantum gate between
two neutral atoms. We note that the Rydberg blockade
gate is inherently optimized for MHz-rate gate operations,
a significant advantage compared to alternative
neutral atom quantum gate proposals.

In  experiments described in Sec. \ref{sec.experiments} it was shown that coherent excitation and deexcitation of single ground-state atoms to
Rydberg levels is feasible\cite{Johnson2008}, and that Rydberg blockade can be observed between two atoms in spatially separated volumes\cite{Urban2009,Gaetan2009}.
The most recent experiments in Madison and Palaiseau\cite{Isenhower2010,Wilk2010}
have shown that the blockade interaction can be used to implement a two-qubit CNOT gate
and to create entanglement between pairs of atoms.
As will be discussed in Sec. \ref{sec.gatesexp} these initial demonstrations  generated relatively weak entanglement and suffered from excess atom loss. Nevertheless
it is realistic to expect significant improvements in the next few years, as  experimental techniques are further refined.

In this review we focus on phenomena that involve quantum states stored in ground hyperfine levels of cold trapped atoms. Interactions between atoms rely on transient excitation of atom pairs to low angular momentum Rydberg states using laser fields. There is never any long term storage of information in the Rydberg levels. There is also an alternative
approach to coherent quantum dynamics  that utilizes strong coupling between Rydberg atoms and microwave photons inside high finesse resonators\cite{Raimond2001,Walther2006}. In this cavity
quantum electrodynamics (CQED) approach relatively long lived microwave cavity fields may be thought of as qubits, with moving atoms serving both to couple the qubits and to control the preparation of non-classical field states\cite{Bernu2008}. Direct interaction between the atoms is never invoked.

 A series of beautiful  experiments  elucidating the interaction between single atoms and single photons\cite{Guerlin2007}, and leading to the creation of two-atom entanglement\cite{Hagley1997}, have been performed using so-called
`circular' Rydberg states with maximal magnetic quantum number
$|m|=n-1$. These states have radiative lifetimes that are close to one-thousand times longer than low angular momentum states of the same $n$. The long coherence times together
with the large dipole moments between states of neighboring $n$ have enabled creation of atom-photon entanglement by shooting Rydberg atoms through microwave cavities. Introducing several atoms sequentially allows  atom-photon entanglement to be mapped onto atom-atom entanglement\cite{Hagley1997}, despite the fact that the atoms have no direct interaction.

The Rydberg CQED approach is quite complementary to the approach based on trapped atoms we focus on in this review. In the CQED experiments the stationary qubits are microwave field states, and two-qubit interactions are  mediated by strong coupling between microwave fields and long-lived circular Rydberg states. In contrast the trapped atom approach uses long-lived hyperfine qubits and direct interaction of relatively short-lived Rydberg states  to mediate the coupling between hyperfine qubits.  
Nevertheless it is quite possible that future developments will lead to a synthesis of elements of both approaches. There are ideas for trapping the circular states\cite{Hyafil2004}  which, if combined with the direct interaction approach, could lead to excellent gate fidelities due to the very large values of $\tau.$

\subsection{Rydberg coupled ensembles}

\begin{figure}[!t]
\centering {\includegraphics[width=5.cm]{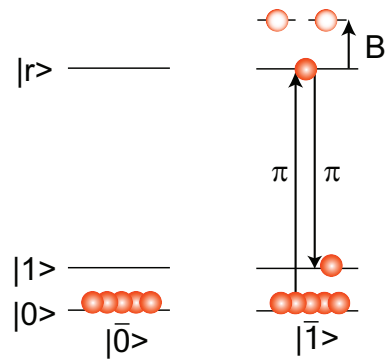}
\caption{(Color online) Ensemble qubits and rotations between the logical basis states $\{\ket{\bar 0},\ket{\bar 1}\}.$ Double excitation of the state $\ket{1}$ is prevented by the two-atom shift $B.$
}\label{fig.ensemblebit}}
\end{figure}

The blockade gate of Fig. \ref{fig.cphase} operates between two individual atoms. However, in a seminal paper
by Lukin and collaborators\cite{Lukin2001} it was shown that this can be directly extended
 to ensemble qubits, each consisting of $N$ atoms. The extension
 relies on the concept of collective Rydberg blockade whereby excitation of
a single atom to a Rydberg state can block the subsequent excitation of not just one, but a large number of atoms within the surrounding volume. As shown in Fig. \ref{fig.ensemblebit}
the blockade effect allows us to manipulate ensemble qubits in a very straightforward manner.
We define an $N$ atom logical $0$ by $\ket{\bar 0}=\prod_{i=1}^N \ket{0_i}$ and a logical $1$ by
the symmetric singly excited state $\ket{\bar 1}=\frac{1}{\sqrt N}\sum_{i=1}^N \ket{0 ... 1_i ... 0}.$ Single qubit rotations in the logical basis
$\{\ket{\bar 0},\ket{\bar 1}\}$ are performed with two-photon transitions via the Rydberg level $\ket{r}$
as seen in Fig. \ref{fig.ensemblebit}. Provided the blockade shift $\bshift$ is sufficiently large double excitation is prevented and we have a closed  two-level system with the collectively enhanced Rabi frequency $\Omega_N=\sqrt{N} \Omega.$
Two-qubit gates can then be performed between two ensemble qubits or between a single atom and an ensemble qubit
in a fashion that is completely analogous to the single atom protocol of Fig. \ref{fig.cphase}.

A requirement for the validity of the  ensemble qubit picture is that a blockade interaction is present for all atoms in the ensemble\cite{Lukin2001}.
The probability of unwanted double excitation is simply determined by the off-resonant excitation to the doubly-excited states that are shifted by an amount $\bshift$ due to the blockade effect.
In the limit of strong blockade, the probability of double excitation is
\be
P_2=\frac{N-1}{N}\frac{\Omega_N^2}{ 2 \bshift^2}
\label{eq.P2}
\ee
where the blockade shift is\cite{Walker2008}
\be
{1\over \bshift^2}=\frac{2}{N(N-1)}\sum_{\varphi}\sum_{i<j}\frac{|\kappa_{\varphi ij}|^2}{\Delta_{\varphi ij}^2}.
\label{eq.bshift}
\ee
Since there are generally a number of possible doubly-excited states $\ket{\varphi}$ for atoms $i$ and $j$, Eq. (\ref{eq.P2}) depends on the dipole-dipole energy shift $\Delta_{\varphi ij}$ of each state, and
on the associated Rabi coupling for exciting a
pair  of atoms to state $\ket{\varphi}$, parameterized by the dimensionless  factor
$\kappa_{\varphi ij}.$
 Equation~ (\ref{eq.bshift}) shows that the blockade shift is dominated by the weakest possible atom-atom interactions; in an electrical circuit analogy the blockade shift can be thought of as an impedance $\bshift^2$ that is formed by a parallel network of individual
impedances, each of size  $\sim (N^2/2) \left[ \sum_{i<j}|\kappa_{\varphi ij}|^2/\Delta_{\varphi ij}^2\right]^{-1}$.
Weak blockade on
any atom-pair state $\ket{\varphi}$ permits efficient two-atom  excitation
and acts to short-circuit the effectiveness of the
blockade process.
 A detailed accounting for the multitude of doubly-excited states and their interactions is therefore crucial to achieving a strong $N$ atom blockade. We review the relevant Rydberg physics in Sec. \ref{sec.interactions} and derive Eqs. (\ref{eq.P2},\ref{eq.bshift}) in Sec. \ref{sec.blockadetheory}.

The powerful idea of Rydberg blockade has led to a large amount of theoretical and experimental  work and has
been further developed far beyond the original proposals\cite{Jaksch2000,Lukin2001}.
Promising ideas exist  for such disparate tasks as deterministic single atom loading\cite{Saffman2002}, spin squeezing of atomic ensembles\cite{Bouchoule2002},   collective encoding of many qubit
registers\cite{Brion2007d,Saffman2008}, nonlocal gates by coupling to microwave
resonators\cite{Sorensen2004,Petrosyan2008}, and   long distance entanglement and quantum communication\cite{Saffman2002,Saffman2005b,Pedersen2009}. Rydberg blockade has also been proposed as a means of generating many particle entanglement\cite{Unanyan2002,Moller2008,Muller2009,Saffman2009b}, as well as a basis for dissipative
quantum many-body simulations\cite{Weimer2010}.

The difficulty of achieving perfect blockade has stimulated renewed theoretical interest in the classical problem of the structure and strength of the dipole-dipole interaction between atoms in excited states\cite{Gallagher1994,Gallagher2008}. Recent work has elucidated the transition from a near-resonant $1/R^3$ F\"orster interaction\cite{Forster1948} at short range to a long range $1/R^6$ van der Waals interaction\cite{Protsenko2002,Walker2005,Li2005}. In addition it was found that pairs of atoms can couple to non-interacting   ``F\"orster zero states" which evade
the Rydberg blockade\cite{Walker2005,Walker2008} unless special care is taken in the choice of atomic states and relative orientation. However, the physics of Rydberg blockade turns out to be subtle and contains some unexpected twists.
For example, recent work has shown\cite{Pohl2009} that the addition of a third atom can create a noninteracting
zero state that can be accessed by a weak
three-photon excitation, even when the two-photon excitation
is well blockaded (see Sec. \ref{sec.3atomblockade}).
Advances in the theoretical description of Rydberg blockade in many atom samples are reviewed in Sec. \ref{sec.collective}.

In parallel with the theoretical developments there has been a great deal of experimental activity in the field.
Experiments performed in the early 80's with atomic beams provided the first direct observations of  many body Rydberg interaction effects\cite{Raimond1981,Safinya1981,Gallagher1982}.
Using modern cold atom techniques
a number of recent experiments have reported the observation of excitation suppression due to Rydberg interactions
in small dense samples\cite{Tong2004,Singer2004,Afrousheh2006,Vogt2006,CubelLeibisch2005,CubelLeibisch2007,vanDitzhuijzen2008}.
 and collective effects in coherent
excitation\cite{Cubel2005,Reetz-Lamour2008a,Reetz-Lamour2008b,Heidemann2007,Heidemann2008,Raitzsch2008}.
Other recent work has explored novel quantum and nonlinear optical effects due to  interaction of light with Rydberg excited samples including  superradiance\cite{Wang2007,Day2008},
electromagnetically induced transparency  \cite{Mohapatra2007,Bason2008,Weatherill2008,Pritchard2009}, and four-wave mixing\cite{Brekke2008}.
These intriguing experiments, reviewed in Secs. \ref{sec.collective} and \ref{sec.quantum_optics}, have  shown the existence of long-range  Rydberg interactions, and in some cases signatures of blockade,  but have not yet reached the strong blockade regime of only a single atomic excitation that is crucial for quantum information applications of ensemble qubits.

\begin{figure}[!t]
\centering {\includegraphics[width=8.5cm]{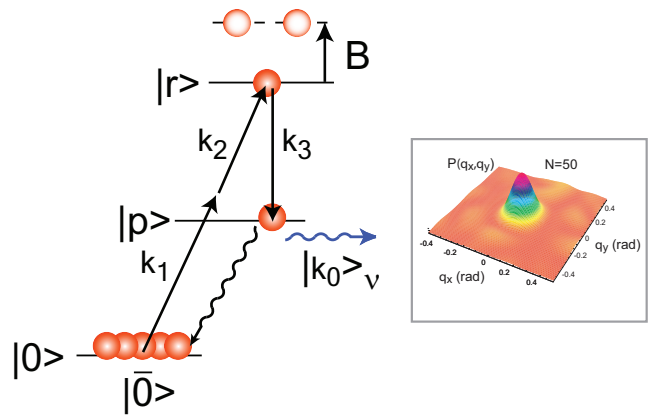}
\caption{(Color online)
Preparation of the symmetric singly excited state (\ref{eq:psi00}).
Radiative decay of this state produces a photon in the phase matched direction $\bf{k}_0$ as shown in the inset for $N=50$ atoms (adapted from \cite{Saffman2002}).
}
\label{fig.ensemble2}}
\end{figure}

Part of the current interest in Rydberg coupled ensembles stems from their use for  creating a light-atom quantum interface and long distance entanglement. It was recognized early on that Rydberg blockade could be used to deterministically create symmetric entangled states with unit excitation\cite{Lukin2001} and with controllable emission characteristics\cite{Saffman2002}. The basic idea is shown in Fig. \ref{fig.ensemble2}
where a multi-photon excitation is used to prepare an ensemble in the state

\begin{equation}\label{eq:psi00}
\ket{\Psi_0} = \frac{1}{\sqrt{N}}\sum_{j=1}^N e^{i\mathbf{k}_0 \cdot
\mathbf{r}_j} \ket{p_j}\otimes \ket{0}_{\nu}.
\end{equation}
Here $\ket{p_j}\equiv\ket{0 ... p_j ... 0}$  is shorthand for the state with atom $j$ at position $\bf{r}_j$ excited, the
other atoms in their ground state $|0\rangle$, and $\ket{0}_\nu$ is the vacuum state of the photon field. This state, containing a single  excitation collectively shared among the atoms, with a position dependent complex phase is  prepared with three laser fields. The first two with wave vectors  $\mathbf{k}_{1}, \bf{k}_2$ drive  a resonant excitation into a Rydberg state, where the blockade interaction prevents transfer of more than a single atom. A resonant $\pi$-pulse with wave vector $\mathbf{k}_3$ hereafter drives the atomic excitation into the excited state $|p\rangle$, producing the state (\ref{eq:psi00}) with  $\mathbf{k}_0=\mathbf{k}_1+\bf{k}_2-\mathbf{k}_3$. As will be discussed in Sec. \ref{sec.quantum_optics} spontaneous decay of this state preferentially creates
\begin{equation}\label{eq:psi01}
\ket{\Psi} = \ket{\bar 0}\otimes \ket{\bf{k}_0}_{\nu}.
\end{equation}
with all atoms in the ground state and a photon emitted at
the phase-matched  wavevector $\bf{k}_0.$

 The use of Rydberg blockade is significant since it would provide deterministic ensemble preparation
and single photon generation
without  the requirement of a deterministic single photon source which  is otherwise  needed
to remove the probabilistic character of non blockade based schemes\cite{Duan2001}.
 In Sec. \ref{sec.quantum_optics} we will discuss the use of Rydberg mediated light-atom interfaces for applications such as single photon on demand generation and entanglement generation between remote ensembles\cite{Saffman2005b,Pedersen2009}.

In the remainder of this review we expand upon the topics sketched above, presenting a comprehensive picture of the current theoretical and experimental state of the field of Rydberg-mediated quantum information processing.
We conclude in Sec. \ref{sec.conclusion} with a brief summary and outlook for the future.

\section{Rydberg atoms and their interactions}
\label{sec.interactions}

As described in the introduction, the success of blockade for quantum manipulation of atoms at large distances depends on the interactions between the atoms in their Rydberg states.  In this section we review the relevant Rydberg physics. Here, and in the rest of the paper apart from Sec. \ref{sec.1000},
we exclusively discuss the  situation for the heavy alkali atoms Rb and Cs. The alkalis in general are the most convenient elements for laser cooling, and the heavy alkalis are best suited for qubit encoding due to the large hyperfine splittings in the excited manifolds of the first resonance lines, which facilitates qubit measurements by light scattering.

\subsection{Properties of Rydberg atoms}
\label{sec.properties}

The properties of Rydberg atoms are nicely reviewed in several books \cite{Gallagher1994,Stebbings1983}
and also in some recent reviews\cite{Gallagher2008,Choi2007}.  We shall consider here those elements of most
importance for current blockade experiments, emphasizing low angular momentum states that are readily accessible via optical excitation from the atomic ground state.  In the absence of perturbing fields, the energy levels are represented by quantum numbers $n$, $l$, $s$, and $j$ denoting the traditional principle, orbital angular momentum, spin angular momentum, and total angular momentum quantum numbers.  The energy levels are accurately represented by
\begin{equation}
E_{nlj}=-\frac{\rm Ry}{ (n-\delta_{lj}(n))^2}
\end{equation}
where ${\rm Ry}=109737.315685$ cm$^{-1}$ is the Rydberg constant and the quantum defect $\delta_{lj}(n)$ is a slowly varying function of the principal quantum number.  The quantum defect is closely related to the scattering phase shift for low energy electron-ion scattering\cite{Fano1986}.  With the exception of low-lying s-states, hyperfine interactions are generally negligible.

\begin{figure}[!t]
\includegraphics[width=8.5cm]{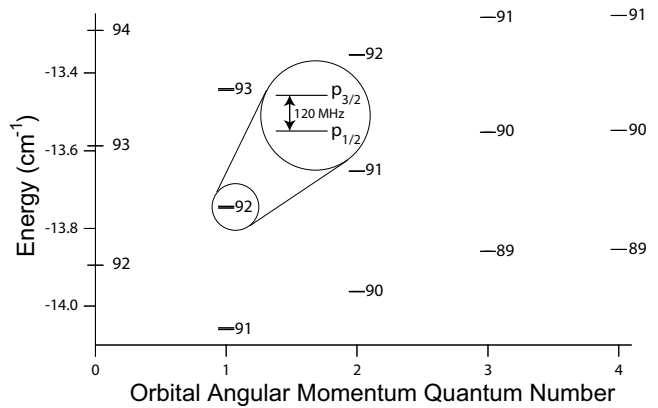}
\caption{Energy levels of Rb Rydberg states near $n=90$. \label{fig.elevels}
}
\end{figure}

High fidelity entanglement at large interatomic separations necessarily involves MHz scale operations due to the finite lifetime of Rydberg levels.  Viewed at MHz resolution, the fine structure splitting is large in the $p-$ and $d-$ states of the heavy alkalis up to $n=100$, while the $f-$state fine-structure levels are resolved below about $n=50$.
A portion of the Rb Rydberg energy level structure around $n=90$ is shown in Fig.~\ref{fig.elevels}.  The energy level structure of Rb and Cs has been recently discussed in detail in \cite{Walker2008}.

The wavefunctions of Rydberg states $\ket{nljm_j}=P_{nl}(r)\ket{ljm_j}/r$ can be conveniently calculated from quantum defect theory \cite{Seaton1958} using hypergeometric functions available in most numerical libraries.  The radial matrix elements
\begin{equation}
\rmat{nl}{n'l'} =\int {rP_{n'l'} (r)P_{nl} (r)dr}
\label{eq.radial}
\end{equation}
are of particular importance because they govern the interaction of the Rydberg atoms with both external and internal electric fields that shift the atomic energy levels. These matrix elements can be calculated
  by numerical integration of the quantum defect wavefunctions, by numerical integration directly from model potentials\cite{Marinescu1994}, or, perhaps most conveniently, using analytical formulae
derived by a semiclassical  WKB analysis\cite{Kaulakys1995}.

The radial matrix elements for dipole allowed transitions with $l'=l\pm 1$ between states with $n\gg 1$ are dominated by transitions between the states with the closest eigenenergies, and hence similar radial wavefunctions.
  These largest  matrix elements are generally of order $0.5-1.5$ $n^2a_0$, with $a_0$ the Bohr radius.  For example, the matrix elements from  $nl=90d$ to ($91p$, $92p$, $89f$, $88f$)  in Rb are $(1.3,0.76,1.3,0.80)n^2a_0$; the matrix elements to all other states are less than 0.17$n^2a_0$.  Thus in most situations the electric field shifts (for small fields) and the interatomic potentials can be understood by focusing on the effects of the 2 (for $s$-states) or 4 neighboring $l\pm1$ energy levels.

\begin{figure}[!t]
\includegraphics[width=3.2in]{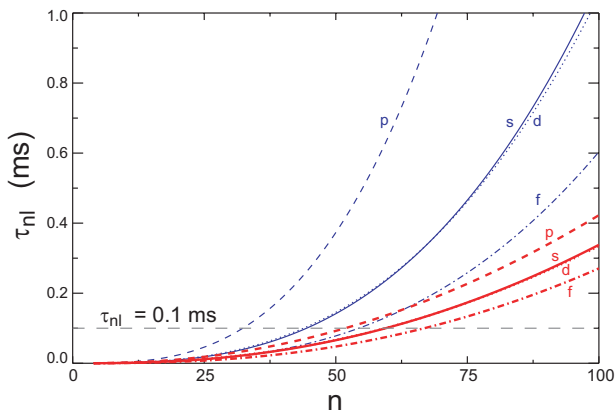}
\caption{(Color online) Excited state lifetimes due to radiative decay  for $T=0$ (blue) and $T=300$ K (red) for s,p,d, and f states of Rb.  From \protect\cite{Saffman2005a}.}
\label{fig.raddecay}
\end{figure}

The radiative and black-body lifetimes of the low angular momentum Rydberg states \cite{Gallagher1994} set fundamental limits on the fidelity of coherent operations.
If the 0 K lifetime is $\tau_{nl}^{(0)}$ the finite temperature lifetime is
\begin{equation}
\frac{1}{\tau_{nl}}=\frac{1}{\tau_{nl}^{(0)}}+\frac{1}{\tau_{nl}^{(\rm bb)}}
\label{eq.tau}
\end{equation}
where $\tau_{nl}^{(\rm bb)}$ is the finite temperature blackbody contribution.
The 0 K radiative lifetime can be parameterized by the expression\cite{Gounand1979a}
\begin{equation}
\tau_{nl}^{(0)}=\tau_l^{(0)} (n^*)^{\alpha_l}.
\end{equation}
with constants enumerated in Gallagher's book \cite{Gallagher1994}.
For all the  alkalis $\alpha_l\simeq 3.$ For large $n$ the blackbody rate can be written approximately as \cite{Gallagher1994}
\begin{equation}
\frac{1}{\tau_{nl}^{(\rm bb)}}=\frac{4\alpha^3 k_BT}{3\hbar n^2}
\label{eq.bbrate}
\end{equation} %
where $\alpha$ is the fine structure constant.
Equation (\ref{eq.bbrate}) includes transitions to continuum states so that it accounts for blackbody induced photoionization. Both black-body transfer and ionization have been recently
considered in detail  by \cite{Beterov2007,Beterov2009,Beterov2009b}.
Figure \ref{fig.raddecay} shows the radiative lifetime for low-$l$ states of Rb.
We see that for $n>50$  the $s$, $p$, $d$ and $f$ states have lifetimes greater than about $50~\mu\rm s$ at room temperature.  Thus, for high fidelity Rydberg manipulations, it is necessary to use MHz-scale operations.

The large dipole matrix elements also imply that Rydberg states are extremely sensitive to small low-frequency electric fields.  This may be a problem or a feature for coherent optical manipulation.  On the one hand, this sensitivity requires that electric fields be well-controlled to avoid frequency fluctuations.  On the other hand, it also makes it possible to tune the strength and angular dependence of Rydberg-Rydberg interactions using such fields.

For small DC electric fields $\sf E$ such that the dipole couplings $e\expect{r}\sf E$ are much less than the energy difference $\Delta E$ of the nearest opposite parity state, the Stark effect is quadratic and the shift is at most of order $-(e\expect{r}{\sf E})^2/\Delta E\sim \hbar^6n^7{\sf E}^2/m^3e^6$.  In fact the shift is often substantially smaller than this due to partial cancelation of states with equal and opposite $\Delta E$, as can be inferred from Fig.~\ref{fig.elevels}, that tend to cause shifts in opposite directions.  Even so, the electric field stability required to hold Stark shifts below 1 MHz is typically of order $0.01 (100/n)^{7/2}$ V/cm.

In higher electric fields, mixing of opposite parity states gives the atom an electric dipole moment of order $n^2 e a_0$ and hence a linear Stark effect.  This may be desirable to get the strongest possible Rydberg-Rydberg interactions (see Sect. \ref{sec.fields}), but stability requirements become more problematic, about $10^{-4}(100/n)^2$ V/cm for a 1 MHz shift.

The response of Rydberg atoms to optical frequency fields is primarily through AC Stark shifts (see Sec. ~\ref{sec.Rydbergtrap}) and photoionization.  Since the Rydberg electron spends the vast majority of its time far from the ionic core, to a good approximation it is a free electron, and in a high frequency field it therefore feels a repulsive force from the ponderomotive potential \cite{Dutta2000}.  While inside the core, however, it can absorb a photon, resulting in photoionization.  In the mK deep optical traps that have been used to date for blockade-based quantum operations, the photoionization rates can approach $10^5$/s \cite{Johnson2008,Saffman2005a}.  A detailed discussion of photoionization, including the wavelength dependence, has been presented by \cite{Potvliege2006}.

\subsection{From F\"orster to van der Waals}
\label{sec.d-d}

Successful Rydberg-mediated entanglement of atoms  is only possible due to the large interatomic 
 potentials that arise from the large dipole moments of Rydberg atoms.  As described below, the Rydberg blockade concept requires that the Rydberg-Rydberg interaction be much stronger than the Rabi coupling of the Rydberg atoms.  When this condition, to be quantified below, is satisfied, the entanglement produced is insensitive to first order in the Rydberg-Rydberg interaction.  Thus the strength of the interaction does not have to be precisely controlled.  This in turn lessens the requirements on atomic position and temperature control.

Due to the extreme electric field sensitivities of Rydberg states, if possible it is desirable to get the strongest possible interactions in the absence of applied fields.  In this section we discuss the properties of dipole-dipole interactions between Rydberg atoms in this limit.  This limit has been discussed in several recent papers \cite{Flannery2005,Singer2005a,Walker2005, Reinhard2007, Stanojevic2006,Stanojevic2008,Walker2008}.

At interatomic distances $R\gg n^2 a_0$ separating two Rydberg atoms $\bf A$ and $\bf B$, the leading electrostatic interaction is the dipole-dipole interaction
\be
V_{\rm dd}=\frac{e^2}{ R^3}\left({\bf a}\cdot {\bf b}-3{\bf a}\cdot  \hat{\bf R} \hat {\bf R}\cdot {\bf b}\right)
\ee
where $\bf a$ and $\bf b$ are the positions of the two Rydberg electrons measured from their respective nuclei. At such large distances, overlap between the atoms can be neglected.

\begin{figure}[!t]
\includegraphics[width=2.in]{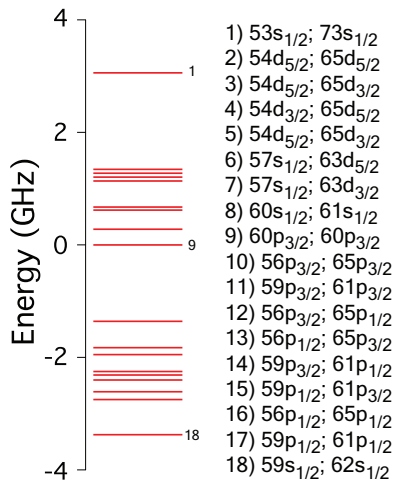}
\caption{(Color online) Two-atom energy levels connected to the $\ket{60p_{3/2}60p_{3/2}}$  state by the dipole-dipole interaction.}
\label{fig.elevels2}
\end{figure}

In most realizations of blockade, the two atoms are excited by light to the same fine-structure level, so that the two atom state for $R=\infty$ can be written
\be
\ket{\psi_2}=\ket{\psi_A\psi_B}=\ket{\psi_{nlj}\psi_{nlj}}
\ee
and the dipole-dipole interactions experienced by atoms in this state are of primary 
interest\footnote{There are, however, cases where excitation of pairs with the same $l,j$ but different $n$ can lead to stronger interactions\cite{Han2009}.}.
In the absence of external fields, this state has a degeneracy of $(2j+1)^2$, where $j$ is the total electronic angular momentum.
The dipole-dipole interaction causes transitions to other two-atom states where the  angular momentum quantum numbers of each electron obey the usual dipole selection rules, namely $l_a,l_b=l\pm 1$, $j_a,j_b=j\pm 0,1$. Including continuum states, there are an infinite number of such states, but in practice the dipole-dipole interaction is dominated by a small number of the closest two-atom states of this type.   To illustrate, Fig.~\ref{fig.elevels2} shows the energy level structure centered around the $\ket{60p_{3/2}60p_{3/2}}$ state of Rb
at zero relative energy.
If we restrict changes in the principal quantum numbers to at most $\pm 8$ there are 18 two-atom states within $\pm 4~\rm GHz$ of the initial state. Despite the large number of states the interactions are strongly dominated by coupling between $\ket{60p_{3/2}60p_{3/2}}$ and $\ket{60s_{1/2}61s_{1/2}}.$ This is because
the energy difference is small and the dipole matrix elements are largest. Other states with larger offsets in the principal quantum numbers have much smaller dipole matrix elements\cite{Walker2008}, and do not play a significant role. The cluster of $\ket{p_jp_j}$ states near $-2$ GHz will only affect the $\ket{60s61s}$ weakly and thus have only a small, indirect effect on the $\ket{60p_{3/2}60p_{3/2}}$ energy.  To an excellent approximation, the long-range behavior of the $\ket{60p_{3/2}60p_{3/2}}$ states is dominated by  interactions with $\ket{60s61s}$.

Thus we can consider the long-range interaction between Rydberg atoms as arising predominantly from two coupled channels $nlj+nlj$ and $n_al_aj_a+n_bl_bj_b$ with an energy defect
$\delta=E(n_al_aj_a)+E(n_bl_bj_b)-2E(nlj)$.  Note that for the purposes of blockade we are not interested in actual population transfer between the channels (which would be detrimental), but rather in the energy shift of the $nlj+nlj$ levels due to the interaction with the $n_al_aj_a+n_bl_bj_b$ manifold of states.

\begin{figure}[!t]
\includegraphics[width=8.5cm]{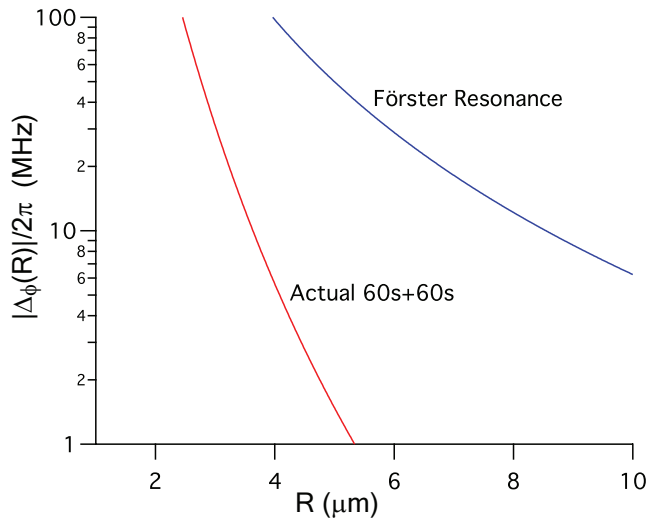}
\caption{(Color online) Dipole-dipole interactions for two 60$s$ Rb atoms are significantly reduced from the resonant \forster\ case
($\delta=0$) for real Rb atoms with $\delta\ne 0$.}
\label{fig.60sForster}
\end{figure}

In this two-level approximation, the corresponding F\"orster eigenstates are linear combinations of states from the different channels, \cite{Walker2008}. Representing the $nlj+nlj$  components of the wavefunction as $\ket\varphi$ and the $n_al_aj_a+n_bl_bj_b$ components as $\ket\chi$, the time-independent Schr\"odinger equation describing the dipole-dipole interaction is
\be
\left(\begin{array}{cc}\delta\cdot I_{\chi} &V_{\rm dd}\\V_{\rm dd}^\dag& 0\cdot I_{\varphi} \end{array}\right)\left(\begin{array}{c}\ket{\chi}\\ \ket{\varphi}\end{array}\right)=\Delta\left(\begin{array}{c}\ket{\chi}\\ \ket{\varphi}\end{array}\right).\label{forsterH}
\ee
Here $V_{\rm dd}$ is a $ \tilde 2 (2j_a+1)(2j_b+1)\times (2j+1)^2$ operator
\footnote{$\tilde 2=1$ if $(n_a,l_a,j_a)=(n_b,l_b,j_b)$ and $\tilde 2=2$ otherwise.}, while $I_{\chi}$ and $I_{\varphi}$ are identity matrices on the $\tilde 2 (2j_a+1)(2j_b+1)$ and $(2j+1)^2$ dimensional Hilbert subspaces of the $\ket\chi$ and $\ket\varphi$ wave function components, respectively.
Solving  for $\ket\chi$ as
$$
\ket\chi={V_{\rm dd}\over \Delta-\delta}\ket\varphi
$$
and substituting this into the second row of (\ref{forsterH}) leads to the (non-linear) eigenvalue equation for $\ket\varphi$:
\be
{V_{\rm dd}^\dagger V_{\rm dd}\over \Delta-\delta}\ket{\varphi}=\Delta \ket{\varphi}.
\label{eq.HF}
\ee
The solutions of this equation are determined uniquely by eigenvalues and eigenvectors of the matrix $V_{\rm dd}^\dagger V_{\rm dd}$. Since all matrix elements of this operator share matrix elements,  $C_3=e^2\rmat{nl}{n_al_a}\rmat{nl}{n_bl_b}$ and a $1/R^6$ dependence on interatomic distance, it is natural to parameterize these Van der Waals eigenstates with eigenvalues $D_{\varphi}$ which in most cases lie between 0 and 1 by
\be
V_{\rm dd}^\dagger V_{\rm dd}\ket\varphi={C_3^2\over R^6}\mathcal D_\varphi\ket\varphi.
\ee
Inserting these solutions, we can proceed and solve Eq.(\ref{eq.HF}) for the F\"orster energy eigenvalues $\Delta$,
\be
\Delta_\varphi(R)={\delta\over 2}-{\rm sign}(\delta)\sqrt{{\delta\over 4}^2+{C_3^2\over R^6}D_\varphi}.
\ee
which constitute the $R$-dependent potential curves between the atoms, correlating asymptotically to the $nlj+nlj$ eigenstates for large $R$. These are the states coupled to the qubit states by the laser fields, and for which we are interested in the energy shifts due to the Rydberg-Rydberg interaction.

It is convenient to define a cross-over distance $R_c$ via $\delta=C_3\sqrt{\mathcal D_\varphi}/R_c^3$ that denotes the region where the energies transition from the van der Waals to the resonant  form. At large distances $R\gg R_c$, the energy shift is of the classic van der Waals form $\Delta_\varphi\approx C_3^2\mathcal D_\varphi/\delta R^6$.
At small distances, $R\ll R_c$, the two channels are effectively degenerate and the energy is
\be
\Delta_\varphi\approx-{\rm sign}(\delta)C_3\sqrt{\mathcal D_\varphi}/R^3
\ee
This gives the largest possible interaction energy between two non-overlapping Rydberg atoms.
A plot of the interaction energy for a pair of 60s Rb atoms ($\delta=1.7$ GHz) is shown in Fig.~\ref{fig.60sForster}, along with the hypothetical interaction energy for $\delta=0$.
  The non-zero energy defect results in a substantial reduction in the interaction energy for the actual case.

The van der Waals interaction typically scales as $n^{11}$ \cite{Boisseau2002,Singer2005b,Walker2008}, excepting cases where the quantum defects give nearly zero \forster\ defect for special values of $n$ \cite{Walker2008}.  Thus it is generally advantageous to work at as high a principal quantum number as is practical.  The resonant \forster\ interaction scales only as $n^4$, so for high principal quantum numbers, often around $n=100$, the atoms are in the resonant limit at 5-10 micron distances.

\begin{figure}[!t]
\includegraphics[width=8.5cm]{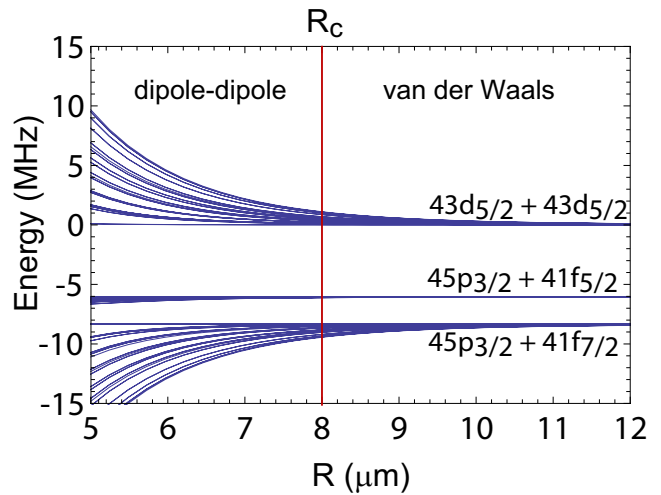}
\caption{Interaction potentials for $43d_{5/2}+43d_{5/2}$ Rb Rydberg atoms. The cutoff radius $R_c$ represents the distance scale for the  transition from resonant dipole-dipole to van der Waals behavior.
}
\label{fig.vdWForster}
\end{figure}

Due to the Zeeman degeneracies of the states at zero external field, there is typically a range of values $\mathcal D_\varphi$ for any given set of angular momenta.  In fact, for most channels there are one
or more   ``\forster-zero" states with $\mathcal D_\varphi=0$ \cite{Walker2005,Walker2008}.  It turns out that \forster-zeros exist for all single channels except those with $j_a=j_b=j+1$, with $j$ the angular momentum of the initial states.  Since \forster-zero states can usually be excited by the lasers, they allow doubly-excited states to be resonantly populated so that the blockade will not work.  Even when multiple channels are considered, there is very often one or more states with nearly zero van der Waals interaction.
As an example, we show in Fig.~\ref{fig.vdWForster} the potential energies for the interaction channel $43d_{5/2}+43d_{5/2}\rightarrow 45p_{3/2}+41f_{5/2,7/2}$ in Rb, which has
 extremely small energy defects of $\delta=-6.0, -8.3 $ MHz for $j_{41f}=5/2, 7/2$ and so might be normally expected to have very promising blockade characteristics \cite{Reinhard2007,Walker2008}. The two ($j=5/2,7/2$) $f$-states break the conditions for the existence of \forster-zero states.  Nevertheless, there are states with extremely small $\mathcal D_\varphi$, resulting in poor blockade.  In Fig.~\ref{fig.vdWForster} these states are the nearly flat ones that correlate to $43d_{5/2}+43d_{5/2}$ at large $R$.

It is important to point out that $s$-states generally do not have \forster-zeros;
in the limit of small fine-structure splitting in the $p$-levels the \forster\ eigenenergies are degenerate, with $D_\varphi=4/3$.  These states are therefore natural choices for blockade experiments.  Unfortunately, excitation Rabi frequencies from the low-lying $p$-states are typically a factor of 3 smaller than for $d$-states in Rb.  This and other technical reasons led to the choice of $d$-states for several recent experiments \cite{Johnson2008,Urban2009,Gaetan2009}.

\subsection{Angular dependence}\label{subsec.angle}

\begin{figure}[!t]
\includegraphics[width=8.5cm]{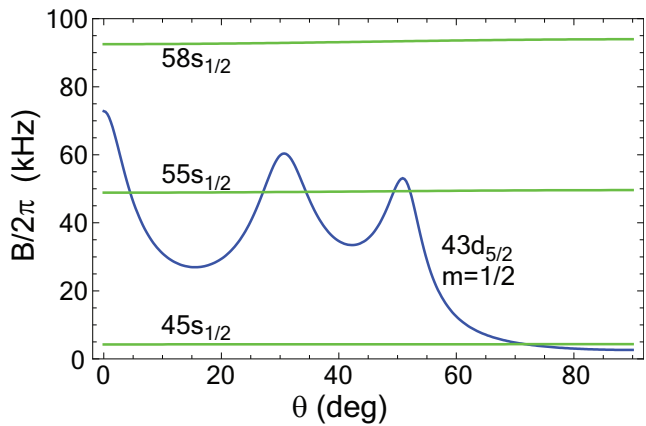}
\caption{(Color online) Angular dependence of blockade shift for $43 d_{5/2}, m=1/2$, $45s_{1/2}$, $55s_{1/2}$, and $58s_{1/2}$ at $R=10~\mu\rm m$
with $\theta$ the angle between the molecular axis and $\hat z.$
}
\label{angle}
\end{figure}

The critical measure of the interaction strength for Rydberg blockade is the blockade shift of Eq.~(\ref{eq.bshift}).  In zero external field, rotational invariance requires that the \forster\ eigenenergies $\Delta_{\varphi ij}$ depend only on the distance $R_{ij}$ between atoms $i$ and $j$, but the coupling of the corresponding eigenstates to the excitation light, $\kappa_{\varphi ij}$, depends on the angle between the interatomic axis and the light polarization.
 A typical three dimensional distribution of atoms includes pairs with arbitrary relative orientations so that laser fields with laboratory fixed polarizations will generally couple to all possible two-atom eigenstates, including those with weak interactions.  The blockade shift is an inverse-square average of the interaction strengths, and so is particularly sensitive to states  with small $\Delta_\varphi$.  For reduced-dimension geometries such as long cylinders, the effects of small $\Delta_\varphi$ may be mitigated by polarization choices that also give small $\kappa_{\varphi ij}$.  However, in a spherical sample the atoms will have random orientations and the blockade strength will be dominated by those orientations with the weakest blockade shifts.  These points are discussed in more detail in \cite{Walker2008}.

Examples of angular dependences of the blockade shifts are given in Fig.~\ref{angle}. Excitation by $\hat{z}$-polarized light has been assumed. For pairs of $s$-state atoms, the blockade is nearly spherically symmetric
(the weak departure from spherical symmetry is due to the fine-structure splitting of the $p$ states), while for  $43d_{5/2}$ atom pairs there is a considerable variation with angle.
The small F\"orster defect of $43d_{5/2}$ produces a much stronger, albeit anisotropic, interaction for most angles than the nearby in energy $45s_{1/2}$ state.  In order to obtain a comparable isotropic interaction strength  it is necessary to use
a $55s_{1/2}$ or higher  state.

\begin{figure}[!t]
\includegraphics[width=8.5cm]{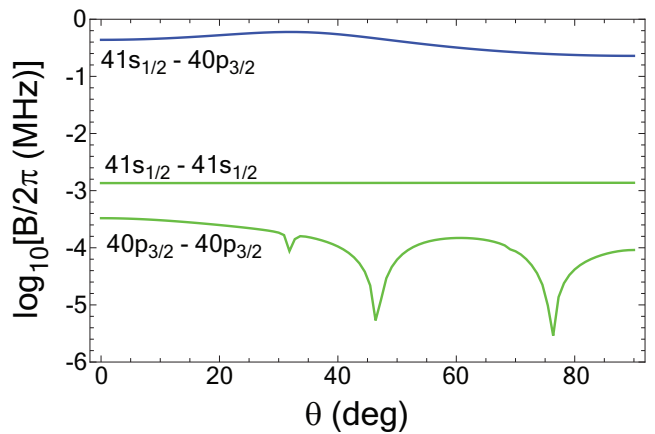}
\caption{Comparison of the blockade shift $\bshift$ at $R=10~\mu\rm m$ for interacting Rb 41$s_{1/2} - 40p_{3/2}$ atoms, as compared to van der Waals blockade for two $41s_{1/2}$ or 40$p_{3/2}$ atoms. All atoms have magnetic quantum number $m=1/2.$.}
\label{fig.asymmetric}
\end{figure}

Rydberg atoms in states with $l$ differing by 1  experience the dipole-dipole interaction in first order so the interaction scales with distance as $1/R^3.$
This case was observed spectroscopically by \cite{Afrousheh2004,Afrousheh2006}.
The van der Waals interaction between states of the same $l$ is much weaker at long range scaling as $1/R^6.$
The use of this interaction asymmetry  for efficient creation of multiparticle entanglement was proposed in \cite{Saffman2009b} (see Sec. \ref{sec.GHZ}).  An example of the blockade shifts for $s+p_{3/2}$ states, including a comparison to the van der Waals case,  is shown in Fig.~\ref{fig.asymmetric}. We see that the asymmetry is more than two orders of
magnitude over the full angular range.

\subsection{Tuning the interaction with external fields}
\label{sec.fields}

The interactions between the Rydberg atoms are generally sensitive to the application of external magnetic and electric fields.  An extreme case occurs when an electric field is applied of sufficient strength to mix states of different parity.  The atoms then acquire a permanent electric dipole moment $\expect{\bf d}$, and the dipole-dipole interaction to first order becomes
\be
V_{\rm dd}={1\over R^3}\left({\bf \expect{\bf d}^2-3\expect{\bf d}\cdot {\bf \hat R}{\bf \hat R}\cdot \expect{\bf d}}\right)
\ee
Since the dipole moments are usually of order $n^2e a_0$, this is a very strong but anisotropic interaction that may work well for blockading anisotropic samples.  A weakness of this situation is that the atomic energy levels are also very sensitive to electric field fluctuations that may inhomogeneously broaden the Rydberg excitation line.

In other cases, it is possible to use electric fields to tune certain Zeeman states into a \forster\ resonance to give dipole-dipole interactions of order $n^4e^2 a_0^2/R^3$. This is illustrated in the early experiment of  \cite{Anderson1998}, more recently in \cite{Vogt2007}, and discussed in the recent review of \cite{Gallagher2008}. Again, this generally results in a strong  but anisotropic blockade shift.  Resonant transfer between spatially separated samples was observed and studied in \cite{vanDitzhuijzen2008,Tauschinsky2008,vanDitzhuijzen2009},
 and simulated in \cite{Carroll2009}.
Double-resonance spectroscopy was used in \cite{Reinhard2008b} to directly observe the difference between van der Waals and resonant dipole-dipole interactions. The effects of DC electric fields on Rydberg-
Rydberg interactions are treated in \cite{Schwettmann2006}.

\begin{figure}[!t]
\includegraphics[width=8.5cm]{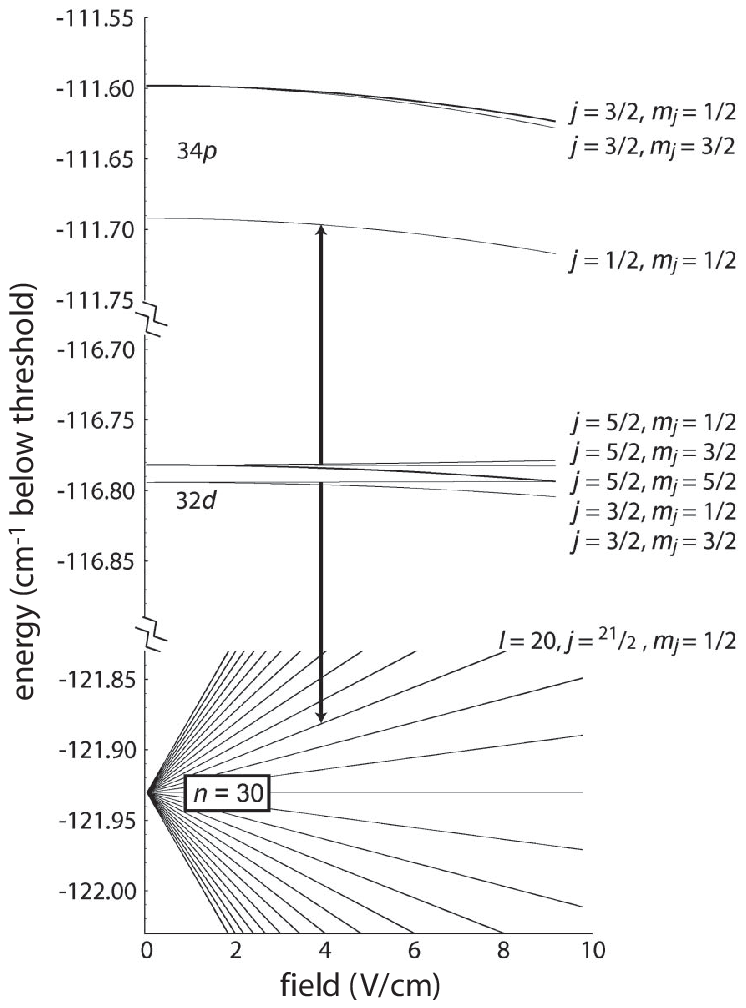}
\caption{Tuning of \forster\ resonances with an electric field.  From \protect\cite{Carroll2004}.}
\label{Carroll2004fig}
\end{figure}

A possible way to DC field-enhance dipole-dipole interactions while minimizing field-broadening of the optical resonance is illustrated by the work of Carroll {\it et al.} \cite{Carroll2004,Carroll2006}.  They used relatively field insensitive $32d$ states for optical excitation, while coupling to a field-sensitive \forster\ resonance.  This is illustrated in Fig.~\ref{Carroll2004fig}.  By exciting a long, narrow tube of atoms, they also observed the angular dependence of the interaction.

Another tool for tuning the atom-atom interactions with external fields is to use resonant or near-resonant microwave fields.  For example, \cite{bohlouli-zanjani:203005}   used the AC-Stark shift from near-resonant microwaves to tune the near \forster\ resonance between Rb 43$d_{5/2}$ atoms illustrated in Fig.~\ref{fig.vdWForster},
enhancing  energy-transfer collision rates.  They then demonstrated resonant $f-g$ microwave coupling to accomplish the same purpose \cite{Petrus2008}.  As with the \cite{Carroll2004} experiment, this coupling has a relatively small effect on the energies of the initial $d$-states.

\begin{figure}[!t]
\includegraphics[width=8.5cm]{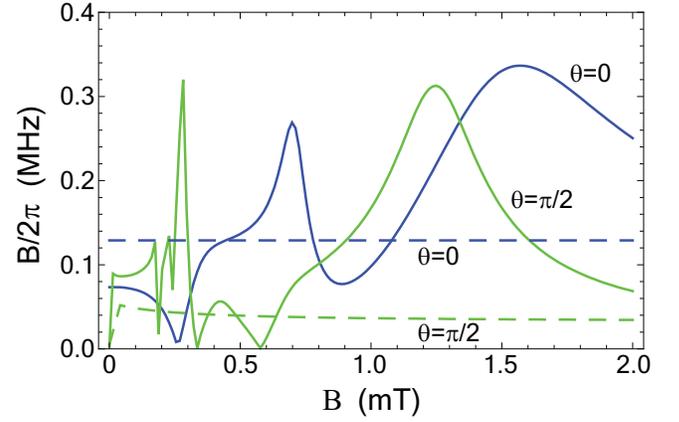}
\caption{Blockade interaction of Rb $43 d_{5/2}$ states as a function of the magnetic field applied along the quantization axis and $R=10~\mu\rm m$. The solid(dashed)
lines are for excitation of $m=1/2(5/2)$ states and $\theta$ is the angle between the applied field and the molecular  axis. }
\label{wiscblockade}
\end{figure}

Modest magnetic fields can also strongly affect the Rydberg-Rydberg interactions, by breaking the Zeeman degeneracy that produces \forster\ zeros.  An example is given in Fig.~\ref{wiscblockade}, which shows the van der Waals interactions of the $43d_{5/2}43d_{5/2}$ states as a function of magnetic field.
The states with maximal magnetic quantum number $m=5/2$  are quite insensitive to magnetic fields since there is only one combination of Zeeman levels with the same total $M=m+m=5$.  On the other hand the $m=1/2$ states are strongly dependent on magnetic field since it breaks the degeneracy between the $M=1$ Zeeman pairs $(1/2,1/2),(-1/2,3/2),(-3/2,5/2).$ The extent to which the laser excited states couple to the molecular eigenstates is also angle dependent, hence the different behaviors seen for $\theta=0$ and $\pi/2.$
Magnetic field effects were demonstrated experimentally in
\cite{Afrousheh2006} who used a small magnetic field to reduce resonant dipole-dipole interactions. Conversely in \cite{Urban2009} a magnetic field was used in a well controlled geometry to increase the interaction strength.

\subsection{Three-body \forster\ interactions}
\label{sec.3atomblockade}

It might be expected that under conditions of strong blockade for atom pairs, three-atom states would be further off-resonance.  However, \cite{Pohl2009} showed that for three two-level atoms interacting via the \forster\ mechanism there are triply-excited states that are non-interacting.  This ``breaks" the blockade in the sense that resonant two-photon transitions can occur between the singly-excited state and the triply-excited state.  However, the two-photon Rabi coupling to the triply-excited states is via the doubly-excited state that is off-resonance by $1/\Delta_{\rm dd}$.  Thus the probability of three-atom excitation is of order $\Omega^2/\Delta_{\rm dd}^2$, similar to the blockade error of Eqs.~(\ref{eq.P2}) and (\ref{eq.bshift}) that will be considered in more detail in the next section.

\section{Rydberg gates}
\label{sec.gates}

In this section we discuss a variety of approaches to implementing  neutral atom gates with Rydberg interactions.
Prior to discussing the various protocols we provide a brief derivation of the blockade shift of Eq. (\ref{eq.bshift}) following the analysis of \cite{Walker2008}.
The original blockade gate of \cite{Jaksch2000} is then examined in Sec. \ref{sec.bgate}
followed by consideration of alternative approaches in Sec. \ref{sec.othergates}. We restrict our attention to gates that operate between single atom qubits. These primitives will find use for ensemble
and collective  gates, and as building blocks for many particle entanglement protocols in Secs.
\ref{sec.col_encoding},\ref{sec.col_gates},\ref{sec.entanglement}. Since the Rydberg interaction
is long ranged (see Fig. 1) it is natural to ask what the limit is on the number of qubits
in a fully connected register. We compare the limits as a function of the space dimensionality of the
qubit array in Sec. \ref{sec.arrayscaling}.

\subsection{Blockade interaction}
\label{sec.blockadetheory}

Consider a  cloud of $N$ atoms with internal ground states $\ket{g}$ and Rydberg states $\ket{\gamma}$ with $g,\gamma$ shorthand for the full set of quantum numbers needed to specify the states.
We are interested in the situation where blockade is active, and assume that at most two atoms can be simultaneously excited to a Rydberg state.
 Thus the atomic cloud can be in the possible states
\begin{equation}
\ket{g},\ket{\gamma k},\;{\rm  and }\; \ket{\varphi kl}
\end{equation}
representing respectively all the atoms in the ground state, the $k$th atom in the singly-excited Rydberg state $\gamma$, and the $k$th and $l$th atoms in the doubly-excited Rydberg state $\varphi$. The label $\varphi$
denotes eigenstates of the F\"orster Hamiltonian of Eq. (\ref{eq.HF}): $H_F\ket{\varphi kl}=\Delta_{\varphi kl}\ket{\varphi kl}.$

The light-atom coupling at atom $k$ is described by an electric dipole Hamiltonian ${\mathcal H}_k$ and an excitation Rabi frequency $\Omega_{\gamma k}=\frac{2}{\hbar}\bra{\gamma k}{\mathcal H}_k\ket{g}.$ We define
an $N-$atom collective Rabi frequency by
\begin{equation}
\Omega_N=\sqrt{\sum_{\gamma k}\left| \Omega_{\gamma k}\right|^2}=\sqrt{N}\Omega
\end{equation}
where $\Omega$ is the rms single-atom Rabi frequency averaged over all the atoms in the ensemble.
With this definition we can write a normalized  singly excited state as
$$
\ket{\sf s}=\sum_{\gamma k}\frac{\Omega_{\gamma k}}{ \Omega_{N}}\ket{\gamma k},
$$
and  the wavefunction for the $N$-atom ensemble as
\begin{equation}
|\psi\rangle=c_{g}|g\rangle+c_{\sf s}\ket{\sf s}+\sum_{\varphi,k<l}c_{\varphi kl}\ket{\varphi{kl}}.
\end{equation}
This expression is valid to first order in $|\Omega/\bshift|$, where $\bshift$ is the blockade shift defined in Eq. (\ref{eq.bshift}).

Using the above definitions, the Schr\"odinger equations for the ground, symmetric singly-excited state, and the doubly excited states are
\bml
i{\dot{c}}_{g} & = &{\Omega_{N}^*\over 2} c_{\sf s} \\
i{\dot{c}}_{\sf s} & = &{\Omega_{N}\over 2} c_{g}+ \frac{\Omega_{N}^*}{N}\sum_{\varphi, k<l}\kappa^{*}_{\varphi kl}c_{\varphi kl}\label{se1}\\
i{\dot{c}}_{\varphi kl} & = & \Delta_{\varphi kl}c_{\varphi kl}+\frac{\Omega_{N}}{N}\kappa_{\varphi kl}c_{\sf s}.
 \label{se2}
\eml
Here the overlap amplitudes between the optically excited states and the F\"orster eigenstates are given by $\kappa_{\varphi kl}=\frac{4}{\hbar^2\Omega^2}\bra{\varphi kl}{\mathcal H}_k {\mathcal H}_l\ket{g}$.

We are now ready to calculate the  effectiveness of Rydberg blockade.
Assume we start in the ground state $\ket{\psi}=\ket{g}$ and apply a $\pi$ pulse: $\Omega_N t=\pi.$
With the assumption that there is a strong blockade, the doubly excited amplitudes are small and $c_{\sf s}(t)\simeq 1.$ Making an  adiabatic approximation to Eq.~(\ref{se2}) we get
$$
c_{\varphi kl}=-\frac{\Omega_{N}\kappa_{\varphi kl}}{ N\Delta_{\varphi kl}}c_{\sf s}
\label{adiab}
$$
and the probability of double excitation is
\begin{equation}
P_2=\sum_{\varphi,k<l}\left|c_{\varphi kl}\right|^2={|\Omega_N|^{2}\over N^{2}}\sum_{\varphi, k<l}\left|{\kappa_{\varphi kl}\over \Delta_{\varphi kl}}\right|^2.
\end{equation}
It is critical to note that, given relatively even excitation of the two-atom Rydberg states,  it is an average of $1/\Delta_{\varphi kl}^2$ that determines the blockade
effectiveness.  This means that Rydberg-Rydberg states with small interaction shifts are much more strongly weighted than those with large energy shifts.  Let us define a mean blockade shift $\bshift$ via
\begin{equation}
{1\over \bshift^2}={2\over
N(N-1)}\sum_{\varphi,k<l}{|\kappa_{\varphi kl}|^{2}\over\Delta_{\varphi kl}^2}.
\label{blockshift}
\end{equation}
Then the probability of double excitation becomes
\begin{equation}
P_2=\frac{N-1}{N}\frac{|\Omega_N|^2}{2\bshift^2}.
\label{pdouble}
\end{equation}
This shows that for fixed $\Omega_N$, the frequency of the {\it collective} oscillation, the probability of double excitation is virtually independent of the number of atoms in the ensemble.

It is  is important to keep in mind that the blockade shift $\bshift$ depends on the polarization of the excitation light as well as the Zeeman structure of the states $\ket{g}, \ket{\varphi kl}$
through the overlap factor $\kappa_{\varphi kl}$.  We do not explicitly indicate these dependences in order to avoid a proliferation of subscripts.  Explicit examples of the angular dependence have been given
 in Sec. \ref{subsec.angle}.

\subsection{Two-atom blockade gate}
\label{sec.bgate}

As has been discussed in connection with Fig. \ref{fig.cphase} Rydberg gates are intrinsically prone to errors due to the finite lifetime of the Rydberg levels that are used. In the strong blockade limit $(\bshift\gg \Omega)$,  the gate fidelity is high, and  we can easily estimate the errors by adding the contributions
from the physically distinct processes of spontaneous emission from Rydberg states and state rotation errors, which are primarily due to imperfect blockade. We note that the gate error depends on the input state applied to the gate.
Referring to Fig. \ref{fig.cphase} the state $\ket{00}$ experiences relatively small errors since Rydberg excitation is off-resonant by a detuning $\omega_{10}$. For hyperfine encoded qubits $\omega_{10}$ is typically several GHz as opposed to the few MHz of $\Omega$ and $\bshift.$ On the other hand the state $\ket{11}$ leads to
the largest errors since both atoms are Rydberg excited and subject to spontaneous decay.
In this section and the following one, we base our analysis on ``square pulse" excitation schemes. The possibility of  improving on the error limits we find using shaped or composite pulses remains an open question.

An average gate error was defined in \cite{Saffman2005a} by simply averaging over the four possible two-atom inputs.
The dominant errors come from imperfect blockade in step 2 with error $E_{\rm bl}\sim \Omega^2/\bshift^2$  and spontaneous emission of the control or target atom with error $E_{\rm se}\sim 1/\Omega\tau $ where $\tau$ is the Rydberg state spontaneous lifetime.
Keeping track of the numerical prefactors, neglecting higher order terms in $\Omega/\bshift, \bshift/\omega_{10}$, and  averaging over the input states, we find the gate error\footnote{Equations (\ref{eq.Ebar}-\ref{eq.Emin}) correct some algebraic errors in  Table V  and Eq. (38) of \cite{Saffman2005a}.}
\be
E\simeq\frac{7\pi}{4\Omega\tau}\left(1+\frac{\Omega^2}{\omega_{10}^2} +\frac{\Omega^2}{7 \bshift^2}\right)
+\frac{\Omega^2}{8\bshift^2}\left(1+6 \frac{\bshift^2}{\omega_{10}^2}\right).
\label{eq.Ebar}
\ee
The first term proportional to $1/\tau$ gives  the spontaneous emission error and the second term
gives the probability of an atom populating the Rydberg level at the end of the gate. 
These expressions assume piecewise continuous pulses. Other assumptions, such as pulses with 
phase jumps\cite{Qian2009}, lead to different coefficients, but it is still the ratio of  the effective bandwidth of the excitation to the blockade strength that governs the magnitude of the errors.

 In the limit of
$\omega_{10}\gg (\bshift,\Omega)$ we can extract a simple expression for
the optimum Rabi frequency which minimizes the error
\be
\Omega_{\rm opt}=(7\pi)^{1/3}\frac{\bshift^{2/3}}{\tau^{1/3}}.
\label{eq.Omegaopt}
\ee
Setting $\Omega\rightarrow\Omega_{\rm opt}$  leads to a minimum  averaged gate error of
\be
E_{\rm min} =\frac{3(7\pi)^{2/3}}{8}\frac{1}{(\bshift\tau)^{2/3}}.
\label{eq.Emin}
\ee

\begin{figure}[!t]
\includegraphics[width=8.5cm]{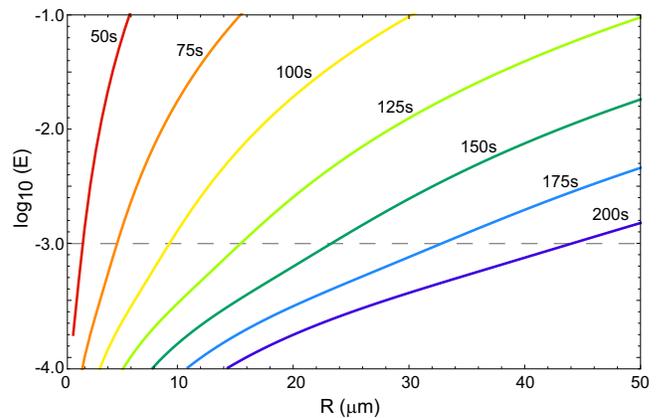}
\caption{(Color online) Intrinsic blockade gate error from Eq. (\ref{eq.Ebar}) for $^{87}$Rb $ns$ states.
The lifetimes for $n=50-200$ were calculated from Eq. (\ref{eq.tau}) to be
$\tau=(70,180,340,570,860,1200,1600)~\mu\rm s$.
}
\label{fig.czerror}
\end{figure}

The gate error for excitation to $^{87}$Rb $ns$ states is shown in Fig. \ref{fig.czerror}
for $50\le n\le 200$ as a function of the two-atom separation $R.$
Despite the use of finite lifetime Rydberg states the interaction is strong enough to allow for fast
excitation pulses which keep the total error small. We see that gate errors below $0.001$, which
 corresponds to recent theoretical estimates\cite{Knill2005,Aliferis2009} for the fault tolerant threshold in a scalable quantum computer, can be achieved over a large range of interatomic separations. At relatively low excitation of  $n=50$,
 an  error of 0.001  requires $R\sim 1~\mu\rm m.$ The atom separation at this error level can be pushed all the way to $R>40~\mu\rm m$ by exciting $n=200$  states. Although coherent excitation of $n=200$ Rydberg states has not been demonstrated, excitation of clusters of states has been reported
up to $n=1100$\cite{Frey1995} and state resolved excitation up to $n=390$\cite{Tannian2000} and $n=500$\cite{Neukammer1987} has been achieved.
In contrast to collisional neutral atom gates\cite{Brennen1999,Jaksch1999} which accumulate large errors if operated too fast, the Rydberg gate would fail if it were run too slowly.
The necessity of achieving sufficiently fast excitation of high lying levels  puts demands on the laser system, but this is a technical, not a fundamental challenge.

In addition to the intrinsic error due to the finite Rydberg lifetime, there are other errors due to technical imperfections. These include Doppler broadening of the excitation, spontaneous emission from intermediate levels when a two-photon excitation scheme is used, pulse area errors due to variations in atomic position, and errors due to imperfect polarization of the exciting laser pulses. All of these effects have been considered in detail\cite{Saffman2005a,Protsenko2002} and will be discussed briefly in Sec. \ref{sec.experiments}. In principle, with proper experimental design, errors due to technical imperfections can be kept below $10^{-3}$ or less. It may be emphasized that the Rydberg blockade operation is not directly sensitive to the center of mass motion and it is therefore not necessary to work with atoms that are in the ground vibrational state of the confining potential. Using blockade to implement a CNOT gate requires additional single qubit pulses. 
These can also be made insensitive to the motional state using two-photon co-propagating Raman pulses\cite{Yavuz2006,Jones2007}. 
This implies an attractive robustness of the quantum gate with respect to moderate vibrational heating of the atomic qubits.

There is also an adiabatic version of the blockade gate which does not require individual addressing of the atoms\cite{Jaksch2000}. However, the adiabatic condition implies that the gate time is long compared to $1/\Omega$ which increases errors due to spontaneous emission. Thus, relaxing the requirement of individual addressing reduces the fidelity compared to the  blockade gate described above.

\subsection{Alternative gate protocols}
\label{sec.othergates}

\subsubsection{Interaction gates}

Already in the original Rydberg gate paper\cite{Jaksch2000} a second type of  gate protocol was proposed.
The `interaction' gate  assumes the excitation of both atoms from the qubit 1-state to the Rydberg state, and makes use of the Rydberg interaction energy to accumulate a phase shift of $\pi$. Thus it works in the opposite regime of the blockade gate, and requires $\Delta_{\rm dd}\ll \Omega.$
The pulse sequence is: 1)  $R(\pi)_{1,r}^{\rm c,t}$, 2) wait a time $T=\pi/\Delta_{\rm dd},$
and 3) $R(\pi)_{1,r}^{\rm c,t}.$ Here $R(\theta)_{i,j}^{c,t}$ is an $X$ rotation by $\theta$
between states $i,j$ on the control and target atoms. The resulting gate is the same $C_Z$ as given in Eq. (\ref{eq.CZ}) apart from an overall  $\pi$ phase shift. This protocol has the advantage that it does not require individual addressing of the atoms.
We can perform an error analysis  along the lines of that in the previous section assuming the orderings
$\omega_{10}\gg \Omega\gg \Delta_{\rm dd}$ and  $\Delta_{\rm dd}\tau\gg 1$.
The interaction energy of two Rydberg excited atoms is $\Delta_{\rm dd}$ which can be calculated from the approximate expression
\be
\Delta_{\rm dd} \simeq \sum_\varphi  \frac{|\kappa_\varphi|^2\Omega^2}{|\kappa_\varphi|^2\Omega^2+\Delta_\varphi^2} \Delta_\varphi.
\label{eq.Deltadd}
\ee
We have dropped the atom labels on $\kappa_\varphi, \Delta_\varphi$ since we are considering only a two-atom interaction.
Note that when $\Omega^2\ll {\rm min_\varphi}|\Delta_\varphi/\kappa_\varphi|^2=|\Delta_{\varphi_{\rm min}}/\kappa_{\varphi_{\rm min}}|^2$ we get
$\Delta_{\rm dd} \simeq |\kappa_{\varphi_{\rm min}}|^2 \Omega^2/\Delta_{\varphi_{\rm min}}$ which verifies that the
two-atom interaction tends to zero in the blockade limit where two atoms cannot simultaneously be excited. In the opposite limit of $\Omega^2\gg {\rm max_\varphi}|\Delta_\varphi/\kappa_\varphi|^2$ we find
$\Delta_{\rm dd} \simeq \sum_\varphi \Delta_\varphi$ which is independent of $\Omega.$ Since the eigenvalues scale as $R$ to a negative power we infer that  $\Delta_{\rm dd}$ has a maximum at an intermediate value of $R$ and tends to zero for $R$ small or large. The implication of this is that
at fixed $n$ the interaction gate has an optimum fidelity at a finite value of $R$.

To see this in more detail consider the leading contributions to the input averaged
gate error which are\cite{Saffman2005a}
\be
E\simeq\frac{\pi}{\tau}\left( \frac{1}{\Delta_{\rm dd}}+\frac{1}{\Omega}\right) + \left(\frac{2\Delta_{\rm dd}^2}{\Omega^2}
+\frac{\Omega^2}{\omega_{10}^2} \right).
\label{eq.Eint}
\ee
For given values of $\omega_{10}, \tau$  the absolute minimum of the gate error can be shown to be
\be
E_{\rm min}=\left[\frac{2^{7/2}\pi}{\tau\omega_{10}} \right]^{1/2}.
\label{eq.Eminint}
\ee
 The error for any particular choices of $\Delta_{\rm dd}(n,R), \Omega$ will always exceed this lower bound.
For $^{87}$Rb we have $\omega_{10}/2\pi=6834~\rm MHz$ and $\tau = 100~\mu\rm s $ gives
$E_{\rm min}\simeq 0.003$.

\begin{figure}[!t]
\includegraphics[width=8.5cm]{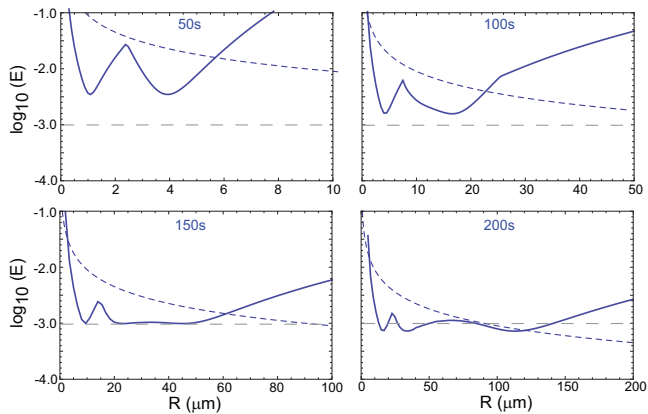}
\caption{(Color online) Intrinsic interaction gate error from Eq. (\ref{eq.Eint}) for $^{87}$Rb $ns$ states.
The lifetimes for $n=(50,100,150,200)$ were calculated from Eq. (\ref{eq.tau}) to be
$\tau=(70,340,860,1600)~\mu\rm s$.
The long dashed line in each figure is  $E=.001$ and the short dashed curve gives the error limit at $\delta R=15~\rm nm$.
}
\label{fig.cz_interaction_error}
\end{figure}

 In order to evaluate the potential of the gate for different
Rydberg levels $n$ and separations $R$ it is useful to find the optimum $\Omega$ for a given $\Delta_{\rm dd}(n,R)$.
We find
\be
\Omega_{\rm opt}\simeq \frac{2^{1/2}}{3^{1/4}} (\Delta_{\rm dd}\omega_{10})^{1/2}
\label{eq.Omegaoptint}
\ee
which leads to
\be
E_{\rm opt}\simeq\frac{\pi}{\Delta_{\rm dd}\tau}+\frac{5\Delta_{\rm dd}}{3^{1/2}\omega_{10}}.
\label{eq.Eopt2}
\ee
Equations (\ref{eq.Omegaoptint},\ref{eq.Eopt2}) are only
 implicit relations for $\Omega_{\rm opt}$ and $E_{\rm opt}$ since Eq. (\ref{eq.Deltadd}) also depends on $\Omega.$  We have therefore used a numerical search to find $\Omega_{\rm opt}$ and $E_{\rm opt}$ at specific values of $n$ and $R$ with the results shown in Fig. \ref{fig.cz_interaction_error}  for Rb $ns$ states.
 As was discussed in connection with Eq. (\ref{eq.Deltadd}) the interaction strength has a maximum at intermediate values of $R.$ However, the interaction can actually be too strong at intermediate $R$ which violates the scaling $\Delta_{\rm dd}\ll \Omega\ll \omega_{10}$ and results in a complex multi-peaked structure for the $R$ dependence of the gate error.
We see that low $n$ states with small $\tau$ do not give particularly low gate errors, even at small separation $R$, due
to  the limit imposed by Eq. (\ref{eq.Eminint}).
For $n>100$ we find errors of $10^{-3}$ or less over a wide range of $R.$  If we are willing to consider very highly excited states with $n=200$ high fidelity gates are possible at $R>100~\mu\rm m$.

There is, however, a caveat to the above discussion since
the interaction gate is sensitive to fluctuations in the atomic separation as was  emphasized in \cite{Protsenko2002}. To estimate the sensitivity  assume we are in the van der Waals limit so $\Delta_{\rm dd}(R)=\Delta_{\rm dd,0} (R_0/R)^6$ with $\Delta_{\rm dd,0}, R_0$ constants. The two-atom phase depends on $R$ as $\delta\phi(R)=\Delta_{\rm dd}(R) T\simeq \Delta_{\rm dd,0} T(1 -6\delta R/R_0)$ with $T$ the gate interaction time.  The fractional phase error is thus $6\delta R/R_0$.
The harmonic oscillator wavefunction of a $^{87}$Rb atom in a
trap with $\omega/2\pi=500~\rm  kHz$, which is a reasonable limit for what is experimentally feasible
in an optical trap, has a characteristic length scale of about 15 nm. Taking $\delta R=15~\rm nm$ gives the short dashed curve in Fig. \ref{fig.cz_interaction_error} which drops below  $10^{-3}$  for $R>100~\mu\rm m$.

These error estimates treat the motion classically. If the atoms are in a pure quantum-mechanical motional state then there is no
position dependent interaction. Nevertheless there are two-body forces when both atoms are excited to a Rydberg level. If these forces are strong enough, then excitation of the motional state will occur which leads to decoherence and gate errors. These errors have been estimated in \cite{Jaksch2000,Saffman2005a} and can again be made small provided
that the atoms are well localized spatially.    We conclude that the interaction mode of operation has potential for remarkably long range gates, extending out to $R>100~\mu\rm m$ with low errors, provided that the atoms are cooled close to the motional ground state or are otherwise trapped with a high degree of spatial localization.

A  variation on the interaction gate described above  was analyzed in \cite{Protsenko2002}. There it  was assumed that  the excitation pulses are purposefully detuned from $\ket{r}$ and
the presence of the two-atom interaction $\Delta_{\rm dd}$ brings the excitation into resonance.
This type of interaction induced resonance was anticipated in an early paper\cite{Varada1992} that predated
the interest in quantum gates and was more recently proposed\cite{Ates2007} and observed\cite{Amthor2010} in an ensemble of  cold atoms.
 This `self-transparency' mode of operation has the same sensitivity as the interaction gate  to fluctuations in atomic position.

A further variation on the interaction gate was described in \cite{Ryabtsev2005}. There it was suggested to excite two atoms to weakly interacting states that are microwave coupled as in Sec. \ref{sec.fields}.
The microwaves are initially detuned from resonance between opposite parity Rydberg levels,  and then the resonance condition is achieved for a controlled time using a Stark switching technique, which was demonstrated in \cite{Ryabtsev2003}.
This approach suffers again from sensitivity to fluctuations in the atomic separation although
  in the limit where the microwave coupling effectuates a $1/R^3$ resonant dipole-dipole  interaction the sensitivity will only be half as large as in the van der Waals regime. In addition a somewhat different gate idea based on Stark switching together with  nonholonomic control techniques can be found in \cite{Brion2006}.

Adapting ideas developed in the context of trapped ion gates\cite{Garcia-Ripoll2003} it was proposed in
\cite{Cozzini2006} to implement a $C_Z$ gate using  controlled atomic motion due to  two-body forces from the
dipole-dipole interaction of simultaneously excited Rydberg atoms. This is reminiscent of
the interaction gate, although here the necessary
conditional phase shift is due to a combination of a dynamic and geometrical phase.
The fidelity of the gate can be 0.99 or better  provided the atoms are again cooled close to the motional ground state.

\subsubsection{Interference gates}

A different type of gate which does not require strongly populating the Rydberg levels can be designed by using the
interference of different multi-photon transition paths. The dipole-dipole interaction suppresses the amplitude of one of the paths\cite{Brion2007b} which creates a conditional phase shift.  This general idea can be used to implement a universal set of gates in a decoherence free subspace of logical qubits, each encoded in two physical
qubits\cite{Brion2007a}. Unfortunately the gate is relatively slow due to the use of multi-photon transitions,  and consequently suffers from spontaneous emission errors despite the fact that the Rydberg states are never substantially populated. Attempts to design entangling
gates that require only virtual  excitation of Rydberg states, and hence do not have any spontaneous emission errors turn out to be futile, since it can be shown that a minimum integrated Rydberg population  of the control and the target atom during the gate execution, $\int (p_r^{(c)}+p_r^{(t)}) dt > 2/\Delta_{\rm dd}$, is necessary for the creation of one bit of entanglement with arbitrary local operations and a Rydberg-Rydberg interaction energy of $\hbar \Delta_{\rm dd}$, cf., the Appendix in  \cite{Wesenberg2007}.

\begin{figure}[!t]
\includegraphics[width=8.5cm]{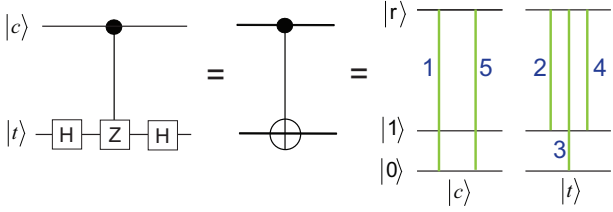}
\caption{(Color online) The CNOT gate (center) can be implemented with  Hadamard gates and a $C_Z$ (left) or
directly with a controlled amplitude swap (right). All transitions are $\pi$ pulses with the indicated ordering.
}
\label{fig.cnot}
\end{figure}

\subsubsection{Amplitude swap gates}

All of the approaches described above implement a $C_Z$ gate which can be converted into a CNOT gate using the quantum circuit equivalence of Fig. \ref{fig.cnot}. An alternative route to the CNOT dispenses with the Hadamard
operations and directly swaps the states of the target qubit, conditioned on the state of the control qubit, as
shown on the right in Fig. \ref{fig.cnot}. When the control qubit is in
state $\ket{0}$ it is Rydberg excited by pulse 1 which blocks the swap action of pulses 2-4, and when the control atom is in state $\ket{1}$ the swap is unhindered.
The result is a CNOT gate with an overall minus sign.
This method  was first proposed in the context of quantum
computing with rare earth doped crystals in \cite{Ohlsson2002} and was used for  experimental demonstration of a two-atom CNOT gate\cite{Isenhower2010}. It turns out that
this approach is particularly well suited to blockade interactions which naturally enable or block the swap operation on the target
qubit. As we will see in Sec. \ref{sec.GHZ} an adaptation of this approach is also efficient for creating
multi-particle entangled states\cite{Saffman2009b}.

\begin{figure}[!t]
\includegraphics[width=8.5cm]{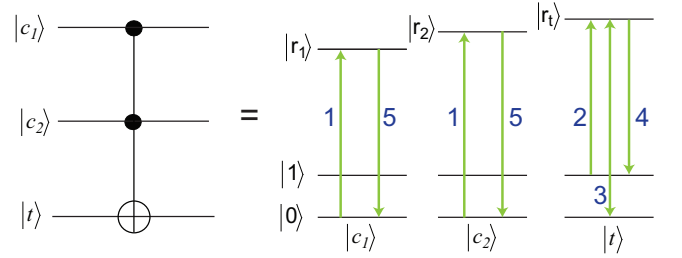}
\caption{(Color online) Implementation of the Toffoli gate as proposed in \cite{Brion2007c}. All transitions are $\pi$ pulses with the indicated ordering.
}
\label{fig.toffoli}
\end{figure}

A variation of the amplitude swap  approach can be used to create three-terminal Toffoli gates which are important for efficient implementations of quantum algorithms. The Toffoli gate switches the value of the target qubit if both control qubits are $1$, and otherwise
leaves the target unchanged.  This can be implemented in several ways using the
population swap primitive. Consider the pulse sequence shown in Fig. \ref{fig.toffoli} where
we now allow for coupling to three Rydberg states $\ket{r_1}, \ket{r_2}$, and $\ket{r_t}.$
It is possible to choose states $\ket{r_1}, \ket{r_2}$ that are weakly interacting, yet both interact strongly
with $\ket{r_t}.$ For example we could set $\ket{r_1},\ket{r_2}$ to be $s$ and $d$ parity Rydberg levels
and $\ket{r_t}$ a $p$ parity level. This case was analyzed in detail for hydrogenic Rydberg levels in \cite{Brion2007c}.
Using these levels the pulse sequence shown in Fig. \ref{fig.toffoli} implements the ideal
Toffoli gate but with an overall minus sign which can be corrected by single qubit operations.
It may also be noted that the requirement of weakly interacting control atoms can also be effectuated
geometrically. If all three atoms are placed on a line with the target atom in the middle then in the van der
Waals limit the control-control Rydberg interaction will be a factor of $2^6$ weaker than either control acting on the target. In this case the three Rydberg states can be
replaced by one state, excited
 by the same laser frequency, as long as the laser can address
the three atoms individually.

 It has recently been shown\cite{Saffman2009d} that the Toffoli gate can be extended
to a Toffoli gate with $k$ control bits, or ${\rm C}^k$-NOT gate.
First, we note that the requirement of weakly interacting control qubits is actually not needed.
If we assume individual addressing of the control and target qubits then the pulse sequence
$R(\pi)_{0,r}^{c_1}   R(\pi)_{0,r}^{c_2} R(\pi)_{1,r}^{t}   R(\pi)_{0,r}^{t}   R(\pi)_{1,r}^{t}     R(\pi)_{0,r}^{c_2}   R(\pi)_{0,r}^{c_1}$
implements the Toffoli gate with only one Rydberg level needed.
If we consider $k$ control atoms and a target atom, all within a blockade sphere of each other, then the $2k+3$ pulse sequence
$$
\left(\prod_{i=0}^{k-1}  R(\pi)_{0,r}^{c_{k-i}}\right)   R(\pi)_{1,r}^{t}   R(\pi)_{0,r}^{t}   R(\pi)_{1,r}^{t}   \left(\prod_{i=1}^k   R(\pi)_{0,r}^{c_i}\right)
$$
 immediately gives the ${\rm C}^k$-NOT gate
with an overall minus sign. The error scaling of this gate is approximately linear in $k$ because of the less than $k$ times larger spontaneous emission error than for the two-qubit blockade gate.
This type of multi-qubit gate is of particular interest in design of efficient
quantum circuits\cite{Beckman1996}.

\subsubsection{Other approaches}

For completeness we would like to mention some alternative paradigms for quantum information processing with Rydberg atoms. In addition to the Rydberg CQED approach described in the Introduction which uses moving Rydberg atoms\cite{Raimond2001} there are additional possibilities working with trapped atoms. Instead of only using Rydberg levels in a transient fashion for achieving a two-atom interaction,
one might consider encoding many bits of information in the  multiplicity of levels of a Rydberg atom.
In this way Grover's search algorithm\cite{Grover1997} was
implemented in a Rydberg atom in\cite{Ahn2000} and other work has considered the implementation of logic gates between bits that are Rydberg encoded\cite{Remacle2001}. These ideas are obviously constrained
by the finite lifetime of Rydberg levels, although some novel approaches
to increasing the coherence time have been studied\cite{Brion2005,Minns2006}. It has also been shown in \cite{Gillet2010} using a density matrix analysis that continuous driving of a two-atom system leads to a stationary degree of entanglement despite radiative decay of the Rydberg levels.

On a more fundamental level it has been argued convincingly  that even if the finite lifetime of excited states was not an issue, the spectrum of energy levels in atomic systems does not provide a scalable approach to quantum computing\cite{Blume-Kohout2002}.
This is due to the fact that a single atom encoding requires an exponential increase in the available physical resources to
provide  a linear increase in the dimension of Hilbert space available for computation.

\subsection{Scalability of a Rydberg gate quantum computer}
\label{sec.arrayscaling}

\begin{figure}[!t]
\includegraphics[width=7.5cm]{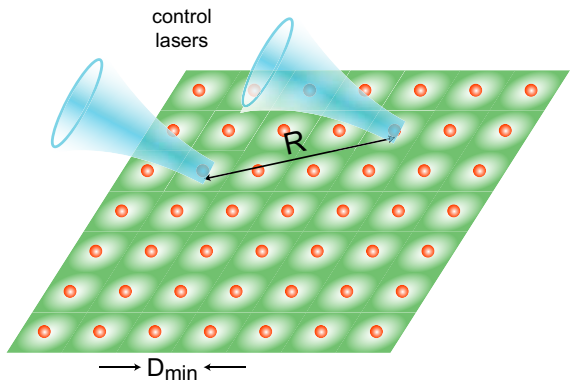}
\caption{(Color online) Neutral atom qubit array in an optical lattice with period $D_{\rm min}$.
A two-qubit gate between sites separated by $R$ is implemented with focused laser beams.
}
\label{fig.array}
\end{figure}

The various gate protocols described in the preceding sections rely on long range interactions between Rydberg excited atoms. Since high fidelity gates are possible at tens of microns of separation (see Figs. \ref{fig.czerror},\ref{fig.cz_interaction_error}) it is natural to ask how many qubits could be directly entangled, without mechanical motion, in an array of trapped atoms, such as that shown in Fig. \ref{fig.array}.
We envision an optical lattice of sites with one atom in each site. The largest
possible separation $R_{\rm max}$ depends
on how high a value of $n$ is feasible. The number of connected qubits then
scales as $(R_{\rm max}/D_{\rm min})^d$ with $d$ the space dimensionality of the array.
The lattice spacing $D_{\rm min}$ is constrained by technical considerations associated with
creating the lattice sites, but also by the requirement that a Rydberg excited electron with characteristic
orbital radius   $\sim a_0 n^2$ should not   collide with a neighboring ground state atom.
An analysis of the scaling of gate errors with $R$ and $n$ in the van der Waals limit, taking these considerations into account,  leads to the results\cite{Saffman2008}
\begin{subequations}
\begin{eqnarray}
N_{\rm max, vdW}^{(2D)}&=& C_{2D} E^{1/3} n^{2/3}\\
N_{\rm max, vdW}^{(3D)}&=& C_{3D} E^{1/2} n.
\end{eqnarray}
\label{eq.scaling}
\end{subequations}
Here $C_{2D},C_{3D}$ are constants that are expected to vary only slightly between different atomic species
and $E$ is the gate error discussed in Secs. \ref{sec.bgate},\ref{sec.othergates}.

A realistic  estimate assuming $n=100$ and $E=.001$ results in\cite{Saffman2008} $N_{\rm max,~vdW}^{(\rm 2D,3D)}=470,\, 7600$. Higher $n$ would of course lead to even larger numbers. These estimates  suggest that a moderately sized, fully interconnected quantum computer with hundreds of qubits in 2D and thousands of qubits in 3D is possible using Rydberg interactions. Actually building such a device will require solutions to several unresolved problems.
A recent analysis of some of the issues related to  scalability can be found in\cite{Beals2008}.
 It is not our intention to engage in a detailed discussion  here.   Nevertheless, a brief overview of what the challenges are
seems appropriate.

A set of  criteria for implementing a quantum computing device have been given by
DiVincenzo\cite{DiVincenzo1998}. In principle all requirements can be met by the type of neutral atom qubit array shown in Fig. \ref{fig.array}. If we limit ourselves to a 2D geometry, site selective addressing of neutral atom qubits with focused laser beams or magnetic field gradients is relatively straightforward, and has been demonstrated in several experiments\cite{Jones2007,Yavuz2006,Urban2009,Schrader2004,Lengwenus2006}. Taking advantage of the higher connectivity of a 3D array makes the problem of site selective
addressing and measurement more challenging, although possible solutions have been proposed\cite{Weiss2004,Vaishnav2008}.

The question of preparing single atom occupancy in a large lattice is more challenging. Stochastic loading
of atoms from a cold background vapor is governed by Poisson statistics giving a maximum success probability of $P_1=1/e\simeq 0.37$ at each site. The probability of loading each of $N$ sites with one atom thus scales as
$P_N=P_1^N=e^{-N}$
which is not useful for large $N.$ The situation  can be improved slightly using collisional blockade \cite{Schlosser2001} or light assisted collisions\cite{Nelson2007} to raise the success probability to $P_1\simeq 0.5.$
Of course it may be sufficient to simply register which sites are occupied, and then use only those sites for performing calculations, although this will imply a departure from the scaling relations of Eqs. (\ref{eq.scaling}).

It is therefore   tempting to seek ways to  approach deterministic single atom loading in all sites of an array.
An elegant solution
involves exploiting the superfluid- Mott Insulator transition in a cloud of ultracold atoms, and then transferring the Mott Insulator to a longer period lattice that is optically resolvable\cite{Peil2003}.
Other solutions involve imperfectly loading a lattice and then removing empty or doubly occupied sites, leading eventually to a zero entropy, perfect lattice  \cite{Weiss2004,Vala2005a,Rabl2003,Wurtz2009}.  There is also the
possibility of using a movable optical tweezer to systematically transfer atoms between sites as needed\cite{Beugnon2007}, or from a reservoir into
lattice  sites. A different type of approach,
to be discussed in Sec. \ref{sec.loading},
uses entanglement to enable deterministic single atom loading\cite{Saffman2002}.
Alternatively, collective encoding of qubits in many atom ensembles\cite{Brion2007d} removes the requirement of single atom loading altogether. We discuss this approach in Sec. \ref{sec.1000}.

Implementation of  extended quantum algorithms requires error correction which depends on the ability to measure the state of a qubit and restore it to an initial state, without atom loss, and  without disturbing
qubits at  neighboring sites. Implementation of multiple  measurement cycles requires further that either the
measurement does not heat the atom, or that recooling at a single site is possible.
 Measurement of the state, and not just the presence, of a neutral atom qubit in free space has until now relied on
mechanical ejection of an atom in one of the qubit states, followed by a state insensitive atom number measurement\cite{Kuhr2003,Treutlein2004,Jones2007}. Although a lost atom
could be replaced from a reservoir  the requirement of moving a replacement atom into the emptied site implies a slow measurement cycle. Recent developments in achieving stronger coupling between single atoms and light in free space with tightly focused optical beams\cite{Tey2008,Aljunid2009}, as well
as resolution of individual sites in short period lattices\cite{Bakr2009}  suggest that the capability of loss free, state selective measurements is not far off.

A related issue in neutral atom systems is that  optical traps have characteristic depths that are small compared to $k_B \times 300~\rm K$ so atom loss due to background collisions with room temperature  atoms cannot be completely eliminated. There will therefore be a requirement for monitoring and correcting qubit loss\cite{Vala2005b}, which could involve replacing lost atoms with fresh ones from a nearby reservoir.
This appears feasible in 2D geometries, but more difficult to implement in the interior of a 3D lattice.
The number of sites that can be maintained for an arbitrary length of time is limited by $N<\tau_{\rm loss}/t_{\rm replace}$ where $\tau_{\rm loss}$ is the time constant for atom loss
and $t_{\rm replace}$ is the time needed to replace a lost atom.
With excellent vacuum pressure of $P\sim 10^{-11}~\rm Pa$ a characteristic collision limited lifetime in an optical
trap is $\tau_{\rm loss}\sim 10^4~\rm s.$ Assuming a replacement time of $t_{\rm replace}\sim 0.1 ~\rm s$
this implies an array size limit of $N\sim 10^5.$ Note that actually reaching this limit is dependent on the ability to rapidly check all the sites for atom loss. This can be done without disturbing the qubit state
at each site  using an ancilla bit at a cost of four gates and one measurement\cite{Preskill1998}. The measurement time should satisfy $N t_{\rm meas} < t_{\rm replace}.$ Even a fast measurement time of, for example, $t_{\rm meas}=100~\mu\rm s$ would limit the
array to a much smaller
$N=1000.$ This highlights the necessity of performing parallel operations in order to build a scalable system.

Ultimately even a thousand qubit computer will not be sufficient to solve hard
problems that are intractable on classical computers. Scaling to even larger numbers may require connecting multiple, smaller processors, using entanglement between stationary matter qubits and photonic qubits. This is also relevant for quantum networking and long distance quantum communication \cite{Kimble2008}. Rydberg interactions present unique opportunities for
quantum interfaces between light and matter, and we defer a discussion of this topic to Sec.
\ref{subsec.communication}.

\section{Experiments}
\label{sec.experiments}

Experimental demonstration of quantum logic with Rydberg atoms builds on
several decades of development of techniques for cooling, trapping, and manipulating atoms with electromagnetic fields. A good overview of this field can be found in   \cite{Metcalf1999}. In this section we
will discuss specific experimental capabilities  needed for neutral atom
logic gates. Much of the discussion in Sec. \ref{sec.groundtraps} is generic to schemes based on atomic qubits. We then turn to requirements specific to the use of Rydberg atoms, and summarize experimental progress in realizing Rydberg mediated quantum gates.

\subsection{Experimental techniques}

\subsubsection{Traps for ground state atoms}
\label{sec.groundtraps}

The starting point for neutral atom logic gate experiments is
cooling, trapping, and detection of isolated atoms. Observation of a single, cold neutral atom was first achieved by Hu and Kimble in a magneto-optical trap (MOT)\cite{Hu1994}. Subsequently several
research groups\cite{Frese2000,Schlosser2001} showed that a convenient setting for studying single atoms is provided by far-off resonance optical traps (FORTs)\cite{Miller1993,Grimm2000}.

FORTs can provide very long atomic confinement times that are in practice  limited only by collisions with untrapped room temperature atoms.   This is because the trap depth scales as $1/\Delta$ while the photon scattering rate,  which leads to heating, scales
as $1/\Delta^2$, with $\Delta=\omega-\omega_a$ the difference between the trapping laser frequency $\omega$ and the relevant atomic transition frequency $\omega_a$.
Trap lifetimes approaching one minute\cite{Frese2000} have been observed for Cs atoms in a 1064 nm Nd:YAG laser FORT. As long as the atoms can be cooled to $k_B T\ll E_{\rm trap}$ there is no fundamental limit to the trap lifetime. However, as mentioned above, optical traps have $E_{\rm trap}\ll k_B\times 300~\rm K$ so trap lifetimes are limited by collisions with hot background atoms. Paul traps for ions can have
$E_{\rm trap} > k_B\times 300~\rm K$ and extremely long lifetimes up to several weeks  have been observed.

With the availability of a single atom in an optical trap a qubit can be encoded in Zeeman or hyperfine ground states. Of particular importance for quantum information applications is
the coherence time of an encoded  quantum superposition state.  In order to maximize the
coherence time in the presence of background magnetic field fluctuations it is advantageous to use pairs of states that exhibit a field insensitive operating point, with a quadratic relative Zeeman shift for field deviations away from the optimum. In this way coherence times of many seconds  have been observed with  trapped ions in Paul traps\cite{Langer2005,Benhelm2008}.

 In alkali atoms with a $^2S_{1/2}$ ground state and nuclear spin $I$   the $f=I\pm1/2$, $m_f=0$ hyperfine clock states have a quadratic relative shift at zero applied field.
There are also pairs of states with $\Delta m_f=1$ that have a quadratic shift, but only at bias fields
greater than $10~\rm mT$, which is problematic in the context of excitation of magnetically sensitive Rydberg states. Alternatively one can find low field quadratic shifts for states that
have $\Delta m_f\ge 2$. However, driving Raman transitions between these states requires multi-photon transitions that tend to be slow. The alkali clock states  therefore appear best suited for qubit encoding at the present time.

In addition to decoherence due to external fields the FORT traps themselves also limit qubit
coherence\cite{Saffman2005a,Kuhr2005,Windpassinger2008b}.
At detunings large compared to the fine structure splitting of the states that are excited by the FORT laser
there is a cancelation of Raman amplitudes, so the rate of state changing photon scattering is very low, scaling as $1/\Delta^4$\cite{Cline1994}.
Hyperfine relaxation times of several seconds were observed in the original work\cite{Cline1994} with Rb atoms.  A subsequent demonstration with an extremely far detuned CO$_2$ laser operating at $10.6~\mu\rm m$
pushed the relaxation time to greater than $10~\rm s$\cite{Takekoshi1996}.
An additional source of decoherence arises from atomic motion in the optical trap which induces differential AC Stark shifts on the qubit states. This issue disappears for atoms in the motional ground state, but is otherwise the limiting factor for qubit coherence in optical traps\cite{Saffman2005a}.
The differential shift can be  reduced using an additional weak laser beam that does not provide trapping, but cancels the trap induced differential shift\cite{Kaplan2002}.

Photon scattering and motional decoherence effects can be significantly reduced by working with dark optical traps where the atoms are localized near a local minimum, instead of a local maximum of the trapping light intensity\cite{Chaloupka1997,Arlt2000}. Several research groups have demonstrated atom trapping in this type of setup\cite{Kuga1997,Ozeri1999,Kulin2001,Terraciano2008,Isenhower2009a}.
We anticipate that trapping and coherence times of at least several seconds will
 be achieved with single atom qubits in optimized, dark optical traps, although a definitive experimental demonstration has not yet been presented. Although long coherence times are important in any
quantum computing device, it is difficult to say whether or not errors in a quantum computer will be dominated by the memory coherence time or by gate errors.
The scaling of memory to gate errors
depends on many factors including the size of the register, the computation
being performed, and other architectural considerations.

Scalability of optical trapping to many qubits relies on either multiplexing  traps using diffractive optical elements\cite{Bergamini2004}, or lens arrays\cite{Dumke2002}, or taking a different route of optical trapping  in lattices\cite{Jessen1996}. Lattices formed from counterpropagating beams in the near infrared have submicron periods and are not readily compatible
with site selective addressing, measurement, and control, although possible solutions\cite{Saffman2004, Zhang2006,  Yavuz2007, Cho2007, Gorshkov2008b,Vaishnav2008}, as well as experimental capabilities\cite{Karski2009, Lundblad2009,Bakr2009} are being actively developed. Alternatives rely on lattices formed from a long wavelength
${\rm CO}_{2}$  laser\cite{Scheunemann2000} or multibeam lattices that have adjustable, longer scale periodicity\cite{Peil2003,Nelson2007}. A new idea, recently demonstrated in \cite{Kubler2010} is to use micron sized  vapor cells, each small enough to enable an effective blockade interaction throughout the volume of the cell,
as a means of defining an ensemble qubit.

Magnetic traps, which do not suffer from photon scattering,  are an interesting alternative to optical approaches\cite{Fortagh2007}. Lifetimes in the range of  10 min. have been achieved with neutral atoms in a cryogenic magnetic trap\cite{Emmert2009} and hyperfine coherence times exceeding 1 s have been
demonstrated\cite{Treutlein2004}. Arrays of magnetic
traps\cite{Weinstein1995,Hinds1999,Grabowski2003,Gerritsma2007} are also a potential setting for
encoding a qubit register.

Finally we note that a large scale atom based quantum computer will require accurate and fast spatial control of several laser beams. Different technologies are suitable for this task  including electro-optic deflectors\cite{Schmidt-Kaler2003b}, acousto-optic deflectors\cite{Nagerl1999,Kim2008}, and microoptoelectromechanical systems\cite{Knoernschild2009}. Scaling to arrays with more than a few tens of qubits will likely require further development of specialized devices.

\subsubsection{Traps for Rydberg atoms}
\label{sec.Rydbergtrap}

Irrespective of the type of trap used to hold the qubits we must also consider the effect of the trapping potential on Rydberg states. Ideally we wish to have  the same trapping potential for
both ground and Rydberg states. If this is not the case excitation to Rydberg levels will result in
motional excitation of the atom, and more importantly, undesired entanglement between the center of mass and qubit degrees of freedom. This can be seen by the following simple argument. Suppose the qubit state
$\ket{\psi}=a\ket{0}+b\ket{1}$ is stored in an atom in the ground state $\ket{0}_{\rm vib}$ of the trapping potential. The total state of the qubit plus atom
is $\ket{\Psi}=\ket{\psi}\otimes\ket{0}_{\rm vib}.$
Vibrational excitation during a Rydberg cycle  will lead to the new state
$\ket{\Psi'}=a\ket{0} \otimes \ket{0}_{\rm vib}+b\ket{1} \otimes ( c\ket{0}_{\rm vib}+d\ket{1}_{\rm vib})$ where, for simplicity,
we have only considered excitation of the first vibrational state with amplitude $d.$
Tracing over the vibrational degrees of freedom gives the reduced density matrix
\be
\rho_{\rm qubit}={\rm Tr}_{\rm vib}[\rho]=
\begin{pmatrix}
|a|^2& ab^* c^*\\
a^*b c&|b|^2
\end{pmatrix}.
\label{eq.qubitvib}
\ee
Since $|c|< 1$ vibrational excitation results in reduced coherence of the qubit.

\begin{figure}[!t]
\includegraphics[width=8.7cm]{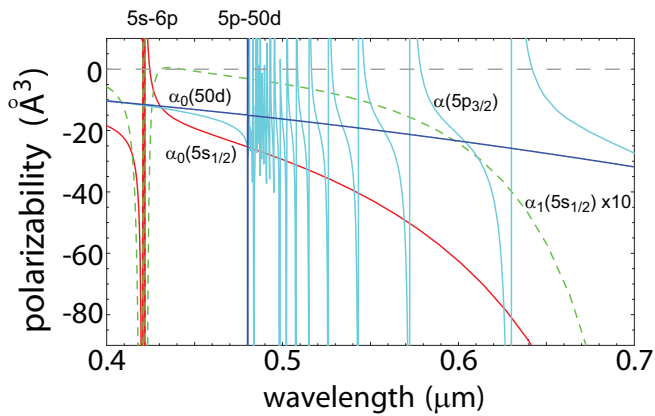}
\caption{(Color online) Polarizability of Rb ground (red curve), first resonance level (aqua curve), and Rydberg 50d (dark blue curve)
states. The vector polarizability of the ground state is shown by the dashed green curve.
}
\label{fig.polarizability}
\end{figure}

In order to have the same trapping potential for ground and Rydberg states in an optical trap the polarizability must be the same for both levels. The polarizability of a highly excited Rydberg state is negative being essentially that of a free electron, $\alpha= -e^2/m\omega^2$, where $-e$ is the electron charge and $m$ is the electron mass.
In a red detuned, bright ground state trap the polarizability is positive. Nevertheless, matching can be achieved at specific wavelengths
by working close to a ground - intermediate level resonance, or a resonance between Rydberg and intermediate levels. Details of specific schemes are given in \cite{Safronova2003, Saffman2005a}.

If we use a blue detuned dark optical trap or dark lattice then
there is a broad region to the blue of the first resonance lines in the heavy alkalis where the ground state polarizability is negative.
As shown in Fig. \ref{fig.polarizability} exact matching between the
Rb $5s$ ground state and the $50 d$ Rydberg state occurs at $\lambda=430~\rm nm$. It turns out that the polarizability of the first resonance level is also approximately equal to the ground state polarizability at this wavelength, which is advantageous for Doppler cooling inside the optical trap. Furthermore the ground state vector polarizability, which determines the rate of hyperfine changing Raman scattering events,  is extremely small. This short wavelength
matching point is thus very attractive for neutral atom optical traps. The matching wavelength does not change significantly with choice of Rydberg level since the polarizability of the
$50 d$ state is already about 95\% of the free electron polarizability.
A similar coincidence point occurs at a longer wavelength for Cs  atoms.  The notion of an optical trap for Rydberg atoms can also be extended to an optical lattice setting as discussed in \cite{Dutta2000}.

An important issue when using optical traps to confine Rydberg atoms is the problem of Rydberg photoionization due to the trapping light. Photoionization rates in mK trapping potentials substantially exceed the radiative decay rate\cite{Saffman2005a,Potvliege2006} and represent the limiting factor for Rydberg trapping.
Thus the experiments to date on coherent excitation in optical traps have relied on turning off the trapping potential during the Rydberg excitation pulse(see Sec. \ref{sec.excitation}). This problem
may be greatly reduced in a dark optical trap where the atom sits at a minimum of the trapping light intensity. However, if we consider excitation of very high lying levels with $n>100$ in order to achieve long range gates, as in Sec. \ref{sec.bgate}, the wavefunction of the Rydberg electron will sample regions of non-negligible trapping light intensity, even in dark optical traps. As is well known, a free electron cannot absorb a photon, and therefore the photoionization cross section tends to be localized near the nucleus. A careful analysis of the photoionization rate in dark traps has not yet been performed.

An alternative to optical traps is to use low frequency electromagnetic traps which can be effective for both ground and Rydberg atoms, see \cite{Choi2007} for an extended discussion.
Rydberg atoms in low field seeking Stark states were loaded into an electrostatic trap in \cite{Hogan2008}.
Magnetostatic trapping of high angular momentum Rydberg atoms was demonstrated in a strong field of several Tesla in  \cite{Choi2005,Choi2006}. Trapping in a high gradient quadrupole field and in combined magnetic and electric traps was studied in several papers  \cite{Lesanovsky2005,Hezel2007,Schmidt2007}.
 It was proposed\cite{Hyafil2004, Mozley2005}
to  use the electrodynamic trap described in\cite{Peik1999} together with conducting planes for inhibition of spontaneous emission to create a long coherence time trap for circular Rydberg states.  A successful demonstration of this idea in two proximally located traps would open the door to long time scale, high precision studies of the dipole-dipole interaction.
It should be remarked that the electromagnetic geometries, although capable of trapping Rydberg atoms, tend to rely on high $m$ states which  are not readily compatible with few photon laser excitation.
This restriction has been relaxed in recent work\cite{Mayle2009,Mayle2009b}  that has shown
theoretically the feasibility of magnetic trapping of $s,p,$ or $d$ Rydberg states in Ioffe-Pritchard geometries, with good qubit coherence.

Approaches based on arrays of magnetic traps\cite{Grabowski2003,Gerritsma2007}, although promising for holding ground state atoms, may be difficult to combine with Rydberg atoms. In order to achieve tightly
confining magnetic traps the surface to trap distance is typically on the order of tens of microns,
which can lead to undesired interactions between the surface and Rydberg atom.
The question of Rydberg-surface interactions may also be problematic for the microcell approach demonstrated in \cite{Kubler2010}.
 Indeed the coupling of Rydberg atoms to conductors forms the basis for hybrid entanglement  schemes to be discussed
in Sec. \ref{sec.hybrid}. A combination of magnetic trapping ideas and hybrid interfaces may
eventually prove fruitful, but remains largely unexplored.

\subsection{Coherent excitation of Rydberg states}
\label{sec.excitation}

\begin{figure}[!t]
\includegraphics[width=2.cm]{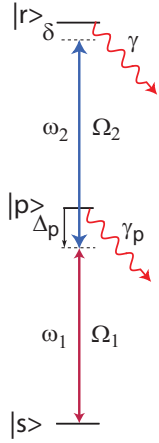}
\caption{(Color online) Two-photon excitation of Rydberg level $\ket{r}.$
The radiative decay rates of $\ket{p}$ and $\ket{r}$ are $\gamma_p=1/\tau_p$ and $\gamma=1/\tau$ respectively. }
\label{fig.twophoton}
\end{figure}

Laser excitation and spectroscopy of Rydberg atoms has a long history dating back to the development of the tunable dye laser in the early 1970's.
Early work is reviewed in \cite{Fabre1983}. Starting from a ground state Rydberg states can be generated with $1,2,3$ or more photon transitions. Using electric-dipole allowed transitions ground $s$ states can be coupled to $p$ states with one photon, to
$s$ or $d$ states with two photons, and $p$ or $f$ states with $3$ photons. These selection rules
can be modified by Stark or Zeeman mixing of the Rydberg states, or by direct excitation via quadrupole transitions\cite{Tong2009}.
The gate protocols discussed in Sec. \ref{sec.gates} require
extremely precise, coherent excitation and deexcitation of Rydberg states.
One photon excitation from the ground state requires a $\sim 297~\rm nm$ photon in Rb
which is possible\cite{Thoumany2009}, but coherence has not yet been demonstrated.

A widely used approach, which does not require deep UV wavelengths,  relies on  two-photon excitation
via the first resonance level as shown in Fig. \ref{fig.twophoton}.
For example in $^{87}$Rb field 1 is at 780 nm for excitation via $5p_{3/2}$ and field 2 is near 480 nm.
The first example of Rydberg spectroscopy with narrow linewidth lasers using this approach was a study
of Autler-Townes spectra in a cold $^{85}$Rb sample\cite{Teo2003}.
Autler-Townes spectra in Rydberg excitation were also studied later in \cite{Grabowski2006}.  Three photon excitation schemes with narrow linewidth lasers have also been used in experiments with Cs\cite{Vogt2006} and Rb\cite{Thoumany2009b}.

 When the intermediate level detuning $\Delta_p=\omega_1-\omega_{ps}$ is large compared to the width of the hyperfine structure of the
$\ket{p}$ level
the two-photon Rabi frequency  is given by $\Omega=\Omega_1 \Omega_2/2\Delta_p$. The one-photon  Rabi frequencies are $\Omega_1=-e{\mathcal E}_1\bra{p} {\bf r}\cdot {\bfepsilon}_1\ket{s}/\hbar$,
$\Omega_2=-e{\mathcal E}_2\bra{r}{\bf r}\cdot {\bfepsilon}_2\ket{p}/\hbar$,
with ${\mathcal E}_j, \bfepsilon_j$ the field amplitudes and polarizations.  The transition matrix elements
can be reduced via the Wigner-Eckart theorem to  an angular factor plus the  radial integral of Eq. (\ref{eq.radial}). For the $s-p$ transition the radial integral is known (for $^{87}$Rb $\langle r\rangle_{5s}^{5p}=5.1\times  a_0$) and for the $p-r$ transition it can be readily calculated numerically.  The following  expressions are accurate to better than 10\%
for $^{87}$Rb:  $\langle r\rangle_{5p}^{ns}=.014 \times (50/n)^{3/2}a_0$ and $\langle r\rangle_{5p}^{nd}=-.024 \times (50/n)^{3/2} a_0$.

There are several potential sources of errors when using two-photon excitation with well defined pulse areas for coherent population transfer between ground and Rydberg states. Partial population of the intermediate $\ket{p}$ level results in spontaneous emission and loss of coherence. The probability of this occurring during a $\pi$ excitation pulse of duration $t=\pi/|\Omega|$ is
$P_{\rm se}=\frac{\pi\gamma_p}{4|\Delta_p|}\left(q+\frac{1}{q}\right)$
where $q=|\Omega_2/\Omega_1|.$ The spontaneous emission is minimized for $q=1$ which lets us write the
Rabi frequency as
$$
\Omega=\frac{P_{\rm se}}{\pi}\frac{|\Omega_2|^2}{\gamma_p}.
$$
We see that fast excitation with low spontaneous emission is possible provided $\Omega_2$ is sufficiently large. This is increasingly difficult as $n$ is raised since $\langle r \rangle_{5p}^{nl}\sim 1/n^{3/2}$.
Put another way, at constant $\Omega$ and $P_{\rm se}$ the required optical power scales as $n^3$.

Another issue is detuning errors due to Doppler broadening. These can be reduced
from $\delta_{\rm max}=(k_1+k_2)v$
to $\delta_{\rm max}=( k_1-k_2)v$ for atomic velocity $v$ using counterpropagating excitation beams. Excitation can be made Doppler free if $|{\bf k}_1|=|{\bf k}_2|$\cite{Lee1978}, or by tuning
close to the intermediate level\cite{Reynaud1982}. However, neither approach is well suited for coherent experiments due to the slow rate of the first approach, which has no intermediate level resonance,  and the large spontaneous emission probability incurred in the second.

Another source of detuning errors arises from the AC Stark shifts caused by the excitation lasers.
The dominant contributions come from the near resonant interactions of $\omega_1$ with $\ket{s}$ and
$\omega_2$ with $\ket{r}.$
The ground state $\ket{s}$ is Stark shifted by $\delta_s=|\Omega_1|^2/4\Delta_p$ while the Rydberg state $\ket{r}$ is shifted by $\delta_r=-|\Omega_2|^2/4\Delta_p.$ We see that the transition shift $\delta_s+\delta_p$ vanishes when $|\Omega_1|=|\Omega_2|$ or $q=1$, which is another reason to work with
equal Rabi frequencies, in addition to  the minimization of spontaneous emission. The cancelation is not perfect since there are additional off-resonant contributions from $\omega_1$ acting on $\ket{r}$
(which tends to be small) and $\omega_2$ acting on $\ket{s}$, which tends to be larger due to the high intensity of the field driving $p-r.$ The intensity dependent Stark shift of the transition frequency has been observed in experiments with trapped $^{87}$Rb atoms\cite{Urban2009b}.

When working with trapped atoms it is also necessary to take account of the differential trap shift between ground and Rydberg levels. Besides the loss of coherence described by Eq. (\ref{eq.qubitvib})
the trap shifts, which depend on the position of the atom in the trap, can easily exceed $|\Omega|$
in a few mK deep trap,  which
would be disastrous for coherent excitation. For this reason, and also because the trapping light in all experiments performed to date rapidly photoionizes the Rydberg atoms, the trap is turned off during the Rydberg excitation pulse.  This is  problematic in the context of many qubit systems, and therefore the development of trap architectures that are insensitive to the internal state, as discussed in Sec. \ref{sec.Rydbergtrap}, will be an important topic for future work.

Transition shifts due to magnetic fields, and Rydberg level shifts due to small electric fields are also of concern. The magnetic sensitivity is manageable in optical traps, but could be problematic
for atoms in thermal motion in a magnetic trap. Ground state polarizabilities of alkali atoms are small enough that typical stray laboratory fields have a negligible effect. However, as has been discussed in Sec. \ref{sec.properties}, the DC Stark shift of a Rydberg state scales as $n^{7}$. This puts severe limits on the field stability needed for excitation of very high lying levels.

A convenient way of measuring and controlling the very small field strengths involved is to use the shift of the Rydberg level itself as a diagnostic\cite{Frey1993,Osterwalder1999}.
More recent work has used electromagnetically induced transparency (EIT)\cite{Fleischhauer2005}
for precise Rydberg spectroscopy. EIT  is a destructive interference effect with a very narrow line width, and it can hence be used to measure very precisely the Rydberg series of energy levels. In \cite{Mohapatra2007}, the fine structure splitting of the Rb $nd$ series with $n$ up to 96 were thereby measured. In the same work, it was pointed out that due to the large dipole moment of Rydberg excited states, they are very sensitive electric field probes. The narrow EIT line width thus makes it possible to detect a very small electric field, or, conversely,  to control the transmission properties of an atomic ensemble with very weak switching fields \cite{Mohapatra2007, Mohapatra2008, Bason2008}. Rydberg spectroscopy using EIT signals in Cs cells has also been demonstrated\cite{Zhao2009}. In  addition EIT  has been shown to be useful for the determination of atom-wall induced light shifts and broadenings in thermal vapor microcells\cite{Kubler2010}.

In order to get a sense of the errors involved consider the following example of excitation of
the $^{87}$Rb $100d_{5/2}$ level via $5p_{3/2}$. Let us assume $\pi$ polarized beams with powers of  $1~\mu\rm W$ at 780 nm and $300~\rm mW$ at 480 nm
 focused to spots with Gaussian waist $w=3~\mu\rm m$.
This gives single photon Rabi frequencies of $225, 210~\rm MHz.$
The light is detuned from $5p_{3/2}$ by $\Delta_p/2\pi = 20~\rm GHz$. These parameters couple  $m=0$ ground states to $m=\pm 1/2$ Rydberg states with a Rabi frequency
$\Omega/2\pi = 1.2~\rm MHz$. The probability of spontaneous emission from the $p$ level during a $\pi$ pulse
is $P_{\rm se}=5\times 10^{-4}. $
 The fractional excitation error after a $\pi$ pulse due to Doppler broadening is $P_{\rm Doppler}=|\delta/\Omega|^2 $. For $^{87}$Rb atoms at $T=10~\mu\rm K$ and counterpropagating   excitation beams we find
$P_{\rm Doppler}=4\times 10^{-4} $. Thus, coherent excitation of a very high lying Rydberg level   with
combined spontaneous emission and Doppler errors
below $10^{-3}$ is within reach of current experimental capabilities.

It is also necessary that the two-photon excitation be performed with well stabilized lasers so that the
detuning $\delta=\omega_1+\omega_2-\omega_{rs}$ is small compared to $\Omega.$ This can be achieved by locking the lasers to stabilized optical reference cavities\cite{Johnson2008,Bohlouli-Zanjani2006}. It is also possible to use the Rydberg atoms themselves as a frequency reference\cite{abel2009}. The relative phase of $\omega_1, \omega_2$ should also be well defined for the duration of a Rabi pulse. Locking the lasers to stable, high finesse resonators readily gives linewidths at the $\sim 100~\rm Hz$ level which is more than adequate for $\mu\rm s$ timescale pulses. Modern frequency comb techniques\cite{Cundiff2003} could also be used for both frequency and phase stabilization of the Rydberg lasers.

Even with the above imperfections under control there  is one more significant issue that must be confronted before coherent Rabi oscillations can be observed. We are coupling hyperfine ground states
characterized by quantum numbers $n,I,j,l,s,f,m_I,m_f$ to highly excited Rydberg fine structure states that have negligible hyperfine structure, and are therefore described by the quantum numbers $n',I,j',l',s,m_I',m_j'.$
In most cases there are two Rydberg Zeeman states with different values of $m_j'$ that have nonzero electric dipole matrix elements with the ground state. Only one Rydberg Zeeman state is coupled to if we start from a stretched ground state $m_f=\pm f$ or use $\omega_1$ with $\sigma_\pm$ polarization coupling via a $np_{1/2}$ level.
Apart from these special cases any difference in energy between the excited states due to $m_j'$ dependent Stark or Zeeman shifts will lead to a complex, nonsinusoidal excitation dynamics since $\Omega$ is also dependent on $m_j'$.
To avoid this problem it is necessary to optically pump the ground state atoms into a specific $m_f$ state, and apply a bias field to separate the $m_j'$ states by an amount that is large compared to $\Omega.$

\begin{figure}[!t]
\includegraphics[width=7.5cm]{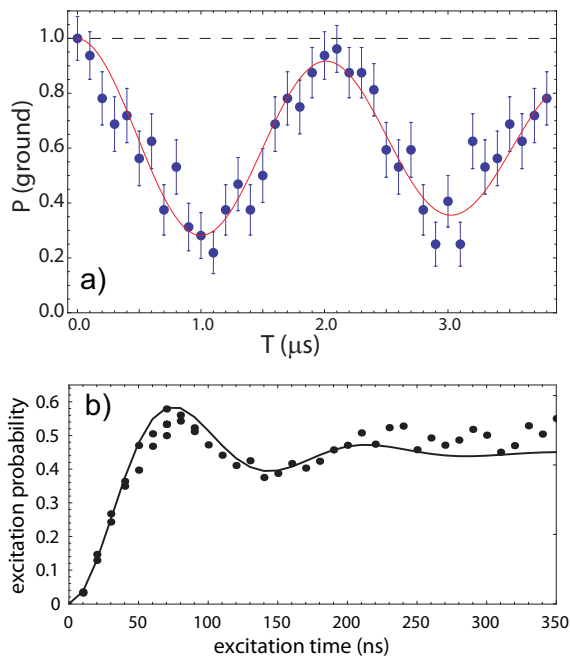}
\caption{(Color online) Rabi oscillations between ground and Rydberg levels: a) using single atoms from \cite{Johnson2008} and b) in a sample with $\sim 100$ atoms from \cite{Reetz-Lamour2008b}.
}
\label{fig.rydbergrabi}
\end{figure}

Taking the above considerations into account the first demonstrations of Rabi oscillations between ground and Rydberg levels were reported in 2008\cite{Johnson2008,Reetz-Lamour2008b}.  Fig. \ref{fig.rydbergrabi}a
shows oscillations of single $^{87}$Rb atoms confined to an optical trap with radius of about $3~\mu\rm m$ and  optically pumped into
$f=2,m_f=2.$ Rydberg excitation to $43d_{5/2},m_j=1/2$ used $\pi$ polarized lasers at 780 and 480 nm.
The excitation laser beams had waists that were a few times larger than the width of the optical trap so the
effects of spatial variation of the Rabi frequency were minimized.
The  confining to an optical trap  was turned off before the excitation lasers were applied. After a variable length excitation pulse the trap was turned on again which photoionized the Rydberg atoms before they could radiatively decay. Loss of a Rydberg atom from the trap therefore provided a signature of successful Rydberg excitation. The less than 100\% probability of exciting a Rydberg atom was attributed mainly to Doppler broadening at $T=200~\mu\rm K$ and the finite Rydberg detection efficiency since the ratio of the photoionization to radiative decay rates was $\gamma_{\rm pi}/\gamma\sim 20.$

Subsequent experiments extended these single atom results to even higher levels: $58d_{3/2}$\cite{Gaetan2009},
$79d_{5/2}, 90d_{5/2}$\cite{Urban2009}, $97d_{5/2}$\cite{Isenhower2010}, and $43d_{5/2}$\cite{Zuo2009}. With sufficient optical power, and careful minimization of stray electric fields, there is no reason why coherent excitation cannot be pushed to even higher $n$. This is attractive as a way of directly entangling many qubits, with the number scaling as $n^{2/3}$ in a 2D array (see Eq. (\ref{eq.scaling})).
A new constraint arises  when the energy separation between states $n,l,j$
and $n\pm1,l,j$ becomes comparable to $\Omega.$ Since the blockade gate spontaneous emission error from Eq. (\ref{eq.Ebar})
has a $1/\Omega\tau\sim 1/\Omega n^3$ contribution, and the error due to excitation of multiple $n$ levels goes as $[\Omega/(E_{n,l,j}-E_{n\pm1,l,j})]^2\sim \Omega^2 n^{6}$ minimization of the
 sum of these errors is independent of $n$ for large $n.$ However, undesired excitation of the noncoupled ground state which normally scales as $\Omega^2/\omega_{10}^2$ will become significant when
$\delta_{n,n\pm1}=|E_{n,l,j}-E_{n\pm1,l,j}|\ll \omega_{10}.$
A conservative estimate of the limit can be deduced by putting $\Omega/2\pi\sim 1~\rm MHz$ so that
a $10^{-4}$ gate error requires $\delta_{n,n\pm1}/2\pi >200~\rm MHz$ which corresponds to $n\sim325$.
We conclude that coherent two-photon oscillations with high fidelity are in principle feasible
up to  $n\sim 300$ given the requisite high power narrow linewidth laser system.
A practical limit may arise at lower $n$ due to the need to control external electric fields to limit Stark shifts which grow $\sim n^7.$

A large amount of work has been done pursuing the observation of  coherent oscillations in a many atom regime which is an essential capability for the ensemble qubit protocols mentioned in the Introduction, and discussed in detail in Sec. \ref{sec.collective}. If the atomic sample is smaller than the range of the Rydberg interaction a $\sqrt N$ collective enhancement of the Rabi frequency is expected. However, even when a full blockade is not achieved, Rydberg interactions serve to dephase the coherent oscillations.
Dephasing without blockade was observed in \cite{Johnson2008} by loading a small number of atoms into the optical trap. The visibility of the Rabi oscillations quickly decayed as the number of atoms  was increased from one to close to ten. As shown in Fig. \ref{fig.rydbergrabi}b
relatively weak oscillations have  also been observed in excitation to $47d_{5/2}$ with much larger samples containing about 100 atoms\cite{Reetz-Lamour2008a,Reetz-Lamour2008b}. In that work the
size of the cold atom sample was larger than the counterpropagating  excitation beams.
In order to reduce the broadening effects associated with a spatially dependent Rabi frequency
the much smaller 480 nm beam was given a close to ``top hat" spatial profile. Inspection of Fig. \ref{fig.rydbergrabi}b shows that  the excitation is only partially returned to the ground state, due to the presence
of dephasing mechanisms, as well as ``excitation trapping" resulting from population transfer to additional Rydberg levels not coupled to the light field\cite{Reetz-Lamour2008a}. 
These dephasing effects are qualitatively well reproduced by model calculations describing mesoscopic samples\cite{Stanojevic2009}.

Other excitation schemes going beyond simple square pulses have also been used in attempts to observe
collective oscillations.
Stimulated Raman adiabatic passage using a counterintuitive pulse sequence where the 480 nm laser is applied before the 780 nm was used
to demonstrate  excitation probabilities as high as $\sim70\%$ in mesoscopic samples of cold atoms\cite{Cubel2005,Deiglmayr2006}.  Excitation of mesoscopic
blockaded samples has been studied using ultracold Rb atoms close to the BEC transition temperature
\cite{Heidemann2007,Heidemann2008}. The $\sqrt N$ scaling was observed, and will be discussed in more detail in connection with collective effects in Rydberg ensembles in Sec. \ref{sec.scaling}.
The dephasing effects of atomic motion and Rydberg interactions can be compensated for using rotary echo techniques, which have been explored in recent calculations\cite{Hernandez2008b} and experiments\cite{Raitzsch2008,Younge2009a}.
Although coherent excitation in many atom samples has been studied in several experiments, and signatures of collective effects are clearly seen, a high visibility time domain record of many body Rabi oscillations has not
been achieved. This remains an outstanding challenge for quantum information applications of Rydberg ensembles.

\begin{figure}[!t]
\includegraphics[width=8.cm]{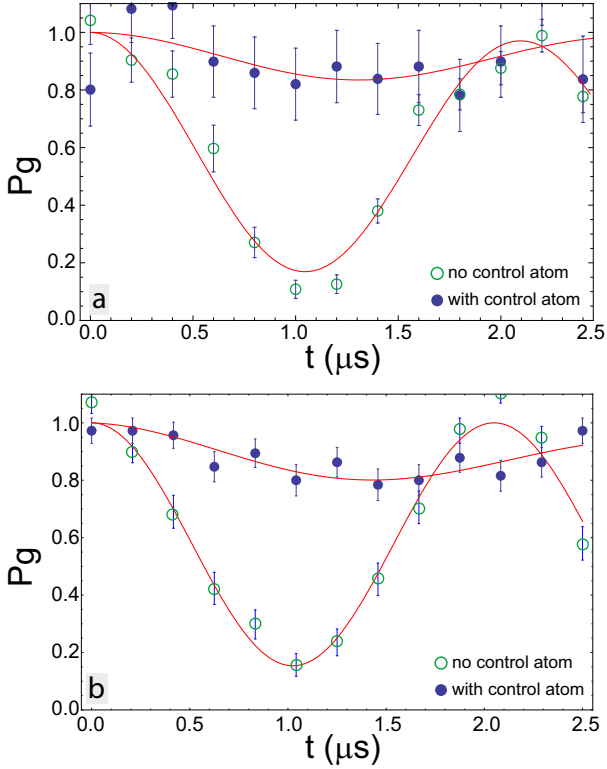}
\caption{(Color online) Two-atom Rydberg blockade
from\cite{Urban2009}. a)  The experimental data for Rydberg excitation of the target atom with and
without a control atom present. b) A Monte Carlo simulation accounting for experimental imperfections. The amplitude of the curve fit to the blockaded
oscillations is $a=0.09$  (experiment) and $a=0.11$ (simulation).
}
\label{fig.blockadeexp1}
\end{figure}

\subsection{Two atom blockade, two-qubit gates, and entanglement}
\label{sec.gatesexp}

It has recently been shown by experimental groups in Wisconsin and at Institute of Optics, Palaiseau that the experimental methods described above can be combined to demonstrate
Rydberg blockade\cite{Urban2009,Gaetan2009},
two-qubit quantum gates\cite{Isenhower2010},
and entanglement generation between two atoms\cite{Isenhower2010,Wilk2010}.

The blockade experiment in Wisconsin \cite{Urban2009} showed that excitation of a Rb atom to $90d_{5/2}$
blocked the subsequent excitation of an atom at $R\ge 10~\mu\rm m$ with a fidelity of about 90\%.
Experimental data, together with a Monte-Carlo simulation taking into account the finite blockade strength and experimental imperfections are shown in Fig. \ref{fig.blockadeexp1}.

\begin{figure}[!t]
\includegraphics[width=8.5cm]{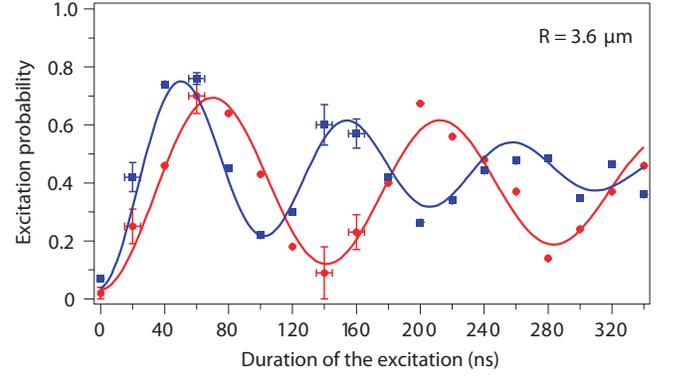}
\caption{(Color online) Collective excitation in the blockade regime
from\cite{Gaetan2009}.
Excitation of one atom versus collective excitation of two
atoms separated by $3.6~\mu\rm m$. The circles represent the probability to excite
atom a when atom b is absent. A fit to the data
yields a frequency of this Rabi oscillation $\Omega/2\pi=7.0\pm 0.2~\rm MHz.$
 The squares
represent the probability to excite only one atom when the two atoms are
trapped and are exposed to the same excitation pulse. The fit gives an
oscillation frequency
$\Omega/2\pi=9.7\pm 0.2~\rm MHz.$ The ratio of the oscillation
frequencies is $1.38\pm 0.03$ close to the value $\sqrt2$ expected for the collective
oscillation of two atoms.
}
\label{fig.blockadeexp2}
\end{figure}

A closely related experiment in Palaiseau\cite{Gaetan2009} used simultaneous excitation of two Rb atoms in traps separated by
$R\sim 3.6~\mu\rm m.$ A very strong blockade shift was obtained by using the
$58d_{3/2}$ F\"orster resonance, which was first identified in \cite{Reinhard2007}. The first stage of the excitation used 794 nm light coupling via the $5p_{1/2}$ level. This is preferable since the  radial matrix elements for excitation of $nd_{3/2}$ states are about $8~\times$ larger when exciting via the $5p_{1/2}$ level rather than the $5p_{3/2}$ level. As shown in Fig. \ref{fig.blockadeexp2} excitation of two atoms
with $\bshift\gg \Omega$ couples the two atom state $\ket{ss}$ to the symmetric singly excited state
$\frac{1}{\sqrt2}(\ket{gr}+e^{\imath\phi}\ket{rg})$ at the collectively enhanced Rabi frequency
$\Omega_c=\sqrt 2 \Omega.$ The $\sqrt2$ speedup is clearly seen in the data which is strong evidence for the creation of a two-atom entangled state.

\begin{figure}[!t]
\includegraphics[width=8.5cm]{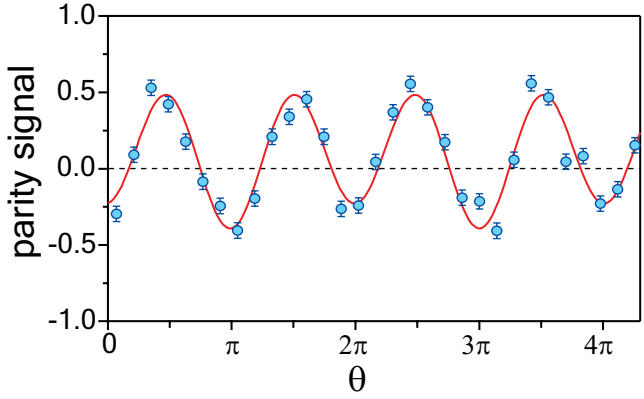}
\caption{(Color online) Measured parity signal from \cite{Wilk2010} for different durations $\theta=\Omega t$
of the analysing Raman pulse. The data are fitted by a
function of the form $y_0 + A \cos (\Omega t) +  B \cos( 2\Omega t)$.  The error bars on the data are statistical.
}
\label{fig.parity}
\end{figure}

The entanglement resides in the Rydberg levels and is therefore very short lived. In a subsequent
experiment\cite{Wilk2010}
the Palaiseau group mapped the state $\ket{r}$ to a different hyperfine ground state to create long lived entanglement of the form $\ket{\psi}=\frac{1}{\sqrt2}(\ket{01}+\ket{10})$. Note that the phase $\phi$ present in the Rydberg entangled state has been canceled by the mapping pulse, provided it is applied fast enough to neglect atomic motion, which is the case in the experiment. To verify the presence of entanglement the coherence of the two-atom density matrix was extracted from  
parity oscillation measurements on the output states\cite{Turchette1998} shown in Fig. \ref{fig.parity}.
The experimental results gave an entanglement fidelity of $F=0.46$ which is just under the threshold of $F=0.5$ for entanglement, whereas a perfectly entangled state would have $F=1.$ In the experiments there was only
a 61\% probability of both atoms remaining in the trap at the end of the entanglement sequence. Correcting for the atom loss  \cite{Gaetan2009b} it was inferred that the remaining atom pairs were entangled with a fidelity $F=0.75(7).$

\begin{figure}[!t]
\includegraphics[width=8.5cm]{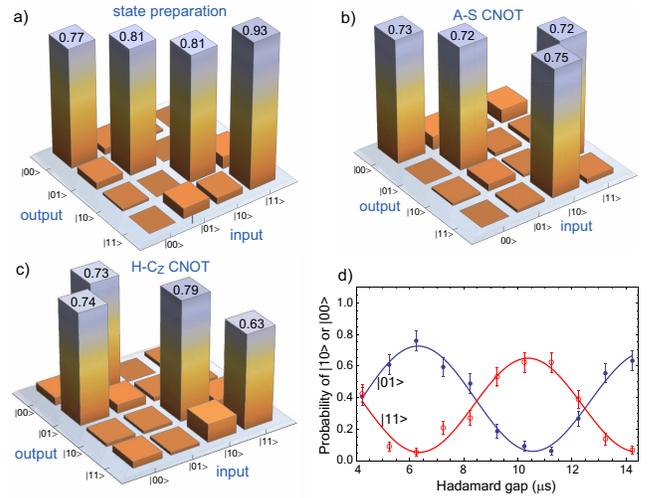}
\caption{(Color online) Rydberg blockade CNOT gate from \cite{Isenhower2010}.
Measured probabilities for a) state preparation, b) A-S CNOT, c) H-C$_Z$ CNOT and d) output states of the  H-C$_Z$ CNOT under variation of the relative phase of the $\pi/2$ pulses.
The reported matrices are based on an average of at least 100 data points for each matrix element and the error bars are $\pm 1$ standard deviation.
}
\label{fig.cnotexp}
\end{figure}

In work completed at the same time the Wisconsin group extended their observation of blockade to demonstration of a  CNOT gate. Taking advantage of their larger $10~\mu\rm m$ atom separation and the ability to apply different pulses to the two atoms they used a version of the
amplitude swap gate (see Fig. \ref{fig.cnot}), as well as the standard Hadamard - $C_Z$ sequence of Fig. \ref{fig.cnot} to acquire the data  shown in Fig. \ref{fig.cnotexp}.
In the same paper they also showed that by putting the control atom in a superposition state $\frac{1}{\sqrt2}(\ket{0}+\ket{1})$ before running the CNOT gate they could create approximations to the  entangled states
$\ket{B_1}=\frac{1}{\sqrt2}(\ket{00}+\ket{11})$ or
$\ket{B_2}=\frac{1}{\sqrt2}(\ket{01}+\ket{10})$
depending on the input state of the target atom.
Using parity oscillations the entanglement fidelity of $\ket{B_1}$ was measured to be $F=0.48\pm 0.06.$ The probability of losing at least one of the atoms during the gate was measured to be $0.17$, and correcting for the atom loss an {\it a posteriori} entanglement fidelity of $F=0.58$ was inferred.

These recent experiments represent the first demonstration of  quantum gates and entanglement  between a single pair of trapped neutral atoms.  The quality of the results is comparable to that obtained much earlier in Rydberg atom CQED experiments \cite{Rauschenbeutel1999,Hagley1997}. In both recent experiments the entanglement fidelity  obtained deterministically, without correction for atom loss, was close to,  but just under the threshold of $F=0.5$. Correcting for atom loss reveals a significant level of entanglement in the remaining atom pairs. This non-deterministic entanglement is not generally useful for quantum computing, but is relevant for other tasks such as Bell inequality experiments\cite{vanEnk2007}.  Although promising, the initial results should only be considered as first steps  as they lag far behind the
high fidelity results obtained with trapped ions\cite{Benhelm2008b}. Both the Palaiseau and Wisconsin experiments suffer from excess atom loss during the gate operation. This is in part due to the fact that pulse or blockade errors which leave the atoms with a non-zero amplitude to be in a Rydberg state at the end of the gate, lead to  a corresponding  probability
for photoionization  when the optical traps are turned back on after the gate operation. Thus, the measured probability to observe two atoms after completion of the gate was 0.61 in the Palaiseau experiments and 0.74 - 0.83, depending on the input state, in the Wisconsin experiments.

There is clearly a large gap between  the theoretical fidelity estimates presented in \ref{sec.bgate}
and the experimental results which show errors in 
the range of 30-50\%. These errors can be largely attributed to technical issues (atom loss due to finite vacuum, laser stability) as well as motional effects since the
atoms were relatively  hot in both sets of experiments($60~\mu\rm K$ in \cite{Wilk2010} and $>200~\mu\rm K$ in \cite{Isenhower2010} ). The ability to obtain quite good results with such hot atoms is a testament to the robustness of the blockade interaction. The Rydberg gate intrinsic errors are potentially 100-1000 times smaller than has been demonstrated.
It is likely that significant progress will be achieved in the coming years, which will pave the way for quantitative comparisons with the theoretical fidelity predictions, as well as demonstrations of   more complex operations with several qubits.   Continued development of the requisite optical and laser systems, combined with improved control of the spatial and momentum distributions of the atoms
will be important ingredients in ongoing work aimed at  approaching  the theoretical limits.

\section{Collective effects in Rydberg coupled ensembles}
\label{sec.collective}

\subsection{Blockade scaling laws in extended samples}
\label{sec.scaling}

The original concept of quantum information processing in atomic ensembles using dipole blockade \cite{Lukin2001} applies to localized samples small enough for the blockade to act across the whole ensemble.  To date, no experiments have been done that satisfy this criterion.  However, a number of experiments, on much larger samples, nevertheless show signs of the blockade effect;  these will be discussed below.  These larger samples are generally not useful for quantum information processing, but are of interest in giving insights into the
blockade effect.  We will refer to these samples as ``extended", reserving the term ``ensemble" for situations where blockade will allow only one Rydberg excitation at a time.

\begin{figure}[!t]
\begin{center}
\includegraphics[width=5.5cm]{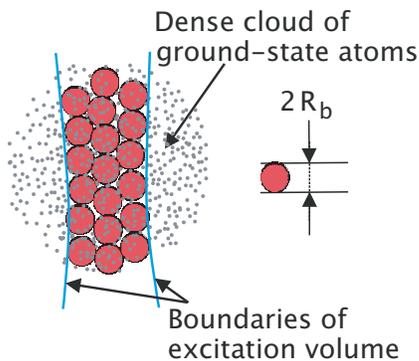}
\end{center}
\caption{In an extended sample under conditions of blockade, only one Rydberg excitation is allowed per blockade sphere, which also contains many ground-state atoms.  These blockaded superatoms fill the excitation volume, saturating the maximum number of excited Rydberg atoms to a small value, and leading to sub-Poissonian Rydberg counting statistics. From \protect\cite{CubelLeibisch2005}.}\label{CubelFig1}
\end{figure}

The study of the behavior of cold Rydberg atoms in MOTs began with the studies of \cite{Anderson1998,Mourachko1998} that showed clear effects of Rydberg-Rydberg interactions.  Subsequent work is reviewed in  \cite{Gallagher2008}.  Many interesting collisional phenomena have been observed in the transition of a Rydberg gas into a plasma and vice versa \cite{Choi2007,Killian2001,Gallagher2003}.  Collective radiative phenomena, such as Rydberg superradiance, have  also been
seen \cite{Wang2007,Day2008}.  A distinguishing characteristic of blockade is a dramatic suppression of excitation, leading to  low Rydberg densities.  Even so, in extended samples Rydberg transport is efficient enough for these effects to still occur, making distinguishing blockade effects from plasma dynamics difficult in many cases\cite{Li2006,Westermann2006}. Other dramatic examples of fast resonant excitation transfer at 100 ns--1 $\mu$s time scales are given in \cite{Mudrich2005,Nascimento2009,Younge2009b}.  In cases where a MOT-sized volume is excited,  superradiance may compete with excitation transfer effects\cite{Day2008}.  These effects will not be present for a single blockaded ensemble where the probability of more than one Rydberg excitation at a time is greatly suppressed.  Thus in this section we will restrict our discussion to those extended sample studies that directly bear on blockade phenomena.

In extended samples subject to strong local Rydberg interactions, it is convenient to introduce the concept of the ``blockade sphere'' \cite{Tong2004}, illustrated in Fig.~\ref{CubelFig1}.   The excitation of a single Rydberg atom prohibits, via the blockade mechanism, subsequent excitations of other ground-state atoms within the radius $R_b$ of the blockade sphere.  Since the $N_b$ atoms within the blockade sphere are indistinguishable, they comprise an effective ``superatom'' \cite{Vuletic2006,Heidemann2008} that interacts with the excitation light via a collective $\sqrt{N_b}$ enhancement of the Rabi frequency.  These basic ideas explain much of the phenomena that have been observed in extended samples.

\subsubsection{Suppression of Optical Rydberg Excitation}

We consider first the optical excitation of Rydberg atoms by a single-frequency laser.
The blockade radius is determined, within a geometrical factor, by the condition that the collective Rabi frequency be comparable to the dipole-dipole shift \cite{Low2009}:
\be
\sqrt{\eta R_b^3}\Omega\approx V(R_b)
\label{scale1}
\ee
where $\eta$ is the atom density.  For a van der Waals interaction, the density of Rydberg atoms therefore saturates at the value
\be
\eta_R\sim {1\over R_b^3}\propto {\eta^{1/5}\Omega^{2/5}\over C_6^{2/5}}
\label{scale2}
\ee
or, equivalently, the excitation fraction is
\be
{\eta_R\over\eta}\propto\left(\Omega\over \eta^2 C_6\right)^{2/5}.
\label{scale3}
\ee
The striking density, intensity, and principal quantum number dependences implied by this relation are key signatures of the blockade effect.

\begin{figure}[!t]
\begin{center}
\includegraphics[width=7.5cm]{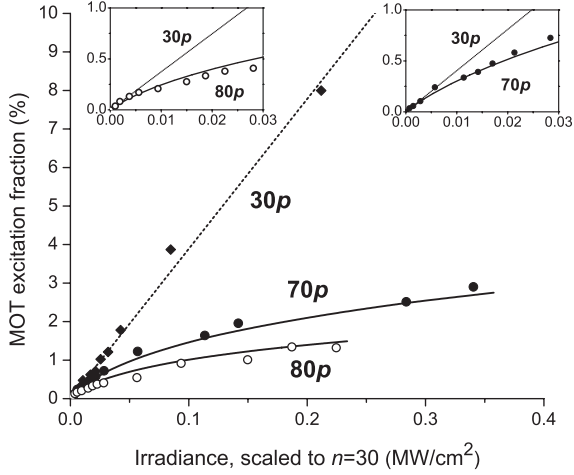}
\end{center}
\caption{Dependence of excitation fraction on principle quantum number.  From \cite{Tong2004}.}\label{TongFig1}
\end{figure}

Equation~(\ref{scale1}) assumes that the collective Rabi frequency is much larger than other line broadening mechanisms such as laser linewidth or inhomogeneous broadening by external fields.  When the opposite limit holds, the density of Rydberg atoms is limited to
\be
\eta_R\propto \sqrt{\Gamma\over C_6}\; \;{\rm or}\;\;{\eta_R\over \eta}\propto \sqrt{\Gamma\over \eta^2C_6}
\ee
where $\Gamma$ is the linewidth of the Rydberg excitation.
In addition to the Rydberg density being independent of the atomic density, the dramatic  $n^{11}$ dependence of $C_6$ on principal quantum number is another strong sign of the blockade effect.

The dependence of the excitation fraction on $C_6$ is illustrated by Fig.~\ref{TongFig1}.
Using a nearly transform-limited 8.6 ns laser pulse, \cite{Tong2004} measured the excitation fraction as a function of pulse intensity.  The roughly factor of 2 ratio of the excitation fraction for the 70$p$ and 80$p$ data is consistent with the blockade density scalings, and the comparison of either with the 30$p$ case dramatically shows the overall blockade effect.  \cite{Singer2004,Singer2005b} used continuous-wave two-photon excitation to $s$-states with similar results, which in addition showed clear lineshape modifications due to interactions.

\begin{figure}[!t]
\begin{center}
\includegraphics[width=7.5cm]{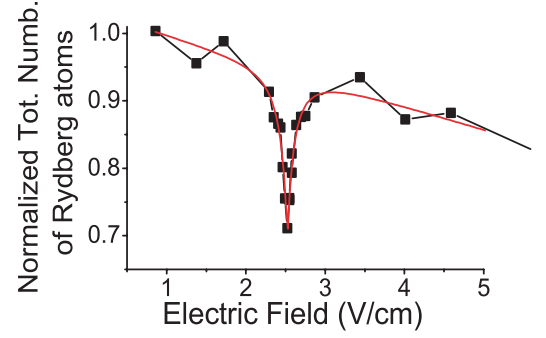}
\end{center}
\caption{(Color online) Enhanced excitation suppression by tuning a \forster\ resonance in Cs.  From \protect\cite{Vogt2006}.}\label{VogtFig3}
\end{figure}

These suppression effects should be considerably enhanced if the atom-atom interactions are made to
be $1/R^3$.  \cite{Vogt2006} showed a greatly enhanced suppression by tuning a \forster\ resonance with an electric field, as shown in Fig.~\ref{VogtFig3}
and \cite{Vogt2007} directly compared van der Waals and \forster-enhanced blockade.  The Stark tuned \forster\ resonance between just two Rydberg atoms has also been observed\cite{Ryabtsev2010}.

\begin{figure}[!t]
\begin{center}
\includegraphics[width=7.5cm]{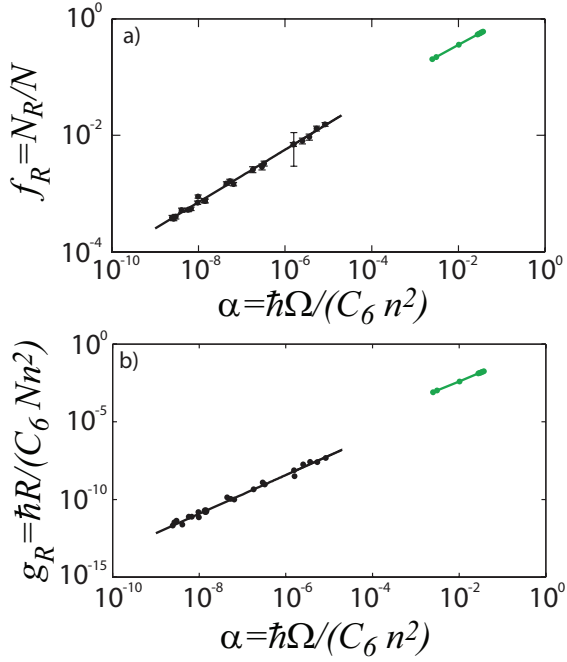}
\end{center}
\caption{\label{fig.scaling}(Color online) Scaling of a) Rydberg excitation probability  and b) excitation rate with density-scaled Rabi frequency.  Adapted from \cite{Low2009}.}\label{Lowscaling}
\end{figure}

The most comprehensive study of the density and intensity-dependence of the scaling relation (\ref{scale3}) is shown in Fig.~\ref{Lowscaling}, which shows  data from \cite{Heidemann2007}  represented according to the universal scaling theory of \cite{Low2009}. The experiment  was done with an evaporatively cooled Rb cloud just above the BEC transition.  The densities, approaching $10^{14}$ cm$^{-3}$, put the experiment well into the fully-blockaded regime where the collective Rabi frequencies were much greater than the single-atom linewidth.  By varying densities and Rabi frequencies, the dimensionless parameter \cite{Weimer2008}
\be
\alpha={\Omega\over C_6 \eta^2}\label{alphaparam}
\ee
was varied by two orders of magnitude.  The cloud geometry, an elongated cylinder, was nearly fully blockaded along the short cloud dimension, so the effective dimensionality was likely somewhat less than 3.  \cite{Low2009} showed that the data are consistent with the scaling relations assuming either 1-D or 3-D.

\subsubsection{Blockade effects on Excitation Dynamics}

The excitation dynamics are also affected by blockade. In extended samples, there will be substantial inhomogeneous broadening that causes strong dephasing of the collective Rabi oscillations.  This inhomogeneous broadening can be due to variation in the excitation laser Rabi frequencies across the sample, as well as Poissonian fluctuations in the number of atoms within each blockade sphere.  Thus the strongly-dephased Rabi oscillations average out to a mean excitation rate $\cal R$  for each superatom that is proportional to the ratio of the square of the collective Rabi frequency to the linewidth. The transition between individual and collective excitation is evident at very low intensities in \cite{Tong2004}.

For monochromatic excitation, the superatom linewidth is determined by power broadening, and is therefore proportional to the collective Rabi frequency.   Thus the superatom excitation rate will be approximately the collective Rabi frequency.  The measured quantity is
\be
{dN_R\over dt}\approx{{\cal R}V\over R_b^3}\approx{\Omega\sqrt{\eta/R_b^3}\over R_b^3}\approx{N\Omega^{6/5}\over \eta^{2/5}C_6^{1/5}}
\ee
where $N$ is the total number of atoms in the sample of volume $V$.  Again, this constitutes a highly non-trivial scaling with accessible experimental parameters. \cite{Low2009} derive a universal scaling of the dimensionless combination
\be
g_R={{dN_R/dt}\over C_6 N\eta^2}\propto\alpha^{6/5}.
\ee
The \cite{Heidemann2007} measurements are presented in this way in Fig.~\ref{Lowscaling} and obey this scaling law quite closely.

\cite{Heidemann2008} studied Rydberg excitation in Bose-Einstein condensates with variable thermal and condensate components.   The dramatic density variations between the condensate and the thermal cloud give rise to multiple time-scales for the blockade dynamics.  Again, the superatom model was successful in explaining the main features of the experiment.

\cite{Raitzsch2008} presented results of a rotary echo experiment, done in a highly blockaded sample.  This involves exciting the atoms for a time $\tau$, phase-shifting the Rabi frequency by $\pi$, followed by deexcitation for time $\tau$.  The echo should, in the absence of dephasing processes, allow reversal of coherent excitation even in the presence of the very large collective Rabi frequency fluctuations due to the fluctuations in atom number for different blockade spheres.  The dephasing rate, measured by the decay of the visibility with pulse time, was found to increase with increasing density.  An EIT experiment, done with larger atom numbers, was also used to deduce the dephasing rate in \cite{Raitzsch2009}. The roughly $N_R^2$ observed dependence of the dephasing on Rydberg atom number is in reasonable agreement with numerical simulations.

\subsubsection{Sub-Poissonian Atom Excitation}

\begin{figure}[!t]
\begin{center}
\includegraphics[width=8.5cm]{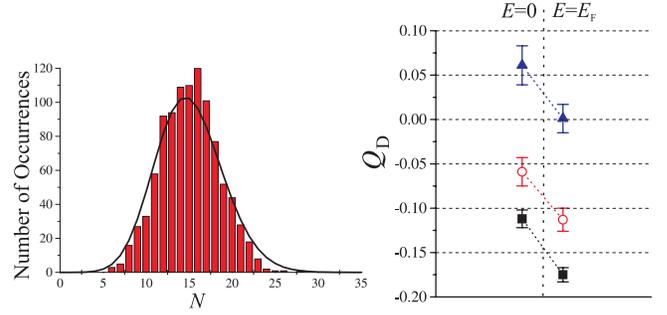}
\end{center}
\caption{(Color online) Left: Experimental Rydberg counting statistics, compared to a Poisson distribution of the same mean. Right: Mandel Q-parameter for atom densities ranging from 1-5 $\times 10^{11}$ cm$^{-3}$ (top to bottom), and for zero electric field and a \forster-resonant field $E_F$.  Adapted from \cite{Reinhard2008c} }\label{ReinhardFig}
\end{figure}

The excitation of Rydberg atoms in a dense extended sample will tend to fill the volume $V$ to a maximum number of Rydberg atoms $V/R_b^3$.  As pointed out by \cite{CubelLeibisch2005,CubelLeibisch2007}, the fluctuations in the number of Rydberg atoms should therefore be sub-Poissonian.
Fig.~\ref{ReinhardFig} shows results from \cite{Reinhard2008c} that demonstrate this effect.  The narrowing effect is clearly seen there, and quantified by the Mandel Q-parameter
\be
Q={{\expect{N_R^2}-\expect{N_R}^2\over \expect{N_R}^2}-1}
\ee
that is zero for a Poisson distribution and negative for a sub-Poissonian distribution.
Note that the experimental Q-values are diluted by finite detection efficiency \cite{Reinhard2008c,Ryabtsev2007,Ryabtsev2007b} so that the actual distributions are more sub-Poissonian than would be indicated from Fig.~\ref{ReinhardFig}.

\subsubsection{Modeling of ensemble blockade}

It is very desirable to go beyond the scaling type arguments we have used above and improve understanding via detailed and quantitative models of how blockade physics plays out in large volume samples.  A variety of approaches have been taken by various groups, as described here.  In general, the various methods are quite successful in accounting for a range of experimental results.

For resonant excitation of Rydberg states, the basic Hamiltonian to be simulated is
\be
H=\sum_i\left({\Omega_i\over 2}\ket{r_i}\bra{g}+{\rm h.c.}\right)+\sum_{j<i}{V_{ij}\ket{r_ir_j}\bra{r_ir_j}}
\ee
The first term is the Rabi coupling of the ground-state atoms to the light, while the second is the interaction between atom pairs.  Three-body interactions are usually ignored, though they may come in in surprising ways as recently pointed out by \cite{Pohl2009} (see also the discussion in Sec. \ref{sec.3atomblockade}).

The first approach from \cite{Tong2004} was a mean-field description where the effective Rydberg-Rydberg van der Waals interaction was averaged to obtain  a non-linear Bloch equation for the Rydberg amplitudes.  The blockade volume was dynamically adjusted to have exactly one atom in it at all times.  Agreement with experiment was obtained when the density or $C_6$ coefficient was scaled by a moderate value of 2.5.

\cite{Robicheaux2005} performed a simulation of a limited volume cube containing 30-160 atoms. Inside the cube the atoms were randomly placed, and treated the effects of atoms outside the cube with a mean-field model.  The still untractable Hilbert space was reduced in dimension by recursively introducing pseudoatoms of the closest atom pairs, with new collectively enhanced Rabi frequencies.  By further limiting the Hilbert space to amplitudes with fewer than six excitations, they were able to make the simulation tractable.  They got similar results to  \cite{Tong2004}, and presented also calculations of the two-atom correlation function, which as expected was nearly zero inside the blockade radius, went through an intermediate peak above 1, and settled to 1 within a couple of microns of the blockade radius.  They pointed out that the pair correlation function is quite sensitive to the detuning of the laser as compared to the sign of the van der Waals shift.  This approach was further extended in
\cite{Hernandez2006}, which also pointed out the effects of anisotropic interactions on the pair correlation functions and studied sub-Poissonian Rydberg excitation statistics.  A further fully quantum study of 1D blockade \cite{Sun2008b} looked particularly at the pair correlation functions and whether they are reflective of entanglement or classical correlation.  \cite{Hernandez2008} compared this general class of methods to  simplified Monte Carlo approaches in the context of the\cite{Heidemann2007} experiment and found them to underestimate the number of excited atoms, for reasons not clear.

Direct simulations using a truncated Hilbert space have been done for up to 100
atoms \cite{Weimer2008,Low2009,Wuster2010}.  By removing states from the full Hilbert space with Rydberg-Rydberg energies greater than some cutoff $E_c$, the size of the space can be made  tractable.  A mean-field theory was also found to compare quite closely, and both results agree with the scaling laws.

\cite{Ates2006} approximated the coherent Rabi interactions by rate equations, thereby greatly reducing the computational difficulties and allowing up to a few thousand atoms to be treated in a Monte Carlo approach.  The justifications for these approximations were described in detail in \cite{Ates2007b}, where master and rate equation solutions were found to be very similar in tractable cases. This approach was used to analyze excitation suppression and sub-Poissonian atom statistics. The results were found to be insensitive to the shape of the potential for distances inside the blockade radius, consistent with the blockade shift concept of Eq.~(\ref{scale1}).  The effects of adiabatic elimination of the intermediate p-state for two-photon excitation were discussed, and in particular \cite{Ates2007} pointed out that when the $s-p$ coupling is sufficiently strong to produce an Autler-Townes splitting of the Rydberg spectrum, there can be an anti-blockade effect where the Rydberg-Rydberg interaction increases the excitation probability by tuning the Autler-Townes peaks into resonance.  Such effects are clearly sensitive to atom-atom spacings, and would be most prominent on
an ordered lattice.

\cite{Chotia2008} present a related Kinetic Monte Carlo simulation of blockade physics, along with a mean-field Hartree-Fock density matrix analysis.  A few thousand atoms are simulated, including effects of field-induced blockade interactions.  These methods impressively reproduce the experimental data of the same group.

\begin{figure}[!t]
\includegraphics[width=8.5cm]{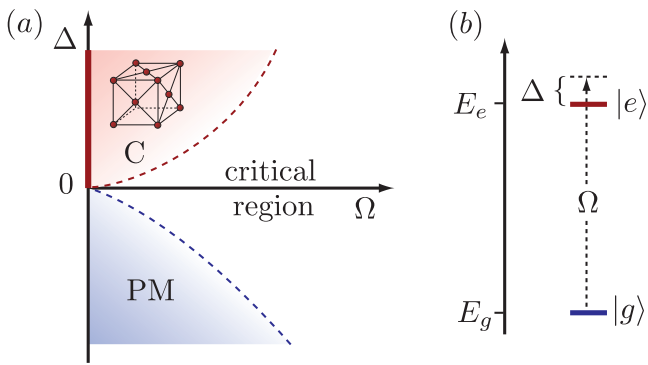}
\caption{(Color online) Phase diagram of blockaded Rydberg excitation.  A crystalline phase, C, is predicted to form for positive detunings when the van der Waals interactions are repulsive.  For negative detunings, the excitation is off-resonance for all atoms and the paramagnetic phase simply constitutes the atom population predominantly in the ground state.  From \protect\cite{Weimer2008}.
}
\label{fig.phasediagram}
\end{figure}

\cite{Weimer2008} reasoned that there is a strong analogy between Rydberg blockade and second order phase transitions for the case of repulsive, isotropic Rydberg-Rydberg interactions.  Thus they approach the problem of blockade from the perspective of statistical mechanics.   They transform the problem into an effective pseudospin representation.  As seen in Fig.~\ref{fig.phasediagram} when the light is tuned below resonance, there is no excitation in steady state.  For positive detuning, Rydberg excitation is possible and the atoms are predicted to form a crystalline lattice.  This appears related to the \cite{Robicheaux2005} prediction of peaks in the two-atom correlation function under similar excitation conditions, as well as a recent calculation of the formation of a 2D lattice structure in a Rydberg excited ensemble\cite{Pupillo2009}.
 At zero detuning, the Hamiltonian of the system depends only on the  single dimensionless parameter
$\alpha={\Omega\over C_6 \eta^2}$ defined in Eq. (\ref{alphaparam}). In this regime the
 system is argued to exhibit universal behavior; the fraction of Rydberg atoms $f_R=\eta_R/\eta$ is a universal function of $\alpha$.  A mean field analysis leads to $f_R\sim\alpha^{2/5}$ for small $\alpha$, in agreement with the scaling laws above.

Extending these ideas, \cite{Low2009} argue that the system should exhibit a quantum critical point.  Drawing an analogy with ferromagnetism,  they predict that for positive detunings (and repulsive van der Waals interactions) the system should condense into a ferromagnetic phase with the Rydberg fraction being a definite, universal function of the parameter $\alpha$ and a similar dimensionless detuning $\Delta=\delta_L/C_6\eta^2$, where $\delta_L$ is the laser detuning.  It is an important experimental challenge to see if the predicted ferromagnetic phase can be observed for positive detuning.

A statistical mechanics approach to blockade was analyzed from another perspective by
\cite{Olmos2009a,Olmos2010}, who compared direct integration of the Schr\"odinger equation of a ring of atoms with perfect nearest-neighbor blockade to a microcanonical ensemble.  The steady-state of both approaches agreed well.

Most of the theoretical approaches  to ensemble blockade described above have used a simplified treatment of the Rydberg-Rydberg interactions.  We note that in the context of fully blockaded ensembles (sample size smaller than the blockade radius), effects such as Zeeman degeneracies and the angular distribution of excitation, discussed in Section~\ref{subsec.angle}, play essential roles in determining the probability of double excitation.  The general success of the theoretical approaches in describing suppression and other blockade-like effects for samples containing many blockade regions is an interesting probe of blockade physics, but the direct implications for the success of quantum information applications of fully blockaded clouds are not evident.
 For experiments and simulations with $s$-states, where the blockade shifts are nearly isotropic and Zeeman degeneracy is not an issue, these complications may be of less importance and extrapolation  to fully blockaded situations should be more reliable.

\subsection{Preparation of single atom states}
\label{sec.loading}

\begin{figure}[!t]
\begin{center}
\includegraphics[width=7.5cm]{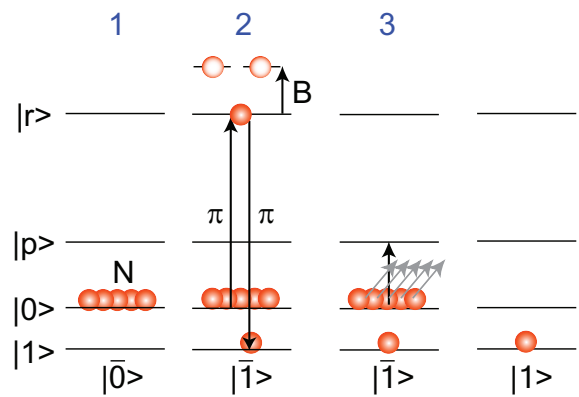}
\end{center}
\caption{\label{fig.loading}(Color online) Three step protocol proceeding from left to right for deterministic preparation of single atom occupancy
following \cite{Saffman2002}. The initial number of atoms $N$ is a stochastic variable. }
\end{figure}

An outstanding challenge for neutral atom quantum computing is controlled loading of single atoms into an optical
lattice that is compatible with site specific  addressing and control. A number of different approaches to this problem have been discussed in Sec. \ref{sec.arrayscaling}. It was proposed in \cite{Saffman2002}
to use many-atom entangled states created by Rydberg blockade as a means of achieving deterministic single atom loading. Entanglement is here a resource that is used, not just for computation, but for preparation of
atomic number states.

The loading protocol is illustrated in Fig. \ref{fig.loading}. We start by loading $N$ atoms into an optical trap and preparing them in state $\ket{\bar 0}.$ Under conditions of Rydberg blockade a $\pi$ pulse to Rydberg level $\ket{r}$ followed by a $\pi$ pulse to $\ket{1}$ creates the singly excited symmetric state $\ket{\bar 1}$.  The remaining $N-1$ atoms
left in $\ket{0}$ can then be ejected by applying unbalanced radiation pressure from
 light that is resonant with an auxiliary level $\ket{p}.$ The trap can also be lowered during this
``blow-away" phase to facilitate ejection after only a small number of photon scattering events.
After the blow-away phase we are left with a single atom in state $\ket{1}$ despite the fact that the initial
number $N$ was random.

The essential question is the probability of success of this protocol. Clearly if the initial atom number is
$N=0$ the protocol fails. With Poissonian loading statistics we can restrict the probability of this happening to $10^{-3}$ by choosing $\langle N \rangle \ge 7$. It is not difficult
to confine such a small number of atoms to a volume of size much less than a blockade sphere.
A more serious difficulty is that the Rabi frequency in step 2 is given by $\Omega_N=\sqrt{N}\Omega$.
Since $N$ is unknown it is not possible with a simple uniform pulse to have a pulse area that is independent of $N.$

One solution is to make  $\langle N\rangle \gg 1.$
It is then easy to show that the error in exciting a single Rydberg atom in the first half of step 2 is
$E=\pi^2/16\langle N\rangle$ which requires $\langle N \rangle \sim 600$ for a $10^{-3}$ error. A more
fruitful approach is likely to take advantage of a composite pulse sequence to reduce the dependence of the
pulse area on $N.$ If we take a moderate $\langle N \rangle = 10$ then the  spread of  pulse areas
due to variations in $N$ is only about 30\%. This spread can be corrected for with high accuracy using
composite pulse schemes.  Even without composite pulses success probability of roughly 80\% is possible at $\langle N\rangle=10$ by optimizing with respect to the single atom pulse area. 

It is also possible to remove the dependence on $N$ entirely by relying on ejection of
atoms from the Rydberg state instead of from the ground state\cite{Molmer2009}. Applying a $2\pi$ pulse to a single atom returns it to the ground state, whereas an ensemble of $N$ atoms  experiences a pulse area of $\sqrt{N}2\pi.$ As long as $N$ is not a perfect square 
there is a nonzero amplitude for an atom to be left in the Rydberg state after the pulse. This atom can then be photoionized before it returns to the ground state, and the sequence repeated until only one atom remains. Also cases where $N$ is a perfect square can be handled by adjusting the  detuning of the Rydberg excitation pulse.

\subsection{Collective qubit encoding}
\label{sec.col_encoding}

At the current stage of development of quantum information processing demonstration experiments have been limited to less than ten qubits. The development of new approaches to encoding and interconnecting many qubits is therefore a central challenge of current research. The long range nature of
Rydberg interactions together with the availability of a strong and controllable blockade interaction enable
a  ``collective" approach to encoding of a multi-qubit register\cite{Brion2007d}.  As we describe in this section collective encoding takes advantage of a multiplicity of stable atomic ground states to encode a multi-qubit register in a many atom ensemble, without requiring separate addressing of the atoms. Quantum information is thereby encoded in a distributed fashion that is an alternative to the usual serial encoding of one qubit for each two-level quantum system. Using blockade interactions one- and two-bit gates  can be performed between any pair of qubits using only globally applied control pulses. This remarkable feature
of the collective encoding approach
 has the potential of greatly simplifying the wiring of a quantum computer.

In order to take full advantage of this approach we will
store information in all of the atomic Zeeman states, not just clock states with optimum
coherence properties.  Effective utilization of collective encoding therefore presupposes excellent control of
the magnetic field environment of the atoms. While this is an outstanding technical challenge, it appears fruitful to nevertheless explore the  collective approach due to the potential for a substantial reduction in complexity of a functioning quantum register.

Figure \ref{fig.ensemblebit} illustrates the proposal in \cite{Lukin2001} to encode a single qubit in symmetric, collective states of an ensemble of atoms, having either all atoms in the same internal state $|0\rangle$ or having precisely one atom transferred to the other internal state $|1\rangle$. If an external perturbation couples these states with exactly the same strength $g$ for all atoms,
\begin{equation}
V_N=\sum_{j=1}^N g(|0_j\rangle\langle 1_j| + |1_j\rangle\langle 0_j|),
\label{eq.colint}
\end{equation}
the collective states $|0^N\rangle \equiv |0_1 ... 0_N\rangle$ and $|1^1 0^{N-1}\rangle \equiv (1/\sqrt{N})\sum_j |0 ... 1_j ... 0 \rangle$, experience an enhanced coupling strength, $g_N=\langle 1^1 0^{N-1}|V_N|0^N\rangle = \sqrt{N}g$.
In Eq. (\ref{eq.colint}), spatially dependent phases of the exciting laser fields have been absorbed in the atomic internal states $|0_j\rangle,|1_j\rangle$. This is a convenient approach for atoms at rest, while time dependent phases for moving atoms translate into Doppler shifts. In the following we shall assume cold, trapped atoms, e.g., in an optical lattice potential.
The interaction (\ref{eq.colint}) also drives further excitation of symmetric states $|1^2 0^{N-2}\rangle$, $|1^3 0^{N-3}\rangle$ ... , with two, three, and more state $|1\rangle$ atoms, and for sufficiently large $N$, these states form the ladder of states of an effective harmonic oscillator. Together with the enhanced coupling strength this constitutes the basis for using atomic ensembles as quantum memories for quantum states of a light pulse \cite{Julsgaard2004}.

As suggested in \cite{Lukin2001}, if the Rydberg blockade applies to the whole collection of atoms, it is possible to restrict the collective states of the ensemble to the pair of states ($|0^N\rangle$ and $|1^1 0^{N-1}\rangle$) with zero and one atom in state $|1\rangle$, only, and thus to implement a logical qubit in the atomic ensemble. To carry out an arbitrary qubit operation on this qubit without accessing states with more than a single atom in state $|1\rangle$ , one applies the following sequence of three pulses: \emph{i)} a resonant $\pi$-pulse on the $|1\rangle-|r\rangle$ internal state transition transfers the collective state $|1^1 0^{N-1}\rangle$ component to the symmetric state with one Rydberg excited atom,
\begin{equation}
|r^1 0^{N-1}\rangle \equiv \frac{1}{\sqrt{N}}\sum_j |0 ... r_j ... 0 \rangle,
\end{equation}
\emph{ii)} the desired qubit operation is implemented as a coherent, resonant transition in the closed two-level system of states $|0^N\rangle$ and $|r^1 0^{N-1}\rangle$, and \emph{iii)} a resonant $\pi$ pulse on the $|r\rangle-|1\rangle$ transition finally transfers the resulting $|r^1 0^{N-1}\rangle$ component back to the qubit level $|1^1 0^{N-1}\rangle$.

Note that these operations act as if resonant lasers are only applied to single atoms (except that the $|0\rangle -|r\rangle$ transitions are collectively enhanced) and during pulse \textit{ii}), the Rydberg blockade interaction takes care of the restriction of the dynamics to the desired collective states of the system. In \cite{Lukin2001}, the authors also proposed to implement conditional quantum gates on several qubits stored in  separate ensembles, either by direct interaction if the ensembles
are within the long range Rydberg interaction of each other, or by transferring the states of the ensembles
into a single intermediate ensemble and carrying out the operation here, before transferring the (now entangled) qubits back to their original ensembles.

\begin{figure}[!t]
\begin{center}
\includegraphics[width=7.5cm]{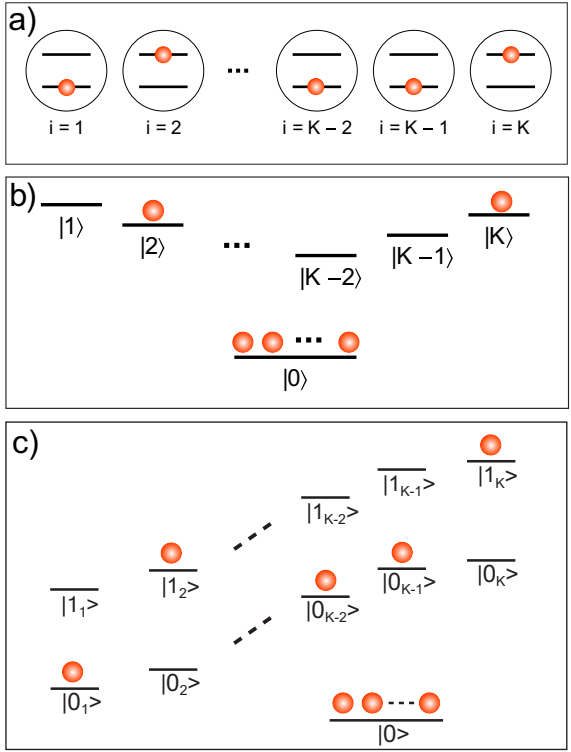}
\end{center}
\caption{\label{FigEnc}(Color online) Three different ways of encoding the $K$-bit state
$\ket{01...001}.$
a) Conventional encoding in $K$ two-level systems.
b) Collective encoding in an ensemble of $(K+1)$-level systems. Filled circles represent
 the number of atoms populating the given
 single particle state $\left \vert i \right \rangle$.
c) Collective encoding in an ensemble of  $(2K+1)$-level systems.
Each qubit value is determined by the population, represented by the filled circles,
within the pair of states $\{|i_0\rangle,\ |i_1\rangle\}$,
and each pair of states has a definite total population of unity.}
\end{figure}

Making use of more states in the internal level structure of the atoms, it is possible \cite{Brion2007d,Brion2008} to encode several quantum bits in the collective states of a single atomic ensemble. Fig. \ref{FigEnc} illustrates the conventional encoding of $K$ quantum bits in $K$ separate two-level systems, part a), and collective schemes making use of a large ensemble of  $(K+1)$-level systems, part b), and $(2K+1)$-level systems, part c), respectively. Part a) of the figure shows the lower and upper state $|0\rangle$ and $|1\rangle$ of $K$ separate particles encoding a full register of $K$ qubits, while parts b) and c) illustrate the multilevel structure of a single atom and the collective population of the individual levels in a collection of  atoms. It is assumed that the collective states in parts b) and c) are symmetric under permutation of the individual atoms, and that the filled circles in the figures merely indicate the number of atoms populating the different internal states. With the definition that a unit population of the $i^{th}$ level implies a bit value of $1$, while a vanishing population implies a bit value of $0$ in Fig. \ref{FigEnc} b), we can encode a register with $K$ qubits in an ensemble of $K+1$-state atoms, provided that the ensemble size $N \geq K$ so that all $K$ qubit states can be populated by a single atom. Part c) of the figure shows a slightly more involved level scheme where $K$ pairs of levels are identified, and where qubit values zero and unity are identified with symmetric atomic states populating one or the other of the states in each pair by a single atom. Observe that parts a), b) and c) of  Fig. \ref{FigEnc} illustrate the different encoding schemes for the same $K$-qubit register state $|01 ... 001\rangle$. In parts b) and c) of the figure, the population in the internal state denoted $|0\rangle$ plays the role of an atomic ´reservoir', being crucial for the exchange of population in one- and two-bit gates and for initialization and implementation of error correction \cite{Brion2007d,Brion2008}.

Let us discuss in a little more detail the collective encoding scheme illustrated in part b) of Fig.\ref{FigEnc}. We formally associate the binary register state $\left \vert n_1,n_2,..., n_K \right \rangle$, $(n_i =0,1)$ with the \emph{symmetric} state of the ensemble with $n_i$ atoms populating the internal states $|i\rangle$.
In this way the binary representation becomes a number state representation of the symmetric states of the ensemble specifying the number of atoms  $n_i=0,1$ populating each register level $|i\rangle$.

Using a notation as in Fig. \ref{fig.ensemblebit}, for a single ensemble qubit, with bars over numbers $0$ and $1$ indicating logical qubit values,  we can for example write the following examples of 3-bit register states,
\begin{eqnarray}
&&|\overline{0},\overline{0},\overline{0}\rangle = |0_10_2 ... 0_N\rangle\nonumber \\
&&|\overline{0},\overline{1},\overline{0}\rangle = \frac{1}{\sqrt{N}} \sum_j |0_10_2 ... 2_j ... 0_N\rangle
\nonumber \\
&&|\overline{1},\overline{0},\overline{1}\rangle =
\frac{1}{\sqrt{N(N-1)}} \sum_{j,k} |0_1 ... 1_j ... 3_k ... 0_N\rangle.
\end{eqnarray}

Including the superposition states, the basis $\{|n_1,n_2, ..., n_K \rangle, n_i=0,1\}$ fully explores the $2^K$ dimensionality, of the register Hilbert space of a $K$ qubit quantum computer.
The full Hilbert space dimension of our ensemble of $N\geq K$ atoms is, indeed, much larger, but the restriction to symmetric states with no register state population exceeding unity, yields precisely the qubit register dimension. Physically, these restrictions are imposed by the interactions: by addressing the system collectively we preserve the symmetry of states with respect to permutations among the atoms, and by application of the Rydberg blockade, we restrict the population of all information carrying states to zero and unity.
In practice, ensemble sizes an order of magnitude larger than the register size, $N\simeq 10\cdot K$, or even more, may improve a number of properties of our proposal, and do not impose major experimental problems: atomic ensembles of thousands of atoms are routinely produced and manipulated in quantum optics laboratories.

The states $|0\rangle$ and $|i\rangle$ can be chosen as the Zeeman sub-levels of atomic hyperfine ground states or metastable excited atomic states. One must ensure that the interaction with the atoms does not entangle their
internal state with their motion - \textit{i.e.}, they must be trapped
by potentials which act identically on all internal states. Far off resonance optical
traps, or small (micron sized) glass cells\cite{Kubler2010} may meet
this demand.  One must also ensure that ground state collisions among the atoms do not perturb their internal states. This can be most conveniently ensured by trapping the atoms in an optical lattice. Note that the distance between the trapping sites in typical optical lattices is of the order of a few hundred nanometers, and hence the Rydberg blockade may be efficient over a volume containing thousands of such sites. A more detailed estimate of the maximum size of a blockaded ensemble can be found in Sec. \ref{sec.arrayscaling}.

\begin{figure}[!t]
\centering {\includegraphics[width=6.5cm]{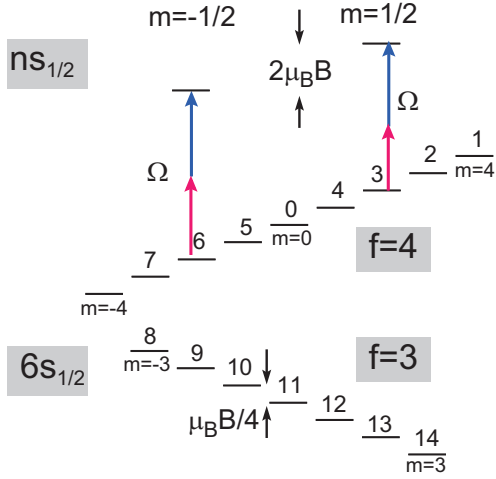}}
\caption{(Color online) Cesium level scheme and identification of qubit register. Encoding of
reservoir state $0$ and $14$ register states in the Zeeman ground states of
Cs. Coupling of $\left\vert 3\right\rangle =\left\vert f=4,m=2\right\rangle $
and $\left\vert 6\right\rangle =\left\vert f=4,m=-2\right\rangle $ to Rydberg
states is shown.
}\label{fig.Cs}
\end{figure}

A specific implementation for Cs atoms is illustrated in Fig. \ref{fig.Cs}. We initially prepare the register
state $|\overline{0},\overline{0}, ... \overline{0}\rangle = |0_10_2 ... 0_N\rangle$, where all atoms are optically pumped into the internal ``reservoir"
state $|0\rangle=\ket{f=4,m_f=0}$. An applied magnetic field Zeeman shifts all hyperfine states, such that state selectivity is obtained through the resonance condition on the optical transitions. For effective Rydberg blockade the atoms are transferred by a two-photon excitation from the hyperfine ground state to high lying interacting $s$ states with $n\stackrel{>}{\sim} 50.$ The hyperfine interaction is very weak for Rydberg excited atoms, and the hyperfine  structure of the Rydberg level is unresolved. The hyperfine ground states are thus coupled to excited fine structure states with good electron spin magnetic quantum numbers, $|ns_{1/2},m=\pm1/2\rangle.$  In Cs, different ground to Rydberg excited state transitions are separated by at least $\mu_B B/4\hbar$, with $\mu_B$ the Bohr magneton, and as long as this quantity is large compared to the two-photon excitation Rabi frequency $\Omega$ the ground states can be selectively excited. Taking into account the finite lifetime of the Rydberg level, optimum parameters are $\Omega/2\pi\sim 1 ~\rm MHz$, and a modest field of $B\sim 15~\rm G$ will provide a sufficient splitting of levels to suppress undesired excitation to the 1\% level. Due to the value of the Land\'{e} factor, however, two transitions: $|6s_{1/2},f=4,m=-4\rangle \leftrightarrow |ns_{1/2},m=-1/2\rangle $ and $|6s_{1/2},f=4,m=4\rangle \leftrightarrow |ns_{1/2},m=1/2\rangle$ are degenerate.  This implies that the ground states of these transitions are not as easily distinguished, and we suggest to exclude one of them from the encoding, leaving 15 readily distinguishable ground states for the register encoding. This implies that with a small cloud of Cs atoms, we can encode a 14 qubit quantum register using only collective addressing. In addition to the use of
external fields to lift degeneracies as illustrated in Fig. \ref{fig.Cs}, techniques from
optimal control theory may serve to identify shaped pulses that may even distinguish degenerate register
states.

Since the individual bits refer to the population amplitudes of different internal states, one- and two-bit operations are carried out as if they are effectively operations on single atoms, as detailed in the following section, but we emphasize that due to the collective nature of the encoding and the blockade, there is no need for addressing of individual atoms.

An effective read-out mechanism can be achieved by coupling the
register levels in a controlled manner to excited states from which ionization can be observed, or by repeatedly transferring  the qubit content by a C-NOT operation to another read-out collective qubit, which may be probed by fluorescence on an optical transition decaying back to the reservoir state (cf. the similar approach applied in ion traps \cite{Rosenband2008}). The directed, collective photon emission of extended atomic samples and the Rydberg blockade may also be utilized for effective read-out, as proposed in \cite{Saffman2002,Saffman2005b}.

\begin{figure}[!t]
\begin{center}
\includegraphics[width=3.5cm]{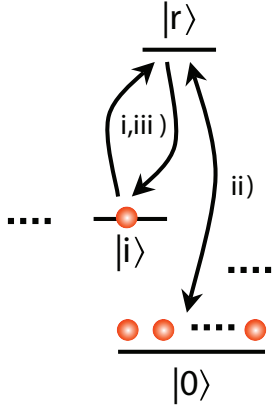}
\end{center}
\caption{(Color online) Single qubit rotation of the $i^{th}$ qubit in the collective encoding scheme. A $\pi$-pulse transfers $i$) the population in $\left \vert i \right \rangle$ to $\left \vert r \right \rangle$, $ii$) a coherent coupling is applied on the two-state system with zero and one atom in $\left \vert r \right \rangle$, $iii$) a $\pi$-pulse transfers level population in $\left \vert r \right \rangle$ to $\left \vert i \right \rangle$. }
\label{FigColGate}
\end{figure}

\subsection{Ensemble gates and error correction}
\label{sec.col_gates}

Let us now show how to implement one- and two-bit
gates on collectively encoded qubits in atomic ensembles, assuming that the atomic ground states are spectroscopically distinguishable and making use of state selective resonant optical transitions and
the Rydberg blockade mechanism.

\subsubsection{Single qubit gates}

In the encoding suggested in Fig. \ref{FigEnc} b), a phase gate on the $i^{th}$ qubit can be implemented straightforwardly, by selectively perturbing the $i^{th}$ energy level, leading to a phase
shift of precisely the components of the states with a single atom
populating that state. In Fig.\ref{FigColGate} we show how to use the Rydberg blockade to perform a selective rotation on the $i^{th}$ qubit, \textit{i.e.}, between the collective states of the ensemble with zero and one atom populating the atomic level $|i\rangle$. As in our explanation of how single qubit gates are performed on the ensemble qubit, shown in Fig. \ref{fig.ensemblebit}, this rotation requires three steps: \emph{i}) a swap of the population between the state $|i\rangle$ and the Rydberg state $|r\rangle$, \emph{ii}) a coherent coupling on the $|0\rangle -|r\rangle$ atomic transition, and \emph{iii}) return of the $|r\rangle$ component to the atomic state $|i\rangle$. None of these processes require individual addressing of the atoms, but while the first and the last process are driven as single atom state selective $\pi$-pulses, the middle process is collectively enhanced due to the population of the  reservoir state $|0\rangle$. Since the occupation of all register states are quantum degrees of freedom, the reservoir population $n_0=N-\sum_i n_i$ may attain a wide range of values, and it is an advantage to assume $N\gg K$, so that the collectively enhanced Rabi frequencies are almost identical, or sufficiently close that composite pulses \cite{Cummins2003} may compensate for their differences.

In the collective encoding, using a pair of states for each qubit, illustrated in Fig.\ref{FigEnc} c), single qubit gates are simpler, as they are obtained by coupling directly the relevant pair of atomic states $|i_0\rangle$ and $|i_1\rangle$, \textit{e.g.,} via an optically excited state. There is no collective enhancement of the transition, which thus proceeds as if we were interacting with only a single atom, and Rydberg
 blockade is not needed to restrict  the total population of the states involved.

\subsubsection{Two-qubit gates}

\begin{figure}[!t]
\begin{center}
\includegraphics[width=5.cm]{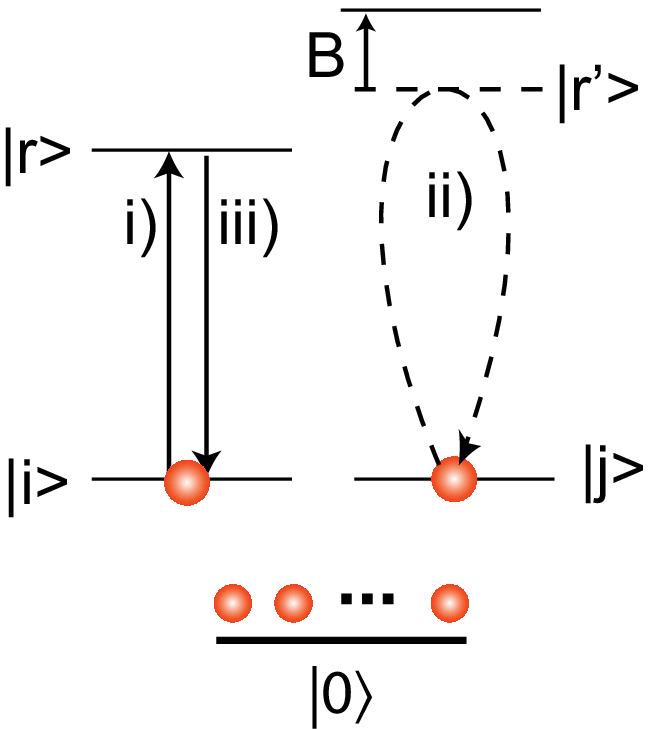}
\end{center}
\caption{(Color online) Two-qubit operation in the collective encoding of qubits.  $i$) transfer of the population in $\left \vert i \right \rangle$ to $\left \vert r \right \rangle$, $ii$) an attempted transfer of the population in
$\left \vert j \right \rangle$ to another Rydberg state $\left \vert r' \right \rangle$ and back, blocked in the presence of an atom in state $\left \vert r \right \rangle$, $iii$) return of the population in $\left \vert r \right \rangle$ to $\left \vert i \right \rangle$. This sequence of operations is equivalent to the sequence shown in Fig. \ref{fig.cphase} for individually encoded qubits and causes a Z-gate on the $j^{th}$ collectively encoded qubit, conditioned on the state of the $i^{th}$ qubit.}
\label{FigColGate2}
\end{figure}

The implementation of two-bit gates in our ensemble scheme is different from the Rydberg blockade gate in the individual atom proposal shown in  Fig. \ref{fig.cphase}, where the excitation of one atom prevents the excitation and accumulation of a phase shift by another atom in the qubit $1$ state. In the collective scheme, illustrated in Fig. \ref{FigColGate2}, we use that the excitation of the Rydberg state from a logical "1" of the $i^{th}$ bit, \textit{i.e.}, from a single atom populating the ´´control" state $|i\rangle$, prevents the resonant driving towards another Rydberg state $|r'\rangle$ of an atom populating the ``target" state $|j\rangle$ , if atoms in these two Rydberg states experience the blockade interaction. Unlike the individual atom proposal \cite{Jaksch2000}, we make use here of two different Rydberg states $\ket{r}$ and $\ket{r'}$, because an atom excited from $|i\rangle$ into $|r\rangle$ may be subsequently driven into $|j\rangle$ if these states are coupled sequentially to the same Rydberg state.\footnote{The requirement of two Rydberg states can in principle be dropped if we use composite Rydberg pulses to discriminate between the $\sqrt 2$ difference in Rabi frequency of singly and doubly occupied states.}
 In the encoding via state pairs, illustrated in Fig. \ref{FigEnc} c), two-bit gates are carried out in essentially the same way, namely by transfer of one of the qubit $i$-states to a Rydberg state $|r\rangle$, followed by a conditional dynamics on the $j^{th}$ qubit via another Rydberg state.
The possibility to couple simultaneously (bright) linear superpositions of the $|i_{0(1)}\rangle$ states and of the $|j_{0(1)}\rangle$ states to the respective Rydberg states makes a wider variety of two-bit gates available in single shot operations \cite{Roos2004}.

\subsubsection{Error correction}

Let us now turn to the issue of errors occurring in the collective encoding scheme for quantum computing. The conventional paradigm for quantum computing is that individual bits are stored in individual physical systems and correction of errors that occur to individual bits is possible  by a suitable redundant encoding of logical qubits in special codewords using several physical qubits. These error correction techniques, which can check and restore the codewords without destroying the quantum content of the states, however, do not apply if errors happening to a single atom do not just affect a single qubit and if we can only collectively and symmetrically address all atoms in the ensemble. In the case of collective qubit encoding via pairs of states, cf. Fig. \ref{FigEnc} c), it is, however, possible to check for errors and repair them by simple encoding schemes. We give here a brief review of the main ideas and refer to \cite{Brion2008} for further details.

Since the ensemble is supposed to consist of a number of particles much larger than
the number of qubits ($N\gg K$), each atom most likely occupies the reservoir state $|0\rangle$, and the loss of a particle therefore most likely leaves the unit population in the qubit state pairs $|i_0\rangle,\ |i_1\rangle$ intact. We therefore propose to monitor the total population in each qubit pair of states and as long as this population is unity, we assume that no error has occurred. If, however, one finds zero occupancy of a qubit state pair, one reverts to a unit occupancy by transferring a single atom from the reservoir state, via the Rydberg state, to the state $|i_0\rangle$. This is very unlikely to be the correct state of the qubit, but we know which qubit position in the register has been thus compromised, and if we use a simple redundant code of two physical qubits per logical qubit, we can reestablish the correct state by a C-NOT gate operation, where the compromised qubit is the target and the uncompromised partner qubit is the control qubit. A more worrisome situation occurs if an atom decays into a qubit state, which is already collectively occupied by other atoms in the sample. This is both a problem, because it leads to a logically meaningless double occupancy of a collective qubit level, and because the erroneous single atom is capable of controlling the other atoms by the Rydberg blockade. We have only the same access to all the atoms, but we recall that the collectively occupied states experience a $\sqrt{N}$ enhanced coupling strength, whenever the internal atomic state is coupled to the macroscopically populated ``reservoir state". Precisely this enhancement distinguishes a single atom populating a given state from a symmetric, collective population, and by driving suitable transitions in the system, it is possible to either dispose of the single atom with the erroneous population, or return it to the
reservoir state \cite{Brion2008}.

\subsection{A 1000 bit collectively encoded computer}
\label{sec.1000}

\begin{figure}[!t]
\begin{center}
\includegraphics[width=8.5cm]{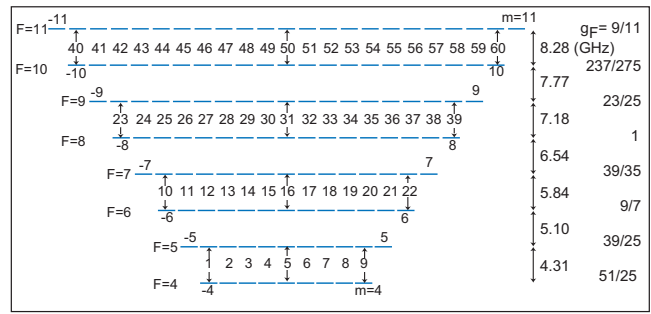}
\end{center}
\caption{(Color online) Hyperfine structure of the Ho $4f^{11}6s^2 (^4I_{15/2})$ ground state.
Assignment of 60 qubits with two atomic states per bit is indicated in the figure together with values of the hyperfine splittings and $g$ factors. }
\label{FigHolmium}
\end{figure}

Using cesium atoms, we achieve 14 collectively encoded qubits, or alternatively 7 qubits with the encoding of qubits in pairs of states.
We are not, however, restricted to alkali atoms, and in Fig. \ref{FigHolmium} we show the hyperfine structure of the ground state in holmium atoms. Holmium has one stable isotope $^{165}$Ho which has a ground electron configuration $4f^{11}6s^2$ with $J=15/2$ and a nuclear spin of $I=7/2$ giving hyperfine levels with $4\leq F \leq 11$, \emph{i.e.}, a total of 128 hyperfine states. In the figure we show all these states with a proposed qubit assignment to each pair of states cf. Fig.\ref{FigEnc}.c), and we indicate the energy splittings and the Land\'{e} factors for the different levels, of relevance to the selective addressing of different transitions, when the atoms reside in a uniform magnetic field. Holmium is a rare earth atom, and like other rare earths, we expect laser cooling and trapping of holmium to be possible based on success with other rare-earth elements\cite{McClelland2006,Lu2010}, and hence implementation of 60 qubits in a small trapped ensemble of holmium atoms may be possible with our proposal.

In \cite{Saffman2008} we further analyze the possibility to trap several ensembles, each providing 60 bits of information, and thus approach a 1000 bit quantum computer with only 16 ensembles in a 2D architecture. With a few $\mu \rm m$ separation between the ensembles it is possible to use the Rydberg `interaction gate' as in Sec. \ref{sec.othergates} to accommodate inter-ensemble gate operations.
Hence neither optical communication nor transport of atoms is necessary to reach a moderately large scale quantum computer.

There are undeniably many challenges associated with
implementing a large scale collectively encoded register.
Since the interaction gate depends sensitively on the separation between atoms (see Fig. \ref{fig.cz_interaction_error}) it appears difficult to reach a gate error of $0.001$, especially considering the finite size of each ensemble, relative to the interensemble spacing. The interensemble gate errors will therefore tend to be higher than for gates between qubits in one ensemble.  There are possible solutions to this difficulty, including implementing gates via a two-step process that uses an intermediary single ensemble (or single atom) located
far enough away to minimize sensitivity to the interensemble spacing.
Rare-earth atoms that are
relatively poorly studied and have not been widely used for
laser cooling present an additional set of challenges. A large number of lasers of different wavelengths
and frequencies are required for the various internal
state manipulations. In addition, as has been mentioned above, excellent magnetic field stability is needed
to achieve long coherence times between  states with a linear differential Zeeman shift. Again, there are possible approaches to mitigating this sensitivity. One possibility is to encode information in logical bits,
each containing four internal states with quantum numbers $\ket{f,\pm m},\ket{f',\pm m}.$ Such a combination has zero linear Zeeman shift, at the expense of
reducing the register size by a factor of two.

In  some sense we have transferred the
complexity of moving quantum information
spatially by implementing gates between qubits in an array,  to the problem of
implementing gates between a multiplicity of nondegenerate internal states.
Although
the overall complexity required to build a large
quantum processor will remain  high, we believe it is
worthwhile to explore a range of approaches. Indeed,
there is no known simple approach to building a 100- or 1000-qubit
scale quantum logic device.

\subsection{Many particle entanglement}

\label{sec.entanglement}

We note that our collective coding of qubits makes explicit use of many particle entangled states, and a single collective qubit attaining the classical, logical value ``1" is equivalent to an entangled, so-called $W$-state. Entangled logical qubits, {\it e.g.}, a Bell state in a two-bit register,  $|\overline{0}\overline{1}\rangle + |\overline{1}\overline{0}\rangle$, are physical states of our ensemble which are not particularly more entangled than the ``classical" logical qubit states $|\overline{0}\overline{1}\rangle$ and $|\overline{1}\overline{0}\rangle$ in our collective encoding of the same quantum register. In this section, we consider how the Rydberg blockade may be used to engineer a variety of entangled states, but our focus here is not on quantum states for quantum computing, but on the few- and many-particle states that find other applications, e.g., in high precision measurements. Internal state energy differences provide the definition of time in atomic clocks and they are sensitive to external perturbations, such as magnetic fields, which can hence be probed more precisely with squeezed states or other kinds of entangled states. With state dependent forces, entangled states may be used in interferometers and yield enhanced sensitivity to external motion and inertial effects. This also implies that we shall explore symmetric states of the atomic ensembles, where the collective population of the internal states is the interesting quantum degree of freedom, and is not restricted to the values of zero or unity.

\subsubsection{Spin squeezing}

In \cite{Bouchoule2002}, it was thus proposed to use Rydberg blockade to squeeze the collective spin variable associated with the effective spin-1/2 description of stable pairs of atomic states. Spin squeezed states show less quantum mechanical spreading of one of the collective spin components at the expense of increased fluctuations in another component. The possibility to squeeze the spin has led to suggestions for the use of spin squeezed states in precision metrology and is described in detail in \cite{Wineland1994}, while the connection between measured values for the mean and variance of different collective spin components and the degree of multi-particle entanglement among the individual particles is quantified in \cite{Sorensen2001,Toth2007}. By analogy with the quantum optical squeezing of light, which is achieved by a Hamiltonian, quadratic in photon creation and annihilation operators,
\begin{equation}
H_{\rm sq}= \xi \hat{c}^2 + \xi^* (\hat{c}^\dagger)^2,
\end{equation}
we obtain spin squeezing by processes that simultaneously transfer pairs of atoms between the internal levels, while processes that transfer the atoms independently of each other, only cause a rotation of the collective spin vector. The spin squeezed state is a superposition of states with different even numbers of atoms transferred to the initially unpopulated state.

In \cite{Bouchoule2002}, it is observed that two-photon transitions between a pair of  ground states and the Rydberg state constitute an effective four-photon Raman transition between the ground atomic states. If lasers are used which accommodate two oppositely detuned transition paths for these Raman transitions, single atom transitions are energetically suppressed, while the eight-photon process where two atoms are simultaneously transferred from one to the other ground state becomes resonant, provided the atoms follow transition paths with opposite four-photon Raman detunings. The high order collective pairwise transition of two atoms with no intermediate single-atom resonances is of course very weak, see for example \cite{Brion2007c}, but as suggested in \cite{Bouchoule2002}, the collective enhancement applies to speed up the transitions. Furthermore, the degree of squeezing, measured by the reduced variance of the squeezed spin component relative to the value in the spin coherent state is given by the total number of atoms transferred, and in very large samples of atoms substantial spin squeezing thus requires only a tiny transfer of ground state population per atom.

\subsubsection{GHZ states}
\label{sec.GHZ}

Stronger, and also more fragile entanglement is shown by the Greenberger-Horne-Zeilinger (GHZ) states, which are superposition states of the form $|0\rangle^N + |1\rangle^N$, where all atoms occupy one or the other internal state $|0\rangle$ or $|1\rangle$.  Several schemes have been suggested for the production of GHZ states by Rydberg blockade, and we will here distinguish between the situation where one has access to a single atom, which is capable of controlling the evolution of an ensemble of $N$ other atoms, and the situation in which we have only access to the collective degrees of freedom of a single ensemble consisting of $N$ atoms.

If a single individually addressable atom in the vicinity of $N$ atoms in the ground state $|0\rangle$ can be excited into a superposition of a ground and Rydberg excited state, the joint system will occupy the state
\begin{equation}
|\Psi_{r,\rm ind}\rangle = \frac{1}{\sqrt{2}}(|0\rangle + |r\rangle)\otimes|0^N\rangle.
\end{equation}
Following the ideas presented in \cite{Lukin2001}, and assuming that all atoms are within the Rydberg blockade distance of each other, a sequence of resonant pulses on the $|0\rangle - |r\rangle$ and the $|r\rangle - |1\rangle$ transition, directed onto the sample of $N$ atoms can then lead to the GHZ state, by producing first the state
\begin{equation}
|\Psi_{r,\rm col}\rangle = \frac{1}{\sqrt{2}}\left(|0\rangle\otimes|r^1 0^{N-1}\rangle + |r\rangle\otimes|0^N\rangle\right),
\end{equation}
where we observe the Rydberg blockade of simultaneous excitation of the control atom and the $N$-atom ensemble (for simplicity, we disregard phase factors associated with Rabi oscillations in the following arguments). The symbol $|r^10^{N-1}\rangle$ denotes the permutation symmetric state of the $N$ indistinguishable atoms with one Rydberg atom and $N-1$ state $|0\rangle$ atoms. A subsequent pulse of light on the $N$-atom sample on the $|r\rangle - |1\rangle$ transition  leads to the state
\begin{equation}
|\Psi_{\rm ent}\rangle= \frac{1}{\sqrt{2}}\left(|0\rangle\otimes|1^1 0^{N-1}\rangle + |r\rangle\otimes|0^N\rangle\right),
\end{equation}
and repeating the pair of pulses on the $N$-atom sample on the two transitions with durations matched to the collective transfer Rabi frequencies, the sample state, correlated with the ground state $|0\rangle$ of the control atom is gradually transferred to the state $|1^N\rangle$. The joint state of all atoms is then equivalent to the GHZ form.

Very recently, in \cite{Muller2009} it was demonstrated, that by a careful choice of field parameters the control atom is capable of blocking a Raman transition via the Rydberg state between $|0\rangle$ and $|1\rangle$ of all $N$ atoms at once, and hence produce the same entangled state in one single step. This process relies on a delicate destructive interference effect like the dark state phenomenon used in electromagnetically induced transparency. Via measurements on the control atom, the interaction with the ensemble offers means to characterize the entanglement in the $N$-atom system \cite{Muller2009}.

It is also possible to produce the GHZ and other many-atom entangled states without separate access to a single atom, which can control the  sample. Let us review two methods relying on adiabatic processes, where the fields are turned gradually on and off on different internal atomic transitions, and another method making use of the possibilities to excite different Rydberg states with different interaction characteristics.

Two adiabatic schemes have been proposed, where the ground states and the Rydberg states are coupled in a Lambda transition and a ladder transition, respectively. In the Lambda configuration scheme \cite{Unanyan2002}, where both ground states couple to the Rydberg state, one uses the blockade to prepare first a superposition state of the form
\begin{equation}
|\Psi_{r,\rm sup}\rangle = \frac{1}{\sqrt{2}}\left(|r^1 0^{N-1}\rangle + |0^N\rangle\right).
\end{equation}
A subsequent Raman adiabatic process from $|0\rangle$ to $|1\rangle$ via the Rydberg state is blocked in the first component of the state, but happens unhindered in the second part without ever populating the Rydberg state, and finally the $|r\rangle$ component can be driven back to $|0\rangle$ to yield the desired state.

The ladder configuration, studied in \cite{Moller2008}, couples the lower ground state $|0\rangle$ directly to $|1\rangle$, which in turn is coupled to $|r\rangle$. Here, the rapid adiabatic passage is used with the ``counterintuitive" pulse sequence \cite{Bergmann1998}, where the coupling on the  $|1\rangle-|r\rangle$ transition is gradually switched off while the coupling on the lower $|0\rangle-|1\rangle$ transition is turned on, having the effect on a single atom to reliably transfer it to the Rydberg state from the lower ground state. When applied to two atoms, the Rydberg blockade prevents the atoms to be both transferred to the Rydberg state, and in fact, the final state contains no Rydberg population at all, but instead leaves the atoms in an entangled superposition state $\frac{1}{\sqrt{2}}(-|0^2\rangle + |1^2\rangle)$. For higher numbers of atoms $N$, one realizes that for even values of $N$, no Rydberg population exists in the final state, which instead becomes a Dicke collective spin eigenstate with zero eigenvalue of the collective spin operator
\begin{equation}
\hat{J}_x \equiv (\hat{a}_0^\dagger \hat{a}_1 + \hat{a}_0 \hat{a}_1^\dagger)\hbar/2,
\end{equation}
where the oscillator raising $(\hat{a}_{0,1}^{\dagger})$ and lowering $(\hat{a}_{0,1})$ operators add and remove a single atom in the two atomic ground states $|0\rangle$ and $|1\rangle$. For $N$ odd, the final state is a permutation symmetric state with a single Rydberg excited atom, and the remaining atoms populating a collective $\hat{J}_x=0$ Dicke eigenstate. The Dicke states are themselves interesting entangled states with possible applications in metrology, and in \cite{Moller2008} it is further demonstrated how to use the Rydberg excitation conditioned on the odd-even atom number parity to accumulate non-trivial phase factors and produce a GHZ state.

Both proposals \cite{Unanyan2002,Moller2008} recognize the presence of an effective Jaynes-Cummings Hamiltonian in the coupling of the $N$ ground state atoms due to the saturation of the two level transition between states with zero and one Rydberg excited states. It is thus possible to perform ``cavity QED" experiments with the atomic sample, and the non-linearity of the Jaynes-Cummings model suggests that with the adiabatic switching between two different Hamiltonian terms, the system may also be used as a quantum simulator to explore phase transition dynamics, as  has also been done with trapped ions \cite{Friedenauer2008}.

As described in Sec. \ref{sec.d-d}, the F{\"o}rster resonance provides strong interactions for particular ``coincidental" degeneracies of pairs of Rydberg excited states while in the absence of F\"orster resonances, the interaction between Rydberg excited states is much weaker and has shorter range. We assume the existence of two Rydberg states $|s\rangle$ and $|p\rangle$ with strong mutual interactions between pairs of atoms in the $|s\rangle$ state, and between a pair of atoms in the $|s\rangle$ and   $|p\rangle$ states, while two atoms in the $|p\rangle$ state feel only a weak interaction. Examples of such states in $^{87}$Rb are $|s\rangle=|41s_{1/2},m=1/2\rangle$ and $|p\rangle = |40p_{3/2},m=1/2\rangle$, for which the $p-p$ interaction is more than two orders of magnitude weaker than the $s-p$ interaction for all relevant parameters \cite{Saffman2009b}, as illustrated in Fig. \ref{fig.asymmetric}. Instead of a definite single atom controlling an ensemble, we can use the Rydberg blockade to introduce a single excitation in an ensemble which can
subsequently  control the collective population in other states. Beginning with all atoms in the state $|0^N\rangle$, a pulse on the $0-s$ transition creates the superposition state,
\begin{equation}
|\Psi_{s,\rm sup}\rangle=\frac{1}{\sqrt{2}}(|s^1 0^{N-1}\rangle+|0^N\rangle).
\end{equation}
Driving subsequently the transition $0-p$, the absence of a strong $p-p$ interaction, allows transfer of all atoms into the $|p\rangle$ state, provided  $|s\rangle$ is not populated,
\begin{equation}
|\Psi_{s,p}\rangle =  \frac{1}{\sqrt{2}}(|s^1 0^{N-1}\rangle+|p^N\rangle).
\end{equation}
Following this process by driving the $p-1$ transition towards the other atomic ground state and inverting the initial pulse on the $0-s$ transition, finally produces the desired state,
\begin{equation}
|\Psi_{\rm GHZ}\rangle =  \frac{1}{\sqrt{2}}(|0^N\rangle+|1^N\rangle).
\end{equation}

In all of the above proposals, the fidelity with which the entangled states can be prepared is an important issue. The fidelity is reduced by atomic decay out of the excited states and by fluctuations in the value of the Rydberg interactions among different pairs of atoms. For the proposals relying on adiabatic transfer, non-adiabatic transitions should be avoided as much as possible, and for the latter proposal the not completely vanishing interaction between atoms in the $|p\rangle$ Rydberg state should be taken into account. For larger atom numbers these errors imply that GHZ states become more and more difficult to produce with high fidelity. In comparison, the  entanglement in spin squeezed states is less critical to losses and noise \cite{Bouchoule2002}, and in samples with many atoms, already a small population transfer and entanglement per atom is associated with considerable squeezing making this a much easier task.

\section{Rydberg excited ensembles and quantum optical effects}
\label{sec.quantum_optics}

Quantum optics is widely defined as the field of physics dealing with the preparation, application and detection of quantum states of the radiation field. The research involves demonstration of quantum mechanical effects and it finds application in precision probing at the limits sets by quantum uncertainty relations and in optical strategies for quantum computing and communication. Single atoms and materials with optical non-linearities have been used to produce a wide variety of field states (squeezed states, Fock states, ``Schr\"odinger kitten" states, --- ), and in this section we will present a few examples of the possibilities offered by the strong Rydberg interaction energy in atomic ensembles to manipulate quantum states of light.

\subsection{Nonlinear and quantum optics in atomic ensembles}

\subsubsection{Optical and electric control of transmission properties of an atomic ensemble}

\begin{figure}[!t]
\begin{center}
\includegraphics[width=8.5cm]{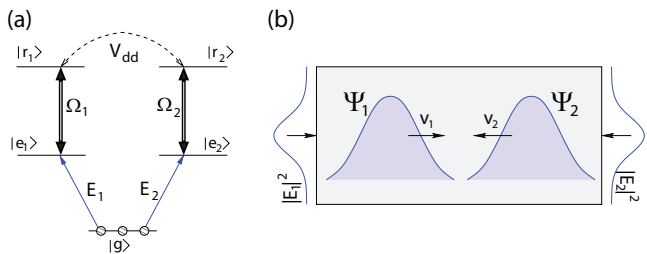}
\end{center}
\caption{(Color online) (a)~Level scheme of atoms interacting with weak (quantum)
fields $E_{1,2}$ on the transitions $\ket{g} \to \ket{e_{1,2}}$
and strong driving fields of Rabi frequencies $\Omega_{1,2}$ on
the transitions $\ket{e_{1,2}} \to \ket{r_{1,2}}$, respectively.
$V_{\rm dd}$ denotes the dipole-dipole interaction between pairs
of atoms in Rydberg states $\ket{r}$.
(b)~Upon entering the medium, each field having Gaussian transverse
intensity profile is converted into the corresponding polariton
$\Psi_{1,2}$ representing a coupled excitation of the field and
atomic coherence. These polaritons propagate in the opposite directions
with slow group velocities $v_{1,2}$ and interact via the dipole-dipole
interaction. From \cite{Friedler2005}.}
\label{fig.friedler}
\end{figure}

In \cite{Friedler2005}, it is suggested to use the Rydberg interaction to effectively provide a phase gate between two single photon pulses propagating from opposite sides through the atomic medium. The  proposal is an elegant application of several aspects of EIT\cite{Fleischhauer2005}. The pulses are assumed to excite the macroscopically populated ground state towards two different optically excited states $|e_1\rangle$ and $|e_2\rangle$, which are already coupled in a ladder configuration by classical control fields towards two different, mutually interacting Rydberg states $|r_1\rangle$ and $r_2\rangle$ (see Fig. \ref{fig.friedler}). The atomic three-level transition transforms each photon wave packet into a polariton \cite{Fleischhauer2000}, i.e., a coherent superposition of a field and an excited atomic component, which travels through the medium with a reduced speed, controlled by the intensity of the classical control field.

At weak control fields, the propagation is very slow, and the polariton is atomic-like in the  form of a collectively shared single excitation of the Rydberg state within a spatial volume which is compressed relative to the initial pulse length by the reduction in propagation speed. Since only single photons are present in the incident quantum fields, there is no effect of the Rydberg interaction on the properties of the individual polaritons. When the two counter propagating polaritons meet in the middle of the ensemble, however, state components with nearby pairs of atoms excited to Rydberg states $|r_1\rangle$ and $r_2\rangle$ are populated in a time-dependent manner, and these components will hence accumulate a phase factor due to the interaction energy. In \cite{Friedler2005}, it is shown, perhaps surprisingly, that rather than a complicated distortion and correlation of the phase fronts of the two polariton modes, the interaction when the two polaritons pass each other, leads to a single uniform phase shift $\phi$ of the state, determined by the polariton transverse width and velocity and the Rydberg interaction strength. As calculated in \cite{Friedler2005} for  parameters 
corresponding to a $100~\mu\rm m$ long sample of cold atoms  this phase may attain the value $\pi$, and hence paves the way to use atomic  samples as an environment for deterministic entanglement of optical fields and photonic quantum gates.

Experimental progress in this direction using a sample of cold Rb atoms is evident in\cite{Pritchard2009} where, with reference to Fig. \ref{fig.friedler}, a control field on  
the $\ket{e}-\ket{r}$ transition is used to strongly modify the transmission of a probe field on 
the $\ket{g}-\ket{e}$ transition via the EIT mechanism. The presence of Rydberg interactions modifies the EIT dark state (in a manner similar to that invoked by \cite{Moller2008} for creation of GHZ states) such that the optical transmission becomes a function of the atomic density and the strength of the probe  field. Although the parameters were not yet in the $\pi$ phase shift regime needed for a single photon phase gate, the experiments do demonstrate mapping of the dipolar Rydberg interaction onto an optical nonlinearity.

\subsection{Light-atom quantum interfaces with Rydberg blockade}
\label{subsec.communication}

A large working register for quantum computing may for practical reasons have to be split into separate physical units each holding part of the register qubits, and flying photonic qubits may be applied for quantum gates between these physical units. Long distance quantum communication suffers from propagation and coupling losses, and it has been proposed to transmit information only over shorter distances and store it in so-called repeaters, while the errors are still small and may be corrected via quantum error correction applied to subsequently transmitted and retrieved flying qubits. By further exchange of entangled qubit pairs with similar distant repeaters, it is thus possible, by error correction and rejection of erroneous qubits, to establish entangled qubit pairs at long distances or between several components of a large quantum computer.

For these reasons, interfacing of stationary and flying qubits constitutes a very active field of
 research\cite{Hammerer2010}.
In this section we will discuss how a sample with a few hundred atoms within a few $\mu$m sized region of space is large enough to provide efficient cooperative absorption and emission of light, and still small enough to ensure strong dipole-dipole interactions when atoms are excited into high-lying Rydberg states. Based on our collective qubit encoding scheme, we propose to build few-qubit quantum registers in such samples which can receive and emit quantum information in the form of single photons. Using the internal atomic level structure to implement logical operations on just a few bits, the samples can then employ entanglement pumping protocols \cite{Dur2003,Jiang2007}to perform ideally in networks for scalable quantum computing and long distance quantum communication. It is furthermore possible to exploit the availability of multiple excitation
paths to prepare superposition states that emit more complex, multi-photon entangled states. 
Protocols for generation of complex photonic states have been developed in several papers\cite{Porras2007,Porras2008,Nielsen2010}.  

The atomic sample may also receive and store subsequent single photon wave packets in separate collective internal states. With the availability of  collective encoding two-qubit gates, it can then be used as a processing unit for gates between photonic qubits. 

\subsubsection{Cooperative emission of single photons}

Spontaneous emission of light from an ensemble of atoms is related to the process of super-radiance and has been extensively studied in the literature. Within quantum information theory, the collectively enhanced coupling strength of atomic ensembles has been proposed as a means to implement effective single photon absorbers \cite{Imamoglu2002,James2002} and to construct quantum repeaters for long distance quantum communication with atomic ensembles \cite{Duan2001}, and, following \cite{Lukin2001}, it was shown in \cite{Saffman2002} that even a fairly small cloud of Rydberg blocked atoms
constitutes a directional source of single photons. The Rydberg blockade ensures that only a single atomic excitation is created in the system, and the directional emission follows from the coherent addition of scattering amplitudes for the individual atoms and the phase matching over the sample.

Assuming an atomic distribution with a  width $w_a$, Saffman and Walker in \cite{Saffman2005b}, estimated that a Gaussian radiation mode with a $1/e^2$ intensity waist of $w_0=\sqrt{2} w_a$ which  radiates into a solid angle of $\Omega_c = 2\pi/(k w_0)^2$, will be populated with a probability related to the cooperativity parameter \emph{C} by \cite{Lugiato1984}

\begin{equation}
P=\frac{\emph{C}}{1+\emph{C}} = \frac{N/2k^2w_0^2}{1+N/2k^2w_0^2}.
\label{eq.coop}
\end{equation}

If we want the photon to function properly as a qubit, not only the directional distribution but also the temporal dependence of the emitted photon field is important. Interestingly the full time-dependent problem of collective emission has very recently received considerable interest  \cite{Das2008,Mazets2007,Porras2008,Mandilara2009},  and in \cite{Pedersen2009} a solution is provided for the light emission from an atomic sample, containing initially a single distributed atomic excitation.

We assume the initial collective atomic state
\begin{equation}\label{eq:psi0}
\ket{\Psi_0} = \frac{1}{\sqrt{N}}\sum_{j=1}^N e^{i\mathbf{k}_0 \cdot
\mathbf{r}_j} \ket{e_j}\otimes \ket{0},
\end{equation}
where $\ket{e_j}$ is shorthand for the state with atom $j$ excited and the
other atoms in their ground state $|g\rangle$. 
 This state, containing a single  excitation collectively shared among the atoms, with a position dependent complex phase is  prepared with three laser fields as shown in Fig. \ref{fig.ensemble2}. The first two fields with wave vectors  $\mathbf{k}_{1}, \bf{k}_2$ drive  a resonant excitation into a Rydberg state, where the blockade interaction prevents transfer of more than a single atom. A resonant $\pi$-pulse with wave vector $\mathbf{k}_3$ hereafter drives the atomic excitation into the excited state $|e\rangle$, producing the state (\ref{eq:psi0}) with  $\mathbf{k}_0=\mathbf{k}_1+\bf{k}_2-\mathbf{k}_3$.

Since the system is restricted to states with only a single excitation, we can expand the time-dependent state of the atoms and the quantized field
\begin{equation}\label{eq:psit}
\ket{\Psi(t)} = \sum_{j=1}^N \alpha_j
e^{-i\omega_0t}\ket{e_j}\otimes \ket{0} + \sum_{\mathbf{k}}
\kappa_{\mathbf{k}} e^{-ickt} \ket{g} \otimes \ket{\mathbf{k}},
\end{equation}
where $\ket{\mathbf{k}}$ denotes the single photon state in the plane wave mode with wave number $\mathbf{k}$, $|g\rangle$ is shorthand for the collective state with all
atoms in the ground state, and where $\alpha_j$ and
$\kappa_{\mathbf{k}}$ are time-dependent expansion coefficients in
the interaction picture.
For simplicity, polarization
of the light and the atomic Zeeman sublevel structures
are omitted from this analysis, and we assume that the interaction between the atoms and the quantized radiation field is described by
$V_{\rm I} = \sum_{j=1}^N \sum_{\bf k} \hbar g_{\bf k} a_{\bf k}^\dagger |g\rangle \langle e| 
e^{-\imath{\bf k}\cdot{\bf r}_j} e^{\imath(ck-\omega_0)t}.$

 The Schr\"{o}dinger equation for the photon state amplitudes
$\kappa_\kbold$ can be formally solved in terms of the atomic amplitudes and substituted back in the Schr\"{o}dinger equation for the atomic amplitudes,
\begin{equation*}
\dot{\alpha}_j = - \sum_{j'=1}^N \sum_\kbold |g_\kbold|^2
e^{i\kbold\cdot(\rjbold-\mathbf{r}_{j'})} \int_0^t
e^{i(ck-\omega_0)(t'-t)}\alpha_j(t')dt'.
\end{equation*}

This expression is the starting point for the Weisskopf-Wigner approximation, which argues that the atomic amplitudes $\alpha_j(t')$ can be taken outside the integral over $t'$ and be simply evaluated at the time $t$, in which case, we only need to solve a simple set of first order differential equations. The number of equations is the same as the number of atoms, and hence the decay problem can be readily solved on a computer for up to several thousand atoms. Singular value decomposition (SVD) provides numerical eigenvalues and eigenvectors which parameterize a formal solution of the equations in terms of exponentially decaying and oscillating terms. The Schr\"odinger equations for the field amplitudes $\kappa_{\mathbf{k}}(t)$ contain the atomic amplitudes as source terms, and by formally integrating the exponentially evolving atomic amplitudes we thus obtain the asymptotic values $\kappa_{\mathbf{k}}(t)$ for $t$ much larger than the atomic excited state lifetime, as simple expressions parameterized by the SVD eigenvectors and eigenvalues.

In the case that the single collective atomic population resides in ground or metastable states coupled by a classical control field to the optically excited state, the atomic emission occurs by a spontaneous Raman transition and $2N$  coupled equations for the atomic amplitudes must be solved\cite{Poulsen2010}, and one subsequently gets the spatio-temporal field distribution in the same way as above.

\begin{figure}[!t]
\centering {\includegraphics[width=8.5cm]{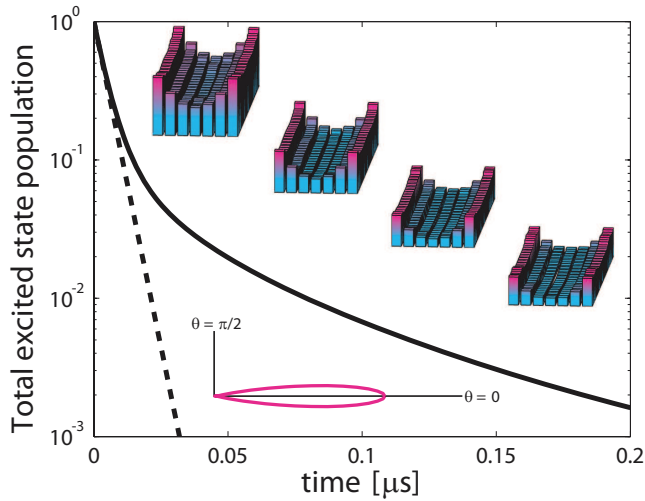}
\caption{(Color online) Excited state population in an $^{87}$Rb
sample with $7\times 7\times 20$ atoms (solid curve). The population
follows an exponential collective decay law (dashed line)
until $t=10^{-8}$ s, where the excitation is no longer uniform in the sample as illustrated by the excited state population in the four top layers of the sample, shown in
the upper part of the graph. The bottom insert shows the directional
photon density at $t=10^{-7}$ s.}\label{fig:excitedpop}}
\end{figure}

In the numerical
simulations presented in Fig. \ref{fig:excitedpop}, we have studied a cubic lattice with an elongated sample
of $7\times 7\times 20 = 980$ atoms. With a lattice spacing of 0.37 \mum
$,$ the maximum distance between any two atoms is 8.3 \mum, short
enough to achieve the Rydberg blockade for the preparation of the initial state and for later qubit manipulation. We use numbers
characteristic for \Rb87 and the 5$P_{1/2}$ excited state with a
spontaneous emission rate of $\gamma_1 = 37~\mu\rm s^{-1}$.

As demonstrated in Fig. \ref{fig:excitedpop}, the excited state population initially
decays as $\exp(-\gamma_{\rm col}t)$ (dashed line), where $\gamma_{\rm col}=5.7 \gamma_1$. The upper insets show the excited state population on individual atoms in the four upper layers of the ensemble at $t=10^{-8}$ s (the sample is mirror symmetric in the central plane of atoms shown as the lowest layer in the figure). As shown by the inset, at this time, the symmetry of the atomic excited state population in the sample is broken, and this explains the slower decay of the remaining few per cent of excitation in the system.

In the superradiant stage, light
emission occurs predominantly within a narrow emission cone. This is
illustrated in the lower inset of Fig. \ref{fig:excitedpop}, showing the total photon
emission probability as function of direction. With more than 95 \%
probability the photon is emitted in a direction within 0.3 radians
off the axis of the sample, which is in qualitative agreement with the estimate (\ref{eq.coop}).

One might worry that the loss or misplacement of a
few atoms from the sample would cause a significant change in the
field mode, and hence unrealistic demands on the ability to trap
atoms would have to be met to reliably produce high fidelity photonic qubits.
We have tested this concern by removing
tens of atoms from random locations in the system studied in Fig. \ref{fig:excitedpop}. We
have then solved the coupled atomic equations and computed the field mode emitted by the modified structure as described in the text and in each case determined the overlap of the resulting field with the one emitted
by the original sample. These overlaps are very robust and in excess
of 99 \% in all our simulations.

\subsection{Quantum communication protocols}

In the previous section, we have reviewed how a cloud of atoms, small enough to enable an efficient Rydberg blockade of the collective excitation, may serve as a directional single photon emitter. For quantum communication it is equally important to be able to receive information as it is to send it, and, indeed, a time reversal argument ensures that a field with a spatial dependence
which is the complex conjugate of the fields emitted by our sample found above will travel
in the opposite direction and converge upon the sample. Moreover, this time or motion reversal also applies to the solution of the coupled atom-light system, and the conjugate field hence becomes extinct by collectively exciting the atomic system. The small atomic cloud is an atomic storage medium for a single photon, if it is incident on the sample in a very particular mode. With controllable classical fields engaged in the Raman processes of light emission from one sample and absorption in another one, we get more degrees of freedom to tailor the single photon mode function so that it may be absorbed with high efficiency. The atomic excited state has finite lifetime, and the population should hence quickly be transferred coherently to a stable atomic state.

The use of much larger atomic ensembles for light storage, e.g., by electromagnetically induced transparency \cite{Fleischhauer2000,Liu2001} has capacity to store a much larger number of modes \cite{Nunn2008}, and the precise shape and arrival time of the weak probe field does not need to be specified for storage of a single incident pulse to work. Note, however, that this apparent robustness to imprecision and noise in the storage and retrieval of light pulses in large samples, does not remove the necessity to address precisely the spatio-temporal photon wave packet, if the coherent qubit space of the photon is to be fully explored.

\begin{figure}[!t]
\centering {\includegraphics[width=7cm]{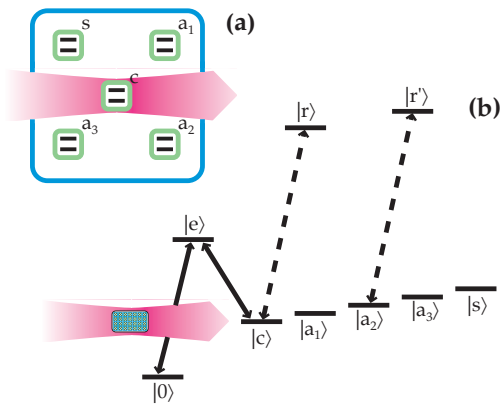}}
\caption{(Color online) (a) A five-qubit register consisting of a
communication qubit ($c$), a storage qubit ($s$) and three auxiliary
qubits ($a_{1,2,3}$) \cite{Gorshkov2008,Jiang2007}. (b) The collective encoding
implementation, with a collective internal state transition
interacting with the field mode, and long lived and Rydberg internal
states used for encoding and coupling of the five
qubits.}\label{fig:qubits}
\end{figure}

A storage fidelity in the range of 95 \% may be enough to demonstrate simple operations, but it is not enough to provide scalability in quantum computing or in long distance quantum communication. Here, however, we shall make use of the fact that this fidelity is obtained for a few hundred atoms within a 10 $\mu$m wide volume. An incident field state with a single  photon may thus be transferred to a state of this ensemble, which is an even superposition of states where each atom is in a stable state $|c\rangle$ while all other atoms are in another specific ground state. This is precisely a state of the form of the collectively encoded qubits described in Sec. \ref{sec.col_encoding}. Using the ensemble Rydberg blockade gates developed in Sec. \ref{sec.col_gates}, it is possible to perform entanglement pumping \cite{Dur2003} via two-bit gates from the information receiving qubit towards other bits in the register, cf. Fig. \ref{fig:qubits}.

 In \cite{Jiang2007}, it is shown that under the assumption of gates being possible between 4-5 auxiliary qubits, measurements and multiple rounds of communication can raise a $90 \%$ transmission fidelity to arbitrarily high degrees of entanglement between two samples. We refer the reader to \cite{Jiang2007} for the algorithmic details and simply note, that by using our collective encoding scheme with 5 internal levels as shown in Fig.\ref{fig:qubits}, we have access to the needed auxiliary qubits and the gates between them. The figure illustrates the five-qubit register designs, proposed in \cite{Jiang2007} (a), and the ensemble qubit proposal of \cite{Pedersen2009} (b). In part (a) of the figure, five separate physical systems take the role of a communication qubit,``c", three auxiliary qubits for temporary storage and entanglement pumping, ``$a_i, i=1,2,3$", and  a storage qubit for the perfected state, ``s". A chain of trapped ions with a single ion residing in an optical cavity for communication, or $^{13}$C atoms in the proximity of an optically addressable NV center in diamond are proposed in \cite{Jiang2007} as candidates for these five physical qubits.  In part b) of the figure is shown a generic single-atom level scheme for our collective encoding with a reservoir state, and five different long lived states playing the same roles as the five physical qubits in part a) of the figure, and an optically excited state and two Rydberg excited states, needed for optical interfacing and one- and two-bit operations, respectively. A thousand atoms should suffice to allow near-perfect one- and two-qubit gate operations, even when the population of the spectator qubits is not definitely zero or unity. We
emphasize that the collective encoding both yields the efficient
coupling to single photons and alleviates the need for addressing of
individual atoms.

\subsection{Hybrid qubit interfaces with Rydberg atoms}
\label{sec.hybrid}

The strong interactions between Rydberg excited atoms makes the Rydberg blockade mechanism an attractive mediator in hybrid proposals for quantum information processing.

In section \ref{sec.entanglement}, we saw several examples where a single Rydberg excitation was able to control collective states of an entire ensemble, and one can easily imagine how this mechanism can be used to interface effectively between the standard encoding scheme of qubits stored in individually addressable
atoms, as in Sec. \ref{sec.arrayscaling}, and the collective encoding in Sec. \ref{sec.col_encoding}. Similarly, an interface between continuous variables associated with the oscillator ladder of multiply excited ensemble states and the discrete qubit scenario may take its starting point in the Rydberg blockade interaction.

In this section we will briefly comment on the possibility to use hybrid physical systems to extend the advantages of the Rydberg mechanism to larger systems of particles. It may be desirable to extend the Rydberg interactions to longer distances, as many atoms can then be more easily addressed in experiments, and with more atoms a photon-atom interface may become more efficient due to the larger optical depth. Also, the dipole moment of a Rydberg atom is sufficiently large, that a single atom or a Rydberg blocked ensemble may provide an adequate strongly coupled non-linear component in various cavity QED set-ups. In \cite{Sorensen2004}, it was thus proposed to locate two Rydberg atoms in suitable vicinity of antenna surfaces and
use a superconducting transmission line to mediate the interaction between the atoms over $\rm cm$ distances. Stripline cavities have been produced and extensively studied in interaction with Cooper-pair boxes, \cite{Schoelkopf2004}, and it seems realistic to trap atoms in the vicinity of such striplines and approach the strong coupling cavity QED regime, both with a single atom, and with the collectively enhanced coupling of an ensemble of atoms to the cavity field.

The stripline cavities have their resonance frequencies in the GHz range \cite{Schoelkopf2004}, and they may be tuned to couple efficiently to transitions among different Rydberg states. In
\cite{Petrosyan2008}, it is thus proposed to use the cavity field to mediate long range interactions between individual atoms, and to obtain an effective blockade effect over samples large enough to provide very effective single photon  emitters and media for single photon quantum gates.
Alternatively a single Rydberg atom may be entangled with the cavity field as described in\cite{Saffman2009c}, and the single atom then entangled with an ensemble atomic qubit to mediate an
interface with photonic qubits.
The stripline cavities interact efficiently with the so-called transmon, Cooper-pair box qubits  \cite{Schoelkopf2004}, which however offer  lifetimes of only a few microseconds. Equivalent to similar proposals using samples of polar molecular \cite{Rabl2007}, it has thus been proposed to transfer the qubit between the transmon and the atomic ensemble via the cavity field \cite{Petrosyan2009} in a truly hybrid proposal making optimum use of the fast gate processing of one and the long storage times of the other component.

Making further use of external laser fields or static fields to address collective spin waves with different wave numbers
\begin{equation}
|1_k\rangle \equiv \frac{1}{\sqrt{K}} \sum_j e^{ikz_j}|0 ... 1_j ... 0\rangle,
\end{equation}
a single atomic ensemble is capable of hosting a large number of collective modes as previously proposed in molecular ensembles \cite{Tordrup2008} and in solid state spin ensembles \cite{Wesenberg2009}. In the collective encoding scheme for qubits each spatial spin wave mode encodes a qubit and the number of modes thus provides the size of the quantum registers. Note that the spin wave modes are only independent for a sufficiently large ensemble, and hence the cavity mediated long range blockade interaction is necessary for this encoding to be effective.

The trapping of cold neutral atoms in the vicinity of a superconducting cavity surely constitutes a major experimental challenge that will have to be solved \cite{Petrosyan2009,Saffman2009c}, but if a successful solution is found, the hybrid system with Rydberg atoms and resonant cavities holds great potential for scalable quantum computing.

\subsection{Rydberg atoms and alternative quantum computing paradigms}

We have in this review paper mainly focused on the possibilities to perform quantum computing with Rydberg excited atoms as described in the circuit model of quantum computation, \textit{i.e.},  quantum computation based on sequences of one- and two-bit operations applied to a register of memory qubits. The collective encoding scheme partly evades the common conception of separate logical qubits being stored in separate physical systems in the circuit model, but the collective modes are, indeed, well characterized degrees of freedom. As emphasized by our explicit presentation of the implementation of one- and two-bit gates, quantum computation with collectively encoded qubits evolves according to the usual circuit model apart from special means needed to counter errors.

Since the emergence of the first proposals for quantum computing other ideas for the implementation and use of quantum computers have appeared. It would take us too far afield  to present a detailed account of these ideas, but let us mention a few examples where the properties of the Rydberg excited atoms have been shown to be particularly interesting.
Cluster state, or one-way, quantum computers \cite{Raussendorf2001,Raussendorf2003} apply physical interactions to establish a multi-particle entangled state, which serves as the initial state for a sequence of one-bit measurements which suffices to produce a final state of the remaining bits of the same universal generality as the output register obtained by a quantum computation in the circuit model. Making use of the Rydberg blockade interaction within atomic ensembles and the ability for atomic ensembles to effectively absorb single photons, it has been proposed that the production of ensemble cluster states has a much higher probability of success than conventional schemes using light and atoms  \cite{Zwierz2009,Mei2009}. Using the few-bit per ensemble and distillation ideas \cite{Pedersen2009} described above, one may achieve an effectively deterministic protocol for cluster state preparation along the same lines, and one may consider tests with cluster states encoded collectively in a single atomic ensemble.

The one-way computer is based on measurements, and thus on a dissipative element. By engineering dissipative channels acting jointly on nearest neighbor qubits in a physical architecture, one can also provide the entanglement capabilities needed for universal quantum computing  \cite{Diehl2008,Verstraete2009}. One may think of the state prepared by this dissipation as a stationary, dark state of the dynamics, and for this state to be an entangled state, the dissipation must be engineered to act in a correlated manner on pairs or larger collections of qubits. In a recent proposal \cite{Weimer2010}, it is suggested to use the Rydberg interaction to engineer these correlated multi-bit dissipative terms by effectively transforming the desired quantum jump operators into the natural jumps, associated with decay of optically excited states of individual atoms in regular lattice structures.

An important process that has gained significant recent interest is the quantum random walk, where a physical parameter walks in opposite directions according to coin tosses, and where a quantum coin in a superposition state causes a ballistic rather than diffusive spreading of the walker caused by a classical random coin. The more rapid exploration of a wide range of parameter values may have implications for search problems, and quantum random walks have been proposed and even studied experimentally in several physical
systems\cite{Perets2008,Karski2009b,Schmitz2009} which may also be used for the usual circuit model of quantum computing. In \cite{Cote2006}, it is thus suggested to implement quantum random walks of a specific collective Rydberg state excitation between an array of atomic ensembles. Using the Rydberg blockade, the ensembles in a number of traps are first excited to have a single $ns$ excited Rydberg atom, except in one trap, which is initially excited to have a single $np$ Rydberg atom. The interaction permits a fast exchange of the excitation degree of freedom between nearest neighbor ensembles, and thus the $np$ excitation becomes the quantum walker, which may diffuse over the entire ensemble and eventually be detected, e.g., by field ionization. For a related study of excitation transport in an irregular trap array of atomic ensembles, see \cite{Mulken2007}.

Last, but not least, quantum simulators present a topic of current interest, with the goal to use engineered quantum systems to simulate complicated many-body physics and read out information, which may not be addressed in the real systems simulated. Quantum magnetism, black hole physics, lattice gauge theories, topological quantum field theories, superconductivity  and a host of phase transition phenomena are among the topics that are thus being addressed. The Rydberg interaction is particularly useful here because it is very strong, and hence allows simulation of steady state dynamics within realistic time scales, and because it is switchable in a way that allows  decomposition of a wide range of effective bipartite and multi-partite interaction operators \cite{Weimer2010}.
Quantum gases of dipole-dipole interacting particles present a major challenge for many-body theory and experiment  \cite{Lahaye2009}, and since polar molecules and Rydberg excited atoms experience this interaction and are experimentally available and may be detected with well-established atomic physics tools, they have been proposed as a test bench for the study of cluster and super-solid formation and non-classical crystal structures\cite{Olmos2010b,Pupillo2010,Pohl2010}, stabilized by quantum fluctuations \cite{Pupillo2009,Olmos2009b}.

\section{Summary and outlook}
\label{sec.conclusion}

As we have emphasized throughout this review there is currently a high level of activity devoted to exploring the use of strong atomic interactions mediated by transient excitation of Rydberg states for a variety of quantum information processing tasks. Approximately one decade after the initial proposal of a Rydberg quantum gate\cite{Jaksch2000}, the achievement of precise control of well localized
single atoms has resulted in demonstration of a two-qubit quantum gate\cite{Isenhower2010,Wilk2010}. At the other extreme collective effects in blockaded ensembles hold promise for multi-qubit registers, for robust light-atom quantum interfaces, and as a tool for simulating quantum many body physics.
The degree of control that has been achieved experimentally still lags behind theoretical expectations which suggest that very high fidelity quantum gates with errors well below $0.001$ are feasible.
Continued development of the requisite optical and laser systems, combined with improved control of the spatial and momentum distribution of single atoms will be important ingredients in ongoing work aimed at  approaching  theoretical limits.

Current experimental work is concentrated on alkali atoms. The alkalis are convenient
as regards the experimental requirements needed for cooling and trapping as well as Rydberg excitation. Single electron alkali atoms are also amenable to a detailed treatment of Rydberg state properties and  interactions using relatively straightforward theoretical tools. Looking to the future other parts of the periodic table
may take on increased importance. Two-electron alkaline earths have been widely used in quantum optics experiments in recent years with applications including  high precision optical clocks, and quantum gases. The  possibility of
incorporating Rydberg mediated effects into alkaline earth systems is a promising, yet unexplored direction.
The rare earths with their complex spectra, and large numbers of hyperfine ground states are another interesting direction which has so far only been touched on theoretically\cite{Saffman2008}. Rydberg interactions have also turned out to be important for creation of long range molecular dimers\cite{Farooqi2003,Bendkowsky2009,Overstreet2009,Stanojevic2006} which may prove useful as part of the current interest in applications of molecules to quantum information processing\cite{Rabl2007}.

A significant aspect of Rydberg mediated  interactions is that they bridge the gap between single atom qubits, with excellent coherence properties, and many atom ensembles, with poorer coherence, yet the potential for establishing a deterministic quantum interface between light and atoms\cite{Hammerer2010}.
Many signatures of Rydberg interactions have been studied in  atomic ensembles, yet a demonstration of controlled excitation of a single quantum in a many body setting with $N>2$ atoms has not yet been achieved. For many (but not all) potential ensemble applications, the number of atoms needs to be known with sub-Poissonian precision.  Thus non-destructive number measurements are important, and these may be challenging due to light-assisted collisions. Furthermore atomic collisions act to reduce the coherence time
of ensemble based qubits.


 It is likely that progress with ensemble based qubits  will be made by moving towards lattice based experiments where there is a natural
minimum barrier to the two-atom separations. The concept of collective encoding has many attractive features but to date has not been pursued experimentally.  Again, successful implementation of this approach seems to favor structuring the atom distribution with a sub-wavelength lattice, as suggested in \cite{Saffman2008}.

The predictions of universal scaling behavior, quantum critical points, and mesoscopic phases of dipolar ensembles\cite{Low2009,Weimer2008,Pupillo2009} are examples of direct realization of non-trivial many-body physics with Rydberg ensembles. Recent  proposals\cite{Weimer2010}  to use  Rydberg atoms for implementing
multi-partite interaction operators point to broad possibilities for simulating quantum many-body physics.
The potential for demonstrating controlled quantum dynamics  has been well established, which will promote continued interest in this rich field for the foreseeable future.

\section*{Acknowledgments}

M. S. and T. G. W.  would like to thank the talented group of experimentalists at UW Madison who have contributed to the demonstration of Rydberg quantum effects over the past many years.

This work was supported  by NSF grant PHY-0653408  and ARO/IARPA under contract W911NF-05-1-0492
and the European Union Integrated Project SCALA.



\end{document}